\documentclass[12pt]{article}
\usepackage{fancyhdr}
\usepackage{amssymb}
\usepackage{amsmath}
\usepackage{oldgerm}
\usepackage{latexsym}  
\usepackage{amsbsy}
\usepackage{float}
\usepackage[dvips]{graphicx}
\usepackage{epsfig}
\usepackage{subfigure}
\usepackage{wrapfig}
\def\slash#1{\setbox0=\hbox{$#1$}#1\hskip-\wd0\hbox to
\wd0 {\hss\sl/\/\hss}}
\def\nd{\noindent}
\begin{document}
\input epsf
\thispagestyle{empty}

\begin{center}{\Large\bf 
Physics of Extra Dimensions
\footnote{Based on a thesis submitted as a partial fulfillment for 
a PhD degree at 
the International 
School for Advanced Studies (SISSA--ISAS, Trieste, Italy), October 2001.}}
\end{center}
\centerline{ \sl
Higher-Dimensional Virtual Environments}

\vskip 1truecm
\centerline{\large Rula
Tabbash\footnote{Present address: {\it
Institute for Theoretical Physics, University of Lausanne, 
CH-1015 Lausanne, Switzerland}; e-mail: Rula.Tabbash@ipt.unil.ch . 
}}
\bigskip
\centerline{\it
SISSA--ISAS,
Via Beirut 4, I-34014 Trieste, Italy$^{3}$}

\vskip 1.2truecm

\tableofcontents
\section{Preface}

The idea of believing in and looking for a unique source to 
existence have been one of the oldest continuously posed questions in 
the history of man kind. The idea goes back to at least 
to the Sumero-Semitic syncretism of the 3rd millennium BC,  
when the process of identification (unification) of deities 
had already taken place before the majority
of records were written. The concept of having 
a mother-creator goddess, {\it Tiamat}, is documented 
in 
the Babylonian Creation story 
``{\it Enuma Elish}''
more than four thousand years ago.\footnote{A translation of 
{\it Enuma Elish} (before 2000 BC) can be found in \cite{history}.}

A search for simplification in understanding an 
obviously sophisticated and detailed phenomena happening in mother Nature
is, in many senses, a search for identification (unification), 
and vise 
versa. Many of the 
human species named under physicists believe in this equivalence.  

The first step towards a better ``understanding'' was 
by introducing the concept 
of a {\it cause} of an action, which was eventually 
translated by Newton 
(1665)
into the more concrete concept of {\it force}.\footnote{Thanks
to gauge theories, now we also know the cause of a force!.}
To date we are aware of the presence of four kind
of distinct forces in Nature
at very low energies; the electromagnetic, weak, strong, 
and gravitational 
forces. The attempt to unify 
these forces at higher energies (though not the only interesting problem 
of physics today) 
by deriving their
equations of motion from 
a single simple renormalizable action was 
pursued by realizing 
the covariance of their physical laws under certain symmetries. 
Those symmetries reduce the number
of free parameters in the theory and therefore enhance
its predictivity; 
this is one aspect of the simplification sought after. 

The present Standard Model (SM) of elementary particles 
(Abdus-Salam, Glashow, and Weinberg; Nobel Prize
1979)
succeeded 
very well in unifying the strong and electroweak
interactions  under
the gauge group $SU(3)_C\times SU(2)_L \times U(1)_Y$ at energy 
scales around $100$ GeV.

No understanding of a similar unification of gravity with 
the other forces at accessible energies to 
accelerators is achieved to date.
The only promising quantum theory
of gravity, so far, is superstring/M theory and its mathematical 
consistency requires the space-time to have ten/eleven dimensions. 
The low energy field theoretical description of string theory
can be used in order to explain our four-dimensional world
using the standard Kaluza-Klein compactification of the 
extra dimensions explained 
in the proceeding chapters.

Coming back to the SM; its main building blocks  
are: 
Action principle,  
Quantum Field Theory,  
Poincar\'{e} and gauge invariances,
and the justified
assumption of living in ($3+1$)-dimensional space-time. 

The community of modern theoretical particle physicists is clearly 
impressed by this model (for an 
obvious reason that its predictions fit the experimental data
with an impressive precision  
\cite{exp}), and believes the assumptions under
which it is made are to be valid at low energies and 
the concepts it is based on are to be generalized to theories 
where gravity is included. 


Despite the 
experimental success\footnote{In the recent past there has  
been dissatisfaction
from the experimental point of view, as well,
since the standard model
is in shortage of explaining, for instance,  
neutrino oscillations \cite{Kajita:1999bw} and the 
recent data of $g-2$ of the Muon \cite{Brown:2001mg}.
Furthermore, there is no hint for the 
existence of an elementary scalar field in Nature, while
the Higgs sector is very essential in the SM (see
\cite{Carena:2000yx} 
for possible future detection of the standard model
Higgs particle).} 
of the standard model as 
a theory describing the strong and electroweak interactions of 
elementary particles, this model 
is not theoretically satisfactory 
for various reasons.  
There are two main sources of 
theoretical
dissatisfaction; the first has to do with the model itself (when 
it comes to explaining, flavor, and charge quantization, for example), 
and
the second arises 
when the standard model is discussed 
within a more general context where the 
fourth fundamental force of nature, gravity, is present; 
leading subsequently to an instability of the 
weak scale at the quantum level (caused by 
the quadratic divergences in the Higgs (mass)$^2$). 

Therefore, 
there are obvious urges to go beyond the standard model: the 
need to explain neutrino oscillations (for instance), 
quantum instability of the electroweak scale, 
and the hope for a quantum theory of gravity to unify with the 
standard model (or its extension). 
Further simplification is yet to be 
worked for.

The electroweak quantum instability, known as the 
{\it hierarchy problem} between the electroweak and gravity scales, 
was the main motivation to start searching 
for new physics beyond the standard model. 
These searches lead to 
the birth of many hoping-to-be-physical
\footnote{I mean by
a ``physical'' theory here the one which is able to 
fit with present experimental data, to reproduce all what the 
standard model is able to reproduce at energies $\sim 100$GeV,
and not to contradict with the standard cosmological 
scenario after Big Bang Nucleosynthesis. In addition, it should
have some extra predictive power to SM preferably testable 
at LHC.}
theories like technicolor \cite{Lane:2000pa}, 
grand unified theories \cite{Kounnas:1984cj}, supersymmetry 
\cite{Nilles:1984ge}, and recently models with 
large 
extra dimensions \cite{add1}.

By coincidence,\footnote{It may be 
realized in the future that it is not 
a mere coincidence. Who knows?.} 
also the most promising theory to quantize gravity
and unify it with the other gauge forces
(through superstring/M theory) seem to require both 
extra dimensions, beyond the known four, and supersymmetry 
as crucial ingredients for its consistency. 

This strongly hints to that,  
hoping for a unique theory, the hierarchy 
problem may be solved in a theory  
in higher dimensions than $4$ with broken supersymmetry. 


This thesis concerns itself to a good extent 
with solving the hierarchy problem within the 
context of extra dimensions.

In string theory, the extra six-dimensional space 
is squeezed and pressed 
(or {\it compactified}) into a manifold
of a tiny volume. The  
original compactification scale of this 
theory is of order $M_P^{-1}$ which is extremely small 
and it would be impossible for any machine to detect 
a modification of the gravitational law at such energies. 
With such a scale, the world will definitely appear four-dimensional 
without any hint to the presence of dimensions beyond the 
known four. Inspired by string theory, though not as mathematically
rigorous, a recent interest in extra-dimensional 
models has been 
revived 
by Dvali {\it et al} \cite{add1,add3},\footnote{
Credits of 
associating a physical meaning 
to extra dimensions prior to \cite{add1} go to 
\cite{sch}-\cite{antoniadis}.} 
who pointed out
that 
the modification to Newton's low by introducing large 
extra dimensions could be a valid possibility since no 
tests of gravity has been carried out to distances much 
below 1 millimeter (see for instance \cite{price,Long:1999dk,kap3}). 

The activities so far fall into two categories:
models as \cite{add1} 
based on the original idea of Kaluza and 
Klein \cite{kaluza,klein1}
which consider a tensor product (factorizable metric) of the 
four-dimensional world with the compact internal space; 
and alternatives to compactification
which consider non-compact extra dimensions,
\cite{Wetterich:1984uc},
with {\it warped} (non-factorizable) metric  
as in  
\cite{sh2,Randall:1999ee}.
The new proposals were all aimed at a solution to the hierarchy 
problem by lowering the gravity scale 
from $10^{19}$GeV 
down to few TeV. 
In Kaluza-Klein (KK) 
like models, the particles are free to propagate inside
the 
internal compact 
space which should be small enough not to 
lead to phenomena contradicting the present knowledge of particle
physics and big bang nucleosynthesis \cite{Groom:2000in}. 
There is an infinite number
of images of the internal space at each point of 
the four-dimensional world.
When the extra dimensions
are non-compact, the only way to avoid 
long-range observable effects is
to localize 
the fields on a thin three-dimensional wall (brane).
In this case, there will not be infinite images of the brane world, 
and in 
principle one can have only one single brane on which ordinary 
matter is localized.\footnote{A special case of only one
image of the 4-dimensional space-time
can also be considered,
within the context of Kaluza-Klein compactification,  
if ordinary matter fields are forced to be 
localized around a specific point in the internal space 
\cite{add1}. We will not discuss this possibility here.}

In both categories, it turns out that lowering
the gravity, or the fundamental, scale of the theory 
does not by itself solve the problem. It rather
addresses it in a different
way as it introduces other sources of fine-tuning into the 
theory like for instance  
explaining the size of the manifold made by
the extra dimensions or the mechanism 
which traps 
the matter fields and gauge forces 
on a brane. Let alone that there is again 
a huge difference between few TeV and $M_P$. However, as we shall 
explain later, the hierarchy problem is not having a low 
energy cutoff close to the weak scale, but rather having no
quadratic or significant quantum correction to the classically
computed physical quantities. 

There are generic difficulties 
which face models with extra dimensions
as upon compactifying down to four dimensions one may 
in general get new degrees of freedom added to the 
spectrum of the Standard 
Model. The new states can be purely from the gravitational
sector, or have Standard Model Kaluza-Klein 
excitations in addition (depending
on
whether the SM interactions are written directly in four dimensions,
using the induced metric, or written fully in $D$ dimensions).
In any case, the new states
might 
lead to detectable modifications of 
the existing accelerator data and cosmological observations
\cite{add2}. This leads to  
imposing 
judicious bounds on the 
parameters of these theories. Whether these
bounds are
implemented
or not,
the theories with large extra 
dimensions experience difficulties
in realizing complementary scenarios
like the standard cosmological one. 
For instance, 
imposing an upper bound on the reheating temperature 
in order to avoid 
overproduction of Kaluza-Klein modes of the graviton, and 
discrete symmetries in order to prevent a fast proton decay make
it difficult to construct a
baryogenesis model. In addition, recovering 
the standard Friedmann-Robertson-Walker Universe in 
$4$ dimensions starting from higher-dimensional 
Einstein's equations is not straightforward 
whether the compactification is standard or warped.


In both factorizable geometry and non, it is important that
the internal space has no effects interfering
with the SM precision tests.
This of course does not happen naturally, and, 
as mentioned above, bounds on 
the parameters of the model should be 
imposed.

So far no full
and consistent
model has been constructed,
neither in tensor compactification nor 
in the warp one, however the phenomenological and cosmological 
implications have been studied extensively,
and attempts towards creating a theoretically appealing model 
persist.   
In this report we will review the tools
and basic concepts in constructing Kaluza-Klein 
theories and 
Brane world scenarios. 

\subsection*{Plan of the thesis work}

This article is organized as follows: In chapter 
\ref{kkp} 
an introduction to the Kaluza-Klein theory
is presented
including the very basic concepts on which the Kaluza-Klein 
picture and Brane-world scenarios are based. 
It is far from being
a complete review, neither in 
topics covered nor in list of references,
however it is meant to serve as
a background material mainly for Chapter
\ref{hssbch}.

In 
Chapter \ref{wned}, the idea of warped extra dimensions
and localization of matter fields on 
a hyperspace in $4+d$ dimensions 
is reviewed with the help of some examples taken from 
the literature.  

Chapter \ref{lwed}
discusses some of the physical implications
of theories with extra dimensions, mainly the 
modification of the Cosmological evolution of 
the Universe and the fast proton decay. 
The last part of this chapter contains 
an example of generating baryon 
number violation in theories with extra dimensions and 
low gravity scales.

Chapter \ref{hssbch}
is devoted to 
a solution to the hierarchy problem, where an understanding
of a single source responsible 
for both electroweak symmetry breaking and
standard model fermion chirality is provided.   

Unfortunately, it was not possible to refer to 
all the interesting and important papers 
written in this subject due to the limitation of space 
and time allowed for writing and submitting
this dissertation. 
Therefore, only
``samples'' of such contributions to the literature
could be provided.

\section{Kaluza-Klein Picture}
\label{kkp}





\subsection{Historical remark}

Although having seven heavens is not a particularly new idea, 
the first  
to be published in a scientific journal, proposing our observed world to 
be an effective theory of a fundamental theory existing in more
than four dimensions,
was 
Nordstr\"{o}m's \cite{nord} back in 1914. Having no general relativity 
at that time, 
Gunnar
Nordstr\"{o}m wrote down Maxwell's equations in 5
dimensional space-time, and by wrapping the fifth dimension on a 
circle, he reduced the equations to Maxwell-Nordstr\"{o}m 
electromagnetic-gravitational theory in 4 dimensions. 

In 1918, Hermann Weyl introduced the concept of gauge 
invariance \cite{weyl}
in the first attempt   
to unify electromagnetism and gravitation in a geometric
context.  
 
Three years later, in 1921, the mathematician 
Theodor
Kaluza
proposed \cite{kaluza} obtaining a four dimensional Einstein-Maxwell
theory starting from Einstein's gravity 
equations in five dimensions. Assuming the five-dimensional manifold,
$W$,
to 
be a product of a 4-dimensional space-time $M_4$
and a circle $S^1$, $W=M_4\times S^1$, 
the fifteen components of the metric can be decomposed 
from 
a
four-dimensional point of view 
into 
10
describing the 
gravity tensor,
four forming the components of a $U(1)$
gauge field, and one degree of freedom representing a scalar
field. By Fourier expanding those fields, retaining 
only the zero 
modes, 
and integrating along the circle $S^1$ one obtains
a theory in four dimensions which is invariant 
under both 
four-dimensional 
general coordinate transformations, 
and a $U(1)$ gauge transformations.

In his original work, Kaluza assumed the zero mode of the 
scalar field to be constant, $
\phi^0=1
$. In any case, the value of $\phi^0$ had to be positive in order to insure
the proper relative sign of the Einstein and Maxwell terms so that the 
energy is positive. This, in turns, means that 
the fifth dimension must be space like. This can also be easily understood
in terms of causality; clearly a compact time-like dimension 
would lead to closed time-like curves.
The abelian gauge symmetry arising in four dimensions upon 
compactification originates
from the isometry of the circle. Those last two point, the requirement that
the extra dimensions must be space-like, and that the 
isometry of the compact space results in a gauge symmetry (generally
non-abelian) of the 
effective action are general arguments \cite{dewitt}. 

Seven years later, Oskar Klein used Kaluza's idea in an 
attempt 
\cite{klein1}
to explain the underlying quantum mechanics of Schr\"{o}dinger 
equation by deriving it from a five-dimensional space-time
in which the Planck constant is introduced in connection with 
the periodicity along the closed fifth dimension. In this paper he 
also discusses the size of the compactified circle, getting 
closer to giving the extra dimensions a physical meaning than his 
predecessors did.
In a separate 
work \cite{klein2}, still in 1926, Klein proposed\footnote{Credit
to this approach goes also to 
V. Fock and
H. Mandel, though others worked for 
and achieved
the same aim 
as well
\cite{kk}.}
a relativistic generalization of Schr\"{o}dinger's 
equation by starting from a massless wave equation in 
five dimensions and arriving at four-dimensional Klein-Gordon equation
for individual harmonics.

Afterwards,
many people adopted Klauza's idea, starting from Einstein early last 
century and continuing 
by several physicists today. During this period, the idea of having
new dimensions to propagate in inspired many to write the first 
complete
models for 
Lagrangians unifying Yang-Mills and gravity theories
\cite{kerner}, supergravity in 11 dimensions 
\cite{Cremmer:1978km}, and superstring theories which has to be 
considered in 10 dimensions for the theory to be 
anomaly free and hence consistent at the quantum level. 

Supergravity in eleven dimensions is special for at least three 
reasons: 
it is unique, the maximum number
of additional dimensions to four
on which a supergravity theory can be used to construct
is seven (otherwise the theory 
would contain massless particles of spin greater than two 
\cite{Nahm:1978tg}), 
and moreover seven extra dimensions is the minimum 
number for 
having the standard model gauge group $SU(3)\times SU(2)\times U(1)$
generated 
from the isometry group of the internal space (which could, 
for instance, be the so called Witten space 
$\mathbb{C}P^2\times S^2\times S^1$).

In 1975, Sherk and Schwarz \cite{sch} attempted to 
put string theory (which is 
consistent only in ten dimensions) in contact with 
the four-dimensional world by assuming that the 
six extra dimensions are curled up into a tiny small size which 
renders them unobserved. 
An alternative to compactification was introduced 
in 1983 by Rubakov and Shaposhnikov,  
\cite{Rubakov:1983bz}, and others 
\cite{Wetterich:1984uc,wit,nie,wett2}.
The idea   
of considering non-compact internal  
spaces 
was used by Wetterich in \cite{Wetterich:1984uc} 
to propose
a possible solution to the chirality problem
\cite{wett3,Witten:ed.ux} in the context 
of Kaluza-Klein theories.

\subsection{Introduction}

In the original Kaluza-Klein framework, the particles are free 
to move in the compact space, as well as the 4-dimensional
one, and hence the volume of the new dimensions should be small
in order not to undesirably interfere with the present 
observations and existing experimental data, since the new 
idea of $4+d$ dimensions has a strong potential to naturally
modify the expansion of the Universe and the cross sections of 
elementary particle interactions. 
\begin{figure}
\centerline{\leavevmode\epsfysize=8cm \epsfbox{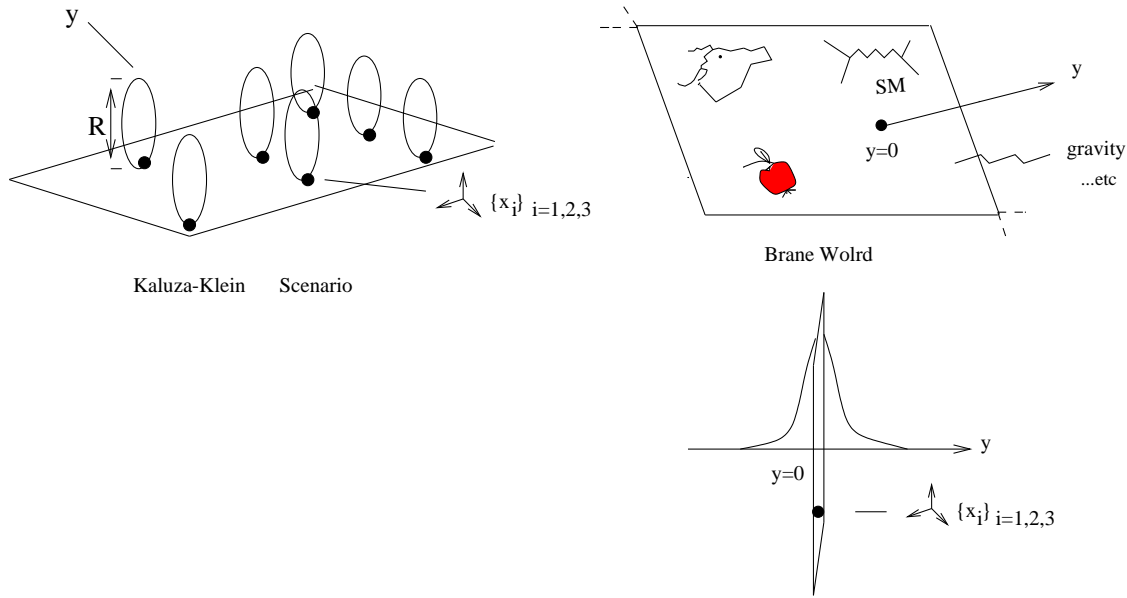}}
\vspace{10pt}
\caption{\small
Kaluza-Klein large extra dimensions versus Brane world.}
\end{figure}
Upon compactification,
the eigenvalues of the Laplacian (or the Dirac operator) 
on the internal space 
will have the interpretation of masses in four dimensions.
An illustrative example is the case of a product of a 
four-dimensional Minkowski space and a circle $M_4\times S^1$.
The wave function of a scalar field can be expanded in Fourier 
series along $S^1$,
$$
\phi(x,y)=\sum_{n\in \mathbb{Z}} \mbox{e}^{in\frac{y}{R}} 
\phi_n(x) 
\;,$$  
where $y$ is the coordinate on $S^1$, of radius $R$,
and $n$ is the eigenvalue of the one-dimensional angular 
momentum operator. 
The Klein-Gordon equation of a massless $\phi$ can therefore
be written as
$$
p^\mu p_\mu =-p_0^2+{\underline{p}}^2=-\frac{n^2}{R^2}  
\;,$$ 
where $p_\mu$ is the 4-dimensional momentum. Clearly, 
from a $M_4$ point of view, each 
Kaluza-Klein mode $\phi_n(x)$ is seen as a separate
particle with mass$^2$ equal to $n^2/R^2$. Below 
energy scales $1/R$, only massless modes with $n=0$ can 
be excited and hence the low energy physics is effectively 
four-dimensional. This is just a na\"\i ve conclusion, 
and we will see later on that there are 
various bounds and conditions 
which apply to realistic Kaluza-Klein type scenarios
in 
order to indeed achieve phenomenologically 
acceptable models in four dimensions.

As discussed in the introduction, 
the modern interest in having extra dimensions is 
more concerned with finding a solution to the hierarchy 
problem than searching for detectable hints 
to string theory at low scales. If the volume 
of the compactified extra space is characterized by 
a scale larger than $M_p^{-1}$, the ($4+d$)-dimensional
gravity  
scale, will be automatically
lowered, could even be few TeV, 
and the difference between the 
fundamental gravity and electroweak scales 
will be smaller.   
As usual, the 4-dimensional Planck scale, 
$M_P=10^{19}$GeV, is related 
to the fundamental gravity scale,
$M$, of the ($4+d$)-dimensional space
via the volume, $V_d$, 
of the internal space.
Having the gravity action in $4+d$ dimensions and integrating
over the coordinates of the internal space, 
\begin{equation}
S=\frac{1}{\kappa^2}\int d^{4+d}z \sqrt{-g(z)}
{\cal R}\Rightarrow 
S_{\mbox{eff}}=\frac{V_d}{\kappa^2}\int d^4x \sqrt{-g_4(x)}{\cal
R}_4
\;,\label{saction}
\end{equation}
where $\kappa^2=M^{-d-2}$ and $\cal R$ is the ($4+d$)-dimensional 
Ricci 
scalar, we get 
\begin{equation}
M_{P}^2=V_d M^{d+2}\;.
\label{mass1}
\end{equation}
$V_d$
is 
usually proportional to the radius of compactification, 
$ V_d\propto R^d$.\footnote{
This relation is not generic to all manifolds, although it is
very common. For example, for a compact hyperbolic manifold,
there will be an additional exponential dependence of the volume 
on the radius of compactification.}  

Having a fundamental gravity scale in $4+d$ dimensions
much different from $M_P$ means, according to (\ref{saction}), 
that Newton's 
low for gravity should be modified
at small distances $r\ll R$ to be 
$$
V(r)=\frac{m_1m_2}{M^{d+2}}\frac{1}{r^{1+d}}\;,
$$
where for larger distances, $r\gg R$, the
potential recovers its usual 
$1/r$ dependence
$$
V(r)=\frac{m_1m_2}{M^{d+2}R^d}\frac{1}{r}\;.
$$
As can be deduced from (\ref{mass1}), or equivalently
$$
R= 10^{-17+\frac{30}{n}} (\frac{1\mbox{TeV}}{M})^{1+\frac{2}{n}}
\;\mbox{cm}\; ,$$
achieving $M\sim 1$TeV
requires assuming
bigger compactification radii than the ones used in string theory, 
{i.e.} $R\gg 10^{-19}$GeV$^{-1}$. 
Having lowered the fundamental gravity scale $M$ to few a TeV, 
the gauge hierarchy problem
will be re-addressed as the problem 
explaining the size of
the internal space. 

For $M\sim $TeV, this radius can be as big as 
$10^{13}$GeV$^{-1}$ for two extra dimensions, 
$10^{8}$GeV$^{-1}$ for three, $10^{5}$GeV$^{-1}$
for four, and so on. However, the radius should 
not exceed the millimeter since we do have tests of 
Newton's low up to this scale \cite{Long:1999dk,kap3}. 
The case of one extra 
dimensions for a Kaluza-Klein type scenario is 
excluded for low $M$ since it requires the compactification
radius to be very big, $10^9$Km!.  
This problem becomes milder as 
the number of extra dimensions increases, however 
it may be desirable to avoid this shortcoming
even for a small number of them.
The primary problem is not explaining the smallness of these radii, 
which is obviously a fine-tuning problem, 
but rather avoiding their undesirable contributions to well 
studied observables (the masses of the KK excitation can be 
very light). 

As said above, the 
Kaluza-Klein idea is that there is a multi-dimensional 
world of the 
low energy physics which could be responsible for the 
internal symmetries that we observe. The scale of this structure
is assumed to be small enough to render it unobservable 
at present energies; below $10^3$GeV. 
The motivation is to obtain a geometric
interpretation of internal quantum numbers, as electric charge, 
and to consider them conceptually the same as the energy and 
momentum.  

In these theories the starting point is a ($4+d$)-dimensional 
space-time and the assumption of general 
coordinate invariance, in other words 
adopting $(4+d)$-dimensional 
Ricci scalar
with  
perhaps some covariant coupling to ``matter'',  
as the Lagrangian.

One argues then that the ground state values of the fields 
(including the metric tensor) must solve the classical 
field equations and must have as much symmetries as 
possible.\footnote{
Since we know from experience that 
more symmetrical states tend to have lower energies.} 
The goal is to find a suitable ground state configuration. 
This ground state should hence have the geometry of 
a product space $M_4\times K$, where $M_4$ is a 
$4$-dimensional space of a Minkowskian signature, 
and $K$ is a $d$-dimensional compact space of 
an Euclidean signature 
and a small size. Both $M_4$ and $K$ should admit groups of 
motion. 

In the case of $M_4$ being a Minkowski space, its group of 
motion is the Poincar\'{e} group in four dimensions. 
The internal space $K$ 
should admit a compact group $G$. 

The masses of the bosonic particle (or fermionic)
will be determined by the eigenvalues of the
Laplacian (or Dirac) operator on $K$, since 
the operator can be written as a 
direct sum of the operators on $M_4$ and $K$.    

The zero modes of those operators, if any, would correspond to 
massless particles and these are important from the 
four-dimensional point of view since  
they will represent the low energy effective Lagrangian. 
The modes (excitations) 
which involve propagation on $K$ will generally have 
large masses, proportional to the scale to which the size of $K$
corresponds.

All excitations of any particle will be classified
in irreducible representations of the ground state symmetry. 
If this symmetry is smaller than the Lagrangian symmetry, we
end up with spontaneous symmetry breaking. 

Finding the eigenvalues of the Laplacian $\nabla^2$ 
(or those of 
$\slash{D}$) 
is not a straightforward problem. In fact, 
there is no analytic expression for the eigenvalues of 
these two operators 
on a generic compact manifold (although lower and upper
bounds on these values exist for a large number of them, 
see \cite{avijit}).
However, things can be made simpler when a specific choice
for $K$
is considered. 

In this chapter the basic concepts on which the 
Kaluza-Klein idea is based 
are briefly reviewed. Topics cover
harmonic expansion on coset spaces, effective 
Lagrangians, 
stable compactification, and the 
issue of chiral fermions.
The chapter is concluded with 
a remark on possible geometric estimates for
masses of Kaluza-Klein fields on a  
generic compact space.

\subsection{Coset spaces}
\label{KK}

Here we will focus on the most symmetrical 
category of compact manifolds, the coset spaces $G/H$,
\footnote{Since, 
after all, the purpose of Kaluza-Klein picture is 
to explain internal symmetries in terms of the 
symmetry of the compactified space.} where 
$G$ is the group (here a Lie group) of motion of $K$
(which leave the metric invariant) and $H$ is a subgroup 
of $O(d)$ (the tangent space 
group) \cite{strathdee}.An example of a coset
space $G/H$ is $S^2=SU(2)/U(1)$. The 
Kaluza-Klein reduction 
procedure for a generic manifold has been worked out
in 
\cite{percacci}-\cite{sch3}.

Let $G$ be generated by 
$$[Q_{\hat a},Q_{\hat b}]= 
{C_{{\hat a}{\hat b}}}^{\hat c} Q_{{\hat c}}\; ,$$
where 
$
Q_{\hat a}= \{ Q_{\bar a},  Q_{a}
\}
$, $Q_{\bar a}$ 
being the 
generators
 of $H$. Note that,

\nd
$d\equiv\mbox{dim}(G/H) = \mbox{dim}G-\mbox{dim}H$.

We assume that $G/H$ is reductive, {\it i.e.}
$$
[Q_{\bar a},Q_{b}]= 
{C_{{\bar a}{b}}}^{c} Q_{c}
\; ,$$
\nd
and symmetric;
$$
[Q_{a},Q_{b}]= 
{C_{{a}{b}}}^{\bar c} Q_{\bar c}
\;.$$
Let $y^i$ parameterize the coordinates
on $K=G/H$, and let $L_y$ be the representative of each 
equivalence class (coset) of $G$ 
with respect to its subgroup $H$ (for example 
one can chose$L_y=e^{y^a Q_a}$). 

A generic transformation by $G$ carries $L_y$
into $L_{y'}$ 
$$
gL_y=L_{y'}h\;\;\;,\;\;\;h\in H
\;.$$
This equation can be solved unambiguously for $y'$
and $h$ as a function of $y$ and $g$.

To find the $d$-bein on $K$, let us construct the 1-form $L_y^{-1}dL_y$
which necessarily belongs to the Lie algebra of $G$ and hence 
can be expanded as
$$
L_y^{-1}dL_y = e^{a}(y)Q_a+ e^{\bar a}(y)Q_{\bar a}
\; .$$
Knowing how $L_y$ transforms under $G$, one can 
derive, in a straight forward way,
the 
the transformation rule for 
$e^{a}(y)$ and show that they 
are the vielbeins on $G/H$ and that   
$e^{\bar a}(y)$ are the connection form on $H$.
They are \cite{salam}
\begin{eqnarray}
e^{a}(y')&=& e^{b}(y)
{D_{a}}^{a}(h^{-1})+ (g^{-1}dg)^{\hat b}
{D_{\hat b}}^{a}(L_y h^{-1}) 
\label{form1}\\
e^{\bar a}(y')
&=& e^{\bar b}(y)
{D_{\bar b}}^{\bar a}
(h^{-1})+ (hdh^{-1})^{\bar a}\nonumber\\
&+&
(g^{-1}dg)^{\hat b}
{D_{\hat b}}^{\bar a}(L_y h^{-1}) 
\;.
\label{form2}
\end{eqnarray}
Firstly, note that $d_yg=0$, and hence $e^{a}(y)$
satisfies the requirement 
for an invariant $d$-bein.
Later on we shall show that the $x$-dependent $g$-transformations
of $G/H$ will give rise to the 
Kaluza-Klein local gauge transformations. For the time being, 
we will assume that $g$ is independent of coordinates. 
The equation \ref{form1} indicates an ordinary tangent space 
rotation of the $e^a$'s, provided $H$ is embedded in 
$SO(d)$, the tangent space group of $G/H$.  

In Kaluza-Klein theory, $g$ can depend on the coordinates 
of $M_4$, and therefore generally $d_xg\neq 0$. To understand
the second term in (\ref{form1}) one has to take into account
the fluctuations of the ground state. Among these there will be 
a Yang-Mills vector, $A^{\hat a}$, which 
undergoes the usual inhomogeneous transformations
of a gauge potential. It turns out that 
the second term in (\ref{form1}) is canceled 
precisely by the inhomogeneous 
term in the transformations
of $A^{\hat a}$ \cite{strathdee} (see the next section). 
\footnote{The metric on $K$ can be 
expressed as usual in terms of the $d$-beins,
$g_{mn}(y)=e_m^a (y) e_n^a (y)\; 
(n,m; a=1,...d)$.
The $d$-beins form a set of linearly independent 
1-forms
$e^a(y) = dy^m e_m^a (y)$. 
When the space admits a group of motion $G$ (leaving $g_{mn}$
invariant), 
to each element $g\in G$ 
there corresponds a transformation 
$y \rightarrow y'$ such that 
$e^{a}(y')
= e^{b}(y)
{D_{b}}^{a}$ where $D\in O(d)$.}

Similarly, the last term in (\ref{form2}) is canceled by the transformations
of $A^{\bar a}$, and  
hence (\ref{form2}) indicates that $e^{\bar a}$
is a connection form on $G/H$. 

In many cases of interest, the ($4+d$)-dimensional 
metric is coupled to a gauge field associated with 
a group $U$. When it is possible to embed $H$ into 
$U$, 
this gauge fields generally 
acquire non zero, but $G$ invariant, values on $G/H$. 

Let $q_{\bar c}$ denote the image of 
$Q_{\bar c}$ in the Lie algebra of $U$. Then on can construct
the 1-form potential 
$$
A= \frac{1}{\rho}e^{\bar c} q_{\bar c}
\;.$$

It can be shown that its associated field strength, 
$$
F =\frac{1}{2\rho} e^a \wedge e^b 
{C_{a b}}^{\bar c}  q_{\bar c}
\;,$$
\nd
satisfied the Yang-Mills equation \cite{percacci}.
The 2-form $F$ is invariant under
the left translations of $G/H$
associated with 
the gauge transformations corresponding to the embedding 
of $H$ in $U$.\\

\nd
{\bf Example: 
$S^4=SO(5)/SO(4)$
}\\

Let $S^4$ be parametrized by coordinates 
\begin{eqnarray}
u^m(y)&=&a\frac{2 y^m}{1+y^2}
\;\;\;;\;\;\;m=1,2,3,4\nonumber\\
u^5(y)&=&a\frac{1-y^2}{1+y^2}\nonumber\;,
\end{eqnarray}
where $y^2=(y^1)^2+...+(y^4)^2$, $u^mu^m+u^5u^5=a^2$.
 
One can chose $L_y$, the $5\times 5$ orthogonal matrix to be
\cite{Randjbar-Daemi:1984ij}
\begin{displaymath}
L_y=
\left(
\begin{array}{cc}
\delta_{m n}-\frac{2y^ny^m}{1+y^2} & 
\frac{2 y^m}{1+y^2}\\
-\frac{2 y^m}{1+y^2} & 
\frac{1-y^2}{1+y^2}
\end{array}
\right)\;.
\end{displaymath}
Now we can construct $L_y^{-1}dL_y$, which is a $5\times 5$
antisymmetric matrix
in the Lie
algebra of $SO(5)$ by decomposing it into a convenient 
basis. Chose $\{Q_{b5},Q_{bc}\},\; b,c=1,...,4$ such that 
$\{Q_{bc}\}$ are the generators of $SO(4)$. Then 
$$
L_y^{-1}dL_y= e^bQ_{b5}+\frac{1}{2}A^{bc}Q_{bc}\;,$$
where 
\begin{equation}
e^b=\frac{2a}{1+y^2}dy^b\;,
\label{viel2}
\end{equation}
\begin{equation}
A^{bc}=y^{[b}e^{c]}\;.
\label{inst}
\end{equation}
The $SO(5)$ invariant metric of $S^4$ can be 
derived from (\ref{viel2}) by identifying the $4$-beins
${e_m}^b$ as the coefficients of $dy^b$. It is
$$
g_{mn}=\frac{4a^2 }{(1+y^2)^2}\delta_{mn}
\;.$$

It is worthwhile to note that projecting
the self-dual part of $A^{bc}$ in (\ref{inst}) yields to 
the $SU(2)$ instanton configuration
\cite{Randjbar-Daemi:1984ij}
\begin{equation}
A^i=-\frac{1}{2}\eta^i_{bc}A^{bc}=
-\frac{2\eta^i_{bm}y^b}{1+y^2}dy^m\;\;\;,\;\;\;i=1,2,3
\label{instanton}\;,
\end{equation}
$\eta^i_{bc}$ being the 't Hooft symbols \cite{tooft}.
$A^i$ is an instanton on an $S^4$ of radius 
$a$. This instanton can be shown to be a solution of the 
Einstein-Yang-Mills equations of motion in a background
metric $M_4\times S^4$ \cite{Randjbar-Daemi:1984ij}. 
We will be using examples of similar instanton/monopole
backgrounds later on.

\subsection{Harmonic expansion}
\label{harmon}

A generic function 
$\phi_{j}(g)$ on $G$ which belongs to 
a unitary irreducible representation of it
can be represented by the expansion
\begin{equation}
\phi_j(g)=\sum_{n}\sum_{p,q}\sqrt{d_n}D^{n}_{jpq}(g)\phi^n_{qp}
\label{expand}\;.
\end{equation}
\nd
$D^n_{pq}(g)$ is a unitary matrix of dimension
$d_n$, and the sum is over all irreducible 
representations of $G$.\footnote{This is a generalized form 
of Fourier transform.} 

Now for functions on a coset space $G/H$, the expansion
is less general, and hence
subject to some restrictions. The concern here is with 
multiplet of functions, $\phi_i(g)$, which have the property
$$
\phi_i(hg)= {{\mathbb{D}}_i}^j(h) \phi_j(g)
\;,$$
\nd
where $h\in H$ and $ {\mathbb{D}} $
is some definite representation of $H$. 
So, for the case of $G/H$, only the terms in (\ref{expand}) 
which satisfy 
$$
D^n(hg)={\mathbb{D}}(h)D^n(g)
\;$$
\nd
should be included. In other words, one should keep
those 
representations of $G$ which upon restriction to $H$
they 
produce the representation $\mathbb{D}$ of $H$. 
Moreover, in Kaluza-Klein theories, the coefficients of expansion
$\phi^n_{qp}$ generally
depend on the coordinates of $M_4$. 

The formula for expanding a function 
on $G/H$ is hence 
\cite{strathdee,salam} 
$$\phi_j(x,y)=\sum_{n}\sum_{p,q}\sqrt{\frac{d_n}{d_{\mathbb{D}}}}
D^{n}_{jpq}(L_y^{-1})\phi^n_{qp}(x)\;,$$
where $d_{\mathbb{D}}$ is the dimension of $\mathbb{D}$. 
The notation means that from the matrix 
$D^n_{ipq}$ 
we take only the rows which satisfy
$$
D^n_{ipq}(hL_y^{-1})= {\mathbb{D}}_{ij}(h) D^n_{jpq}
(L_y^{-1})
\;.$$

To see how the four-dimensional fields transform, and 
hence to find out the internal symmetry of the 4-dimensional 
effective action, we should look at the 
transformations of 
$\phi^n_{qp}(x)$.
Under the transformation
$y\rightarrow y'$ induced by the $G$-action on $G/H$,
$$\phi_j(x,y)\rightarrow {\phi '}_j(x,y')= {\mathbb{D}}_{ji}(h)
\phi_i(x,y)\;.$$
Using $L_y$ transformations we find that 
$$
\phi^n_{qp}(x)\rightarrow {\phi '}^n_{qp}(x)= 
D^{n}_{jpq}(g)\phi^n_{qp}(x)
\;.
$$

Therefore the 
$4d$ fields transform under $G$, which is the Kaluza-Klein gauge group
(referred to by physicists as the isometry group)

\subsection{Effective Lagrangians}
First of all, let us fix the notations which we will 
use here onwards, unless otherwise indicated.
We denote the coordinates in $4+d$ dimensions 
by 
$$
z^M=(x^\mu, y^m)
\;,$$ 
where the Greek middle alphabet indices take the values 0,1,2,3 
and their Latin counterparts take the values 4,5,...d. The internal 
space is parametrized by $y^m$, and the four-dimensional one 
by $x^\mu$. The tangent space metric is $\eta_{AB}=
\mbox{diag}(-1,1,1,...,1)$.

The metric in $4+d$ dimensions is 
$$
g_{MN}(z)= {E_M}^A(z){E_N}^B(z) \eta_{AB}
\;,$$
where ${E_M}^A$ are the vielbeins. 

To see how the isometry group result in an internal gauge
symmetry of the effective action, let us start with the following
ansatz for the vielbeins in $4+d$ dimensions 
\cite{luc,witten3}
\begin{displaymath}
{E_M}^A(x,y)=
\left(
\begin{array}{cc}
{e_\mu}^\alpha(x) 
&-{A_\mu}^{\hat b} (x) {D_{\hat b}}^a(L_y)\\
0 & {e_{m}}^a (y)  
\end{array}
\right)\;.
\end{displaymath}
This ansatz is compatible with the 4-dimensional general coordinate
transformations $x^\mu\rightarrow {x'}^\mu$, and the Yang-Mills 
transformations $y^m\rightarrow {y'}^m$ with the associated 
frame rotations
${D_a}^b$

\begin{eqnarray}
{e_\mu}^\alpha(x')&=& \frac{\partial x^\nu}{\partial {x'}^\mu}
{e_\nu}^\alpha(x)\;,\nonumber\\
{E_\mu}^a(x',y')&=& 
\left(
\frac{\partial x^\nu}{\partial {x'}^\mu}
{E_\nu}^b(x,y)+
\frac{\partial y^n}{\partial {x'}^\mu}
{e_n}^b(y)\right)
{D_{b}}^a(h^{-1})\label{ndd}\;,\\
{e_m}^a (y')&=& \frac{\partial y^n}{\partial {y'}^m}
{e_n}^b(y) {D_{b}}^a(h^{-1})\;,\nonumber 
\end{eqnarray}
${e_\mu}^\alpha(x)$ and
${e_n}^b(y)$ are the vielbeins on $M_4$ and 
$K=G/H$ respectively. 
The equation (\ref{ndd}) implies, in the view of the 
ansatz above,
\begin{equation}
-{A'}_\mu^{\hat b}(x') 
{D_{\hat b}}^a(L_{y'})
=\left(
- \frac{\partial x^\nu}{\partial {x'}^\mu}
A_\nu^{\hat c}(x){D_{\hat c}}^b(L_y)
+ \frac{\partial y^n}{\partial {x'}^\mu}
{e_n}^b(y)
\right)
{D_{b}}^a(h^{-1})\;.
\label{gauge}
\end{equation}

Extracting the coefficient of $d{y'}^m$ and $d{x'}^\mu$
from the formulae (\ref{form1}) and (\ref{form2})
one gets
$$
\frac{\partial y^m }{\partial x^\mu}
=-(g^{-1}\partial_\mu g)^{\hat b}{K_{\hat b}}^m(y)
\;,$$
where ${K_{\hat b}}^m(y)$ is the Killing vector defined by 
\cite{salam}
$$
{K_{\hat b}}^m(y) = {D_{\hat b}}^c(L_y) {e_c}^m(y)  
\;.$$
After finding the 
transformation rule for 
$K$, it is simple to derive the formula for 
${\partial {y'}^m }/{\partial x^\mu}$
$$
\frac{\partial {y'}^m }{\partial x^\mu}
=-(g^{-1}\partial_\mu g)^{\hat b}{K_{\hat b}}^m(y')
\;.
$$
Going back to (\ref{gauge}) one finds that 
$$
{A'}_\mu (x') 
=\frac{\partial x^\nu}{\partial {x'}^\mu}
\left( g {A}_\nu(x)g^{-1}
-  g^{-1}\partial_\nu g\right)
\;,$$
where $A_\mu={A_\mu}^{\hat a}Q_{\hat a}$. This is precisely
the transformation rule to be 
expected for a Yang-Mills potential. 

Now we arrive to the point where we 
derive the effective four-dimensional 
gravitational action in $4+d$ dimensions,
$$
S=\int d^{4+d}z\det{e(x,y)}\;{\cal R}
\;,$$
where $\cal R$ is the Ricci scalar curvature
of the ($4+d$)-dimensional space-time. Upon substituting the 
metric ansatz into $\cal R$ (in the case of a zero torsion) 
we get 
$$
{\cal R}=
{\cal R}_4-\frac{1}{4} {F_{\alpha\beta}}^{\hat a}
{F_{\alpha\beta}}^{\hat b} {D_{\hat a}}^c (L_y)
{D_{\hat b}}^c (L_y)+{\cal R}_d
\;,
$$
where ${\cal R}_4$ is the usual 4-dimensional Ricci scalar, and 
${\cal R}_d$ is the constant curvature of $G/H$. 
Upon integrating over the internal space, the 
conventional 
Yang-Mills term
emerges
since
$$
\frac{1}{V_d}\int d^dy \;\mbox{det}[e(y)]\; {D_{\hat a}}^c (L_y)
{D_{\hat b}}^c (L_y)=d\;\delta_{{\hat a} {\hat b} }
$$
and therefore we get massless Yang-Mills fields and 
graviton.


\subsection{Compactification \& Stability}
\label{com}
As said earlier, 
Scherk and Schwarz first attempted in
\cite{sch} to put string theory in contact with
the observed world by compactifying the extra dimensions.
The important factor at this point was to actually show that the 
product of a four-dimensional Minkowsky space-time 
with a compact  
internal space is a solution to the equations of 
motion. This
was done again by Scherk and Schwarz 
\cite{sch2} by proposing the concept of
{\it spontaneous compactification}. 
Until then, compactification seemed an 
arbitrary condition imposed on the Kaluza-Klein picture. 

Spontaneous compactification, proposed in \cite{sch,sch2}, occurs
when the
ground state of an Einstein gravity theory
coupled to some matter fields (could be 
Yang-Mills fields \cite{Horvath:1977st,palla1}, 
non-linear sigma models
\cite{Omero:1980vx}, or anitsymmetric tensor gauge fields 
\cite{Freund:1980xh}) 
is described by a product of 
a four-dimensional space-time of a constant curvature,
with a two-dimensional sphere in the presence of
a magnetic monopole background.
In this solution, the field strength of the Yang-Mills 
potential contributes to the stress-energy tensor in the
right-hand side of the six-dimensional  
Einstein equations and generates the curvature of $S^2$.

Remarkably, it has been long realized that topological considerations
are central to the question of stability.

The existence of the solution is again not enough to consider it 
physical. It should be moreover stable, in the sense that there
should not be tachyons or ghosts in small fluctuations around the 
background solution. The issue of classical 
stability of the ($4+d$)-dimensional Einstein-Yang-Mills
theories with an arbitrary gauge group was explored by Randjbar-Daemi, 
Salam and strathdee \cite{sss1} and Schellekens 
\cite{Schellekens:1984dm}.

All stable compactifications on $d$-dimensional spheres 
$S^d=SO(d+1)/SO(d)$ 
with a symmetric topologically non-trivial classical gauge field
which is embedded in an $H$-subgroup (here $SO(d)$) of the 
Yang-Mills gauge group have been classified by 
Schellekens in \cite{Schellekens:1985ks} and they occur only for
$d=$2,4,5,6,8,9,10,12, and 16. 

In the following we briefly 
review
\cite{Randjbar-Daemi:1983hi} which provides a 
explanation of the procedure used to probe 
the classical stability 
through an example of spontaneous 
compactification of a six-dimensional 
Einstein-Maxwell theory. 
More general treatment
of spontaneous compactification on generic symmetric coset spaces
and Yang-Mills theories 
is provided in \cite{percacci}. 

The procedure
for checking the classical stability of compactification
against small perturbations, 
in words, is done by
\begin{itemize}
\item[-]{checking if the ansatz of the desired form $M_4\times K$
is a solutions of the bosonic equations of motion.}
\item[-]{performing small perturbations around the background solution,
and computing the physical states (the ones which couple to 
conserved currents).}
\item[-]{showing that the spectrum does not contain tachyons or
negative metric states.}
\end{itemize}
There may be other ways, however we 
restrict ourselves in this
section to the above one. 

More discussion regarding 
the size of the internal 
space is in section \ref{stab}. 

\subsubsection{The vacuum solution}
Consider an action of gravity in $D$ dimensions coupled to 
an $U(1)$ gauge field, $F_{MN}$, and a cosmological constant,
$\Lambda$,
\begin{equation}
S= \int d^D x \sqrt{-G} \left(\frac{1}{\kappa^2}
{\cal R} -\frac{1}{2g^2}\mbox{Tr}F^2 +\Lambda \right)
\label{action}\;,
\end{equation}
where $\kappa^2=M^{-d-2}$ is the $d$-dimensional Newton's constant.
The field equations of 
(\ref{action}) 
are
\begin{equation}
\frac{1}{\kappa^2}{\cal R}_{MN}=\frac{1}{g^2}
\mbox{Tr}F_{MR}{F_N}^R-\frac{1}{D-2} G_{MN}
\left(\frac{1}{2g^2}\mbox{Tr}F^2 + \Lambda \right)
\label{bosonic1}\;,
\end{equation}
\begin{equation}
\nabla_MF^{MN}= 0\,.
\label{bosonic2}
\end{equation}
Now consider solutions in $D=6$ 
of the form
$M_4\times K$, where $M_4$ is the flat $4$-dimensional
Minkowski space and $K$ is a compact manifold. 
Throughout this section 
$K$ will be taken to be $S^2$. 
Furthermore we shall
assume that the gauge field configuration $A$ will be
non-vanishing only on $K$. One can of course think of
many other choices for $K$.

The flatness of the Minkowski space implies
\begin{eqnarray}
&&\frac{1}{2g^2}\mbox{Tr}F^2 + \Lambda=0\;,\nonumber\\
&&{\cal R}_{{ m}{n}}=\frac{\kappa^2}{g^2}
\mbox{Tr}F_{{m}{r}}{F_{n}}^{r}\;,
\label{ink}
\end{eqnarray}
\nd
where $m$, $n$ are indices in $K$. 
Our
problem is now to find solutions of Yang-Mills
equations in $K$ which also solve the Einstein
equation (\ref{ink}).

The ansatz for solutions to (\ref{ink}), for the 
case $K=S^2$, are
\begin{equation}
ds^2=\eta_{\mu\nu}dx^\mu dx^\nu+
a^2\left( d\theta^2+\mbox{sin}\theta\;d\varphi^2\right)
\label{metricback}\;,
\end{equation}
where $a$ is the radius of $S^2$ to be determined, 
and 
\begin{equation}
{\bar A}_\varphi=-
\frac{n}{2}(\mbox{cos}\theta\mp 1)
\label{back1}\;.
\end{equation}
Here $n\in \mathbb{Z}$, because of the proper 
patching on the upper
and lower hemispheres, and $- (+)$ 
indicate the expression on the 
upper (lower) hemispheres. $(\theta, \varphi)$ 
are the coordinates on 
$S^2$. 

Substituting the ansatz back into (\ref{ink}) reduces the equations
into algebraic equations between $\kappa$, $a$, $g$, and $\Lambda$:
$$
g^2=\frac{1}{8}n^2\kappa^4\Lambda
\;,$$
\begin{equation}
a^2=\frac{1}{8}n^2\frac{\kappa^2}{g^2}
\;.
\label{radius}
\end{equation}

\subsubsection{Fluctuations and gauge fixing}
The spectrum of the 
four-dimensional theory is obtained by looking at the small
fluctuations 
around the backgrounds (\ref{metricback}) and (\ref{back1}),
$$
g_{MN}(x,y)={\bar g}_{MN} +\kappa h_{MN}(x,y)\;,
$$
$$
A_M={\bar A}_M+V_M(x,y)\;,    
$$
where ${\bar g}_{MN}$ should be read off (\ref{metricback}), and 
${\bar A}_M$
is (\ref{back1}). 

To study the stability of the above configuration, one must study 
the response of the system to some external physical disturbance. 
This is usually done by coupling the perturbations $h_{AB}$ and $V_A$
to appropriate sources $T_{AB}$ and $J_{A}$ respectively. The sources
are constrained to respect the symmetries of 
(\ref{action}). The new action (\ref{action}) is
$$
S'= S+\int d^6z \sqrt{-\bar g}
\{\frac{1}{2}T_{AB}h_{AB}+J^AV_A\}
\;.$$
The constraints on the sources read
$$
\nabla^B T_{AB}-\kappa{\bar F}_{AB} J^B=0\;\;\;,\;\;\;
D^AJ_A=0$$
We chose the gauge fixing
$$
\nabla^B(h_{AB}-\frac{1}{2}\eta_{AB}h_{CC})=0\;\;\;,\;\;\;
D^AV_A=0
\;.$$
The next step is to solve 
the linearized equations of motion, taking into account the constraints
on the sources and the gauge fixing,
for $h_{AB}$ and $V_A$ in terms of the sources
$T_{AB}$ and $J_A$.

The linearized equations of motion then take the form 
\cite{Randjbar-Daemi:1983hi}
\begin{eqnarray}
&&\nabla^2 h_{AB}-\frac{1}{4}\eta_{AB}\nabla^2 h_{CC}+
{\bar {\cal R}}_{BC}h_{AC}+{\bar {\cal R}}_{AC}h_{BC}
\;\;\;\;\;\;\;\;\;\;\;\;
\nonumber\\
&&\;\;\;\;\;\;\;\;\;\;\;\;\;
+\kappa{\bar F}_{BC} \nabla_{[A} V_{C]}
+\kappa{\bar F}_{AC} \nabla_{[B} V_{C]}
-\kappa\eta_{AB}{\bar F}_{CD} 
\nabla_{[C} V_{D]} +T_{AB}=0
\nonumber\;,
\end{eqnarray}
$$
\nabla^2 V_A +
{\bar {\cal R}}_{AB}V_B
+\kappa \nabla_C 
(h_{AB}{\bar F}_{BC}
+{\bar F}_{AB}h_{BC}+ 
\frac{1}{2}h_{BB}{\bar F}_{CA})+J_A=0\;.
$$

\subsubsection{Solutions to the linearized equations}
\label{stabspec}
To find the solutions for the above equations, 
it is useful to expand 
the fluctuations and sources in spherical harmonics of $S^2$ 
using the procedure 
given in section \ref{harmon}.

Firstly, all fields should be decomposed into irreducible representations
of the $H$-subgroup, in this case represented by the $SO(2)$ rotations 
(or a $U(1)$). Those representations are labeled by the 
$\lambda$ (named {\it isohelicity} \cite{Randjbar-Daemi:1983hi}). Let 
$\phi_\lambda (x,\theta,\varphi)$ be a typical 
one. 
$$
\phi_\lambda (x,\theta,\varphi)
=\sum_{l\geq |\lambda|} \sqrt{2l+1}\sum_mD^l_{\lambda m}
(L_{\theta \varphi}^{-1}) \phi^l_{\lambda m(x)}
\;,$$
where $D^l_{\lambda m} $ belong to the ($2l+1$)-dimensional unitary
irreducible representation of $SU(2)$. $\lambda$ can be an integer
or half an integer, and $m=-\lambda,...,+\lambda$.
One can chose
$L_{\theta \varphi}= \mbox{e}^{\frac{i}{2}
\varphi \sigma_3}\mbox{e}^{\frac{i}{2}
\theta \sigma_2}
\mbox{e}^{-\frac{i}{2}
\varphi \sigma_3}
\in SU(2)$ ($\sigma_{1,2,3}$ are Pauli matrices), 
in other words
\begin{equation}
L_{\theta \varphi}^{-1}=\left(\begin{array}{cc}
\mbox{cos}\frac{\theta}{2}&-\mbox{e}^{i\varphi}
\mbox{sin}\frac{\theta}{2}\\
\mbox{e}^{-i\varphi}
\mbox{sin}\frac{\theta}{2}
&\mbox{cos}\frac{\theta}{2} 
\end{array}\right)
\;.\label{form}
\end{equation}

To see an 
example of how the summation in the harmonic expansion is done, 
consider the representation
$l=1/2$. The coefficients of 
$\phi^{l/2}_{1/2,-1/2}$, for instance, 
can be read from (\ref{form}) and it is 
$-\mbox{e}^{i\varphi}
\mbox{sin}\frac{\theta}{2}$. The general rule have been 
explained in section \ref{harmon}.

Now we are left with the search for tachyons or negative norm states. 
Among the fields $h_{AB}$ and $V_A$ there will be following 
fields (the $SO(2)$ irreducible pieces)
which have a definite isohelicity:
\begin{displaymath}
\begin{array}{ll}
h_{++}=\frac{1}{2}(h_{55}-h_{66}-2ih_{56})
 & \lambda= 2, \\
h_{+-}=\frac{1}{2} (h_{55}+h_{66})
& \lambda=0,\\
h_{\mu+}=\sqrt{\frac{1}{2}} (h_{\mu5}-ih_{\mu6})
& \lambda=1,\\
V_{+}=\sqrt{\frac{1}{2}} (V_{5}-iV_{6}) &\lambda=1,\\
h_{\mu\nu}, V_{\mu\nu}&\lambda=0,\\
h_{--}=h_{++}^*, h_{-+},...etc.
\end{array}
\end{displaymath} 
 
The non-zero masses are the following labeled by their
spin $0,1,2$ (
the technical details of the calculations were neatly worked out in 
\cite{Randjbar-Daemi:1983hi}),
\begin{displaymath}
\begin{array}{ll}
M_{0}^2=(l-1)(l+2)/a^2, & l\geq 2\\
M^2_{0\pm}= [2l(l+1)+1+\pm \sqrt{1+12l(l+1)}]/{2a^2}  
,& l\geq 0\\
M^2_{1\pm}= [l(l+1)+\pm \sqrt{2l(l+1)}]/{a^2},& l\geq 1\\ 
M^2_2= l(l+1)/a^2 ,&  l\geq 0
\end{array} 
\end{displaymath} 
and it can be checked that no tachyons are present in this model, 
and hence it is expected to be stable against small perturbations. 
There will be in addition six massless states; the graviton 
($\lambda =\pm 2$), a massless ``photon'' ($\lambda=\pm 1 $), 
and a Yang-Mills triplet 
($\lambda=\pm1,l=1$).


It is worthwhile to emphasis as a final point in this
section that 
there is a mixing of the six-dimensional metric and Maxwell fields
as pointed out in \cite{Randjbar-Daemi:1983hi}. 
This mixing is contrary to the common assumption that Kaluza-Klein 
gauge fields originate purely from the metric.
This argument is expected to hold  
whenever matter fields which participate in the
spontaneous compactification carry the appropriate 
quantum numbers \cite{Randjbar-Daemi:1983hi}.

\subsection{Chiral fermions}
\label{chiralint}

Upon compactifying a Kaluza-Klein type theory with fermions, 
one expects in general three kinds of difficulties.    
In order for the resulting fermions in 4 dimensions to be 
chiral, they should be massless to start with. 
Hence the first requirement
for the Dirac operator on the compact internal manifold
is to have at least one zero mode, while not all 
manifolds admit harmonic spinors.
Secondly, the index of the 
Dirac operator is often (not always) zero
and hence the 4-dimensional theory will 
be non-chiral.

Moreover,
the irreducible 
spin representation in $4+d$ ($d\geq 1$) dimensions 
is always, greater than the irreducible 
one in four dimensions.\footnote{Which is 
$2^{\frac{n}{2}}$ for $n\in 2\mathbb{Z}$ and
$2^{\frac{n-1}{2}}$ for $n\in 2{\mathbb{Z}}+1$ .}  
Also the number of zero modes, 
though finite \cite{lawson}, 
can lead to more massless fermions than desired
in four dimensions. 
So, one generally expects more degrees of freedom, than the ones present
in the standard model, to result in the four-dimensional 
effective action after compactification. Many 
of those extra degrees
of freedom can be eliminated once the issue of achieving chiral
fermions 
is settled. 

One way to eventually achieve a chiral Lagrangian in 
four dimensions 
stemming from a Kaluza-Klein field theory
(another way is be by using 
by orbifold compactification)
is by coupling the fermions in $4+d$ dimensions
to a stable non-trivial background, like a magnetic monopole,
as
was proposed by Randjbar-Daemi, Salam  and 
Strathdee \cite{Randjbar-Daemi:1983hi,
Randjbar-Daemi:1983qa}. 
The problem of obtaining left-right 
asymmetry of fermion 
quantum numbers was discussed by Witten \cite{Witten:ed.ux}.  


The necessity for a non-trivial background to get chiral 
fermions from extra dimensions can be explained in the 
following simple example. Consider a manifold, $W$, 
in 6 dimensions
where $W=M_4\times S^2$. 
The Dirac equation on a generic manifold is 
$$
\slash{D}_W\psi(x,y)=0\;.
$$
With the appropriate choice of Dirac matrices, 
the above equation can be 
written as  
$$
\slash{D}_4\psi(x,y)+ \slash{D}_{S^2}\psi(x,y) =0 
\;.$$
In the absence of a background gauge field
\begin{itemize}
\item[-]{$\slash{D}_{S^2}\psi=0 $ has no
{\it regular} solutions.}
\item[-]{ $\mbox{ Index}\slash{D}_{S^2}=0$.}
\end{itemize}

The first point can be understood easily, with and without 
going into explicit 
computations, since, 
according to Lichnerowicz's \cite{L} 
theorem, 
all positively curved 
smooth compact manifolds, including $S^2$,
do not admit harmonic spinors. 
In order for the resulting fermion in 4 dimensions to be 
chiral, it should be massless to start with. 
Hence the first requirement
for the Dirac operator on the compact internal manifold
is to have at least one zero mode. The 
existence of fermion zero modes
by couplings to gauge fields with non-trivial 
topology was pointed out in \cite{Horvath:1977st,palla}.
Also, the index of the 
Dirac operator is often (as in the case of 
$S^2$), however not always, zero.

Now let us couple $\psi(x,y)$ to a magnetic monopole background
described in equation (\ref{back1}).
\footnote{For simplicity, we will do the computations on the 
upper hemisphere only, while taking into account 
the consistent patching.}
Consider the coupled Dirac operator on $S^2$ 
$$
\slash{D}_{S^2}=
\Gamma^mE_m^\alpha(\partial_\alpha
-\frac{1}{2}\omega_{\alpha [k,l]} \Sigma^{kl}-i
{\bar A}_\alpha(y))\;.
$$
As can be seen from 
(\ref{back1}),
the background solution is proportional to the 
spin connection $\omega_\alpha$ 
($\omega_\theta =0$, and $\omega_\varphi = -(\mbox{cos}
\theta -1)$).

Now let $\psi(x,y)$ be a Weyl spinor in 6 dimensions.
Using the chirality matrix, $\gamma_5$,
in four dimensions, $\psi$ can be written as: 
\begin{equation}
{\psi}= \frac{1+\gamma_5}{2}\psi+\frac{1-\gamma_5}{2}\psi
={\psi_R+\psi_L}\;.
\label{ferm}
\end{equation}
The Dirac equation in a monopole background on the upper 
hemisphere hence simplifies to
$$
\left( \partial_\theta +\frac{i}{\mbox{sin}\theta}
-\frac{n+1}{2\mbox{sin}\theta}
+ \frac{n+1}{2} 
\;\mbox{ctg}\theta \right){\psi_R}
=0
\;,$$

$$
\left(
-\partial_\theta +\frac{i}{\mbox{sin}\theta}
-\frac{n-1}{2\mbox{sin}\theta}
+\frac{n-1}{2} 
\;\mbox{ctg}\theta \right) {\psi_L}
=0
\;.$$

It can be easily  
checked that regular solutions exist only for one of the 
chiralities, which is here
$
{\psi_L}
$ for $n>0$,
and for
$n\neq 0$.\footnote{And {\it vise versa} for $n<0$.}

\subsection{Mass estimates for KK excitations}

As we mentioned in the introduction, the eigenvalue
problem of the Laplacian, $\nabla^2$, 
is not known 
on a generic compact space. 
However, we know  
that the spectrum 
of $\nabla^2$ 
on a compact Riemannian manifold 
is discrete, bounded from below, 
and the eigenvalues (counted with 
multiplicity) are ordered: 
$0=\lambda_0\leq\lambda_n\leq \lambda_{n+1}$. 
In the mathematics literature there are many 
results on the lower bounds on
the first eigenvalue,  
$\lambda_1$, of the Laplace-Beltrami operator, 
$$\nabla^2\equiv \frac{1}{\sqrt{g}}
\partial_m (\sqrt{g}g^{mn}\partial_n),$$ 
\nd
(Laplacian acting 
on scalars)
on a compact 
manifold.\footnote{ 
Upper bounds on those eigenvalues have also 
been worked out (for a review 
see \cite{avijit} and references
thererin),
however these are
not much of a physical relevance since the upper
bounds depend generally on the level of excitation.}

These general bounds can be helpful 
in estimating the masses of Kaluza-Klein excitations, 
mainly of the graviton and scalar fields, in the cases where 
it is difficult to perform
explicit computations of the specturm. 

The lower bounds on the first eigenvalue of 
the Laplacian
translate into lower bounds on 
the 4-dimensional 
tree-level
masses of the bosonic particles arising from compactification.

Lichnerowicz theorem enables us to 
to impose 
similar lower bounds on fermion masses, and also
to exclude tree level massless fermions, unless coupled to
a non-trivial background (as explained in section 
\ref{chiralint}),
for positively curved internal spaces.

The lower bound on the first non-zero eigenvalue
of $\nabla^2$ acting on scalars, 
$\nabla^2_K \phi_n=-
\lambda_n \phi_n$, 
on a generic compact manifold of a scalar curvature bounded from below
by $(d-1)\zeta$ ($\zeta\in \mathbb{R}$)
is \cite{yau} 
\begin{equation}
\lambda_1 + \mbox{max} \{  -(d-1)\zeta, 0
\} \geq \frac{\pi^2}{4L^2}
\;,
\label{eq:w}
\end{equation}

where $L$ is the diameter of the manifold (the longest 
distance). 
The fundamental parameter for masses arising from compactification
is hence $L$ (this can be
understood by observing that 
it is possible to change the spectrum of
the Laplacian by deforming the manifold, and yet keeping its 
volume 
fixed). Usually $L$ is greater or equal 
than the characteristic scale
of the volume of $K$ (see for example \cite{Kaloper:2000jb}), 
and hence the estimates of the Kaluza-Klein excitations
depending on mere dimensional 
analysis 
(on complicated 
manifolds where explicit computations can not be done)
has to be somehow
lowered.
  
In most cases, specially for a space of a constant curvature, 
one 
can relate
$L$ to the volume of the manifold, and hence 
rewrite 
the bounds in terms of the volume instead (e.g. in $S^d$ and 
compact 
hyperbolic manifolds).
The inequality (\ref{eq:w}) translates
effectively into a statement about the bounds on the 4-dimensional
masses
of the lowest
excitations, $m_1^2$.
It is obvious
that when the Ricci curvature, ${\cal R}_d$, of $K$ 
is non-negative, then one recovers
the standard scenario:
$$\lambda_1\geq \frac{\pi^2}{4L^2},$$
where in the standard Kaluza-Klein scenario, as in \cite{add1}, 
$L$ is identified with the compactification 
radius, $R$ ($e.g.$ for a circle, $L =2\pi R$).

When ${\cal R}_d<0$,
\begin{equation}
\lambda_1 \geq \frac{\pi^2}{4L^2} -(d-1)|\zeta|
\;,
\end{equation}
the lower bound  
may not in this case always hold \cite{rula}, 
specially if some particular 
tuning between $L$ and the volume scale of $K$
is needed (in order to address the hierarchy problem, for 
instance \cite{Kaloper:2000jb}). 

The observed massless fermions in four dimensions
are nothing but the zero modes of $\slash{D}_K$ 
(which lie in $\mbox{ker}
\slash{D}_K$). 
The argument is based on the relation
between the squared Dirac operator and the scalar curvature,
\begin{equation}
{\slash{D}}_K^2= 
{\nabla}^*\nabla +\frac{1}{4}\;{\cal R}_d
\;,
\label{square}
\end{equation}
where 
${\nabla}^*{\nabla}$
is the connection Laplacian and is a positive operator.
It has been pointed out by Lichnerowicz \cite{L} using 
(\ref{square})
that manifolds with a positive curvature do not
admit harmonic (massless) spinors. This can be easily
seen  
by sandwiching (\ref{square}) for $\slash{D}^2=0$ 
inside an complete 
orthonormal set of wave functions and considering a constant
curvature case, one gets  
$$
|\nabla \psi(y)|^2  
+\frac{1}{4}
{\cal R}_d |\psi(y)|^2=0\;, 
$$ 
obviously the above equation has no solutions for 
${\cal R}_d>0$ on a compact space.

As it is the case for the Laplacian, the eigenvalues of the Dirac
operator on a compact space are discrete. Therefore, the eigenvalues
of the squared Dirac operator are discrete and positive, and in 
addition any eigenvalue, $\nu_q^2$,is 
bounded from below by the curvature \cite{hijazi}, 
including $\nu_1^2$, 
\begin{equation}
\nu_1^2\geq \frac{d}{4(d-1)}{\lambda '}_1
\label{nu}\;,
\end{equation}
where ${\lambda '}_1$ is the first eigenvalue of the Yamabe operator, 
$$
L\equiv \frac{4(d-1)}{d-2}\nabla^2_K+{\cal R}_d
\;.
$$
with $\nabla^2_K$ being the positive Laplacian acting on functions.
This means that the lower bounds on the massive spin 0 and 
spin 1/2  
excitations are related. For constant curvature ${\cal R}_d=
(d-1)\zeta \geq 0$

$$
\nu_1^2\geq \left(\frac{d}{d-2}\right)\frac{\pi^2}{4L^2}+
\frac{d}{4(d-1)}
{\cal R}_d,
$$
and for ${\cal R}_d< 0$
$$
\nu_1^2\geq \frac{d}{d-2}
\frac{\pi^2}{4L^2}-
\frac{d(5d-6)}{4(d-1)(d-2)}|{\cal R}_d|
.$$

\section{Warped Non-compact Extra Dimensions}
\label{wned}



\subsection{Introduction}

The notion of warped product of Riemannian manifolds was first
introduced by Bishop and O'Neill in 1969 \cite{oneil}, and
it was shown that such solutions can be of a physical  
relevance in \cite{Rubakov:1983bz,Wetterich:1984uc,wit,nie,wett2}, 
and was recently 
revived 
in \cite{Randall:1999ee,randal2,lukas}.
The recent literature
is rich of new scenarios
in non-compact spaces 
\cite{cohen}-\cite{Ito:2001gk}.\footnote{For  
recent reviews see \cite{Nath:2001kr,rubn, Dick:2001sc}.}

The non-direct product spaces (non-factorizable geometry)
of relevance are solutions to the equations of motion in which the 
metric of space-time is a product of an internal space and 
a scalar function of the extra coordinates (the warp 
factor), and the metric ansatz representing the
space $M \otimes_f K$ is 
$$
ds^2= f(y)^2g_{\mu\nu}dx^\mu dx^\nu+ g_{mn}(y)dy^mdy^n 
\;,$$
where the {\it warp factor}, $f$, is a smooth 
positive function $f:K\rightarrow {\mathbb R}^+$. 
The Ricci 
scalar is:
\begin{equation}
{\cal R}=\frac{1}{f^2}\left\{
{\cal R}_n -\frac{n}{2}f\nabla^2f-
\frac{n}{4}(n-1)|\nabla f|^2\right \}+ {\cal R}_d
\;,\label{sing}
\end{equation}
where ${\cal R}_n$ and ${\cal R}_d$ are the intrinsic
curvatures of $M$ and 
$K$ respectively. $n\equiv \mbox{dim}M$ and is usually taken to  
to be equal to $4$.

The space $M$ is hence an ($n-1$)-dimensional 
extended object inside the full mother manifold 
with a tension $\lambda$ and
has generally the action
$$
S_M=\lambda\int_{M}d^{n}x \left[-\mbox{det}(\partial_\mu
z^M(x)\partial_\nu z^N(x)\eta_{MN})
\right]^\frac{1}{2}
\;.$$
The warping space is usually taken to be non-compact
with an infinite volume. \footnote{A topological space  
is said to be non-compact if it can not be represented
by a union
of a countable number of compact subsets.
A compact subset is a set which is closed 
(contains the limits of all its convergent sequences), 
and bounded (is contained in a sphere of ${\mathbb{R}}^d$).
Non-compact spaces can also have a finite volume.} 

However, using a damping warp factor as
a solution of the equations of motion
leads to a 
finite volume of the internal non-compact space
\cite{wett2}
by ``effectively'' changing the measure
as we shall see later on (as in 
\cite{Randall:1999ee,randal2}). Therefore, it is perhaps more
natural to use warped geometry in the presence of 
non-compact internal spaces. 

The idea \cite{sh1,sh2,Rubakov:1983bz}
of living on a distributional
{\it brane} source embedded 
in a five-dimensional (or higher) 
manifold with non-compact transverse direction   
has in principle a welcoming 
environment in superstring theory, 
$D$-branes \cite{pol}. 
A $D$-brane is a BPS state carrying Ramond-Ramond charge, 
nicely stable, being the lowest energy state, and naturally admits
localized modes on its world-volume.  
There are also BPS flat domain wall solutions  
\cite{stelle}-\cite{agata}, as well as curved
\cite{card,lop}, in five-dimensional 
gauged supergravity.

A classification of all (non D-brane) 
possible
generalizations of Green-Schwarz superstring action to 
$p$-dimensional extended objects
is given in 
\cite{DeAzcarraga:1989vh} 
where it was also shown
that topological 
extenions may arise in the $p$-brane supersymmetry algebra
\cite{Achucarro:1987nc}. 
In performing the classification, the symmetries of the 
world-volume of the $p$-brane (reparametrization
invariance
and Sigel symmetry) were used in order to show that  
gauge fixed version of the brane world-volume action 
coupled to a ($p+1$)-form field 
will have some amount of supersymmetry of the mother 
manifold, depending on the 
dimension of the embedding space and the number 
$p$. In this classification 
the supersymmetric 
$3$-brane ($4$-dimensional world-volume) 
can occur only in $2$ and
$4$ extra dimensions. The supermultiplets will certainly
be
confined to the brane. 
Making a non-supersymmetric gauge theoretical 
analogy to such a classification 
is obviously a difficult task,
however it remains desirable until proven impossible!.

This chapter discusses some examples
of warped solutions to Einstein and
Einstein-Yang-Mills systems, as well 
as the 
conventional methods
for localizing gravity and 
fermions on a domain wall with a warped
non-compact
transverse space.

\subsection{Junction conditions}
\label{junc3}
Consider a simple case of  
one extra dimension $d=1$, 
\begin{equation}
S=\int d^4x\;dy\sqrt{g(x,y)}\left(
\frac{1}{\kappa^2}{\cal R}_5+\Lambda\right)+
\lambda \int d^4xdy\; \sqrt{\bar g(x,y)}\delta(y)
\;,
\label{act}
\end{equation}
where the last term represents a  
$\delta$-function like 
co-dimension 1 sources, the simplest form of a 
{\it brane}, which is a
hypersurface in $5$-dimensions characterized by 
the equation $y=0$. 
The brane has
the induced metric
${\bar g(x)}_{\mu\nu} =g_{\mu\nu} (x,y=0)$, 
and a tension $\lambda$. 
The ansatz for the background solution is 
$M_4\times 
\mathbb{R}$ as in \cite{Randall:1999ee} 
\begin{equation}
ds^2= {\mbox{e}}^{-2\sigma(y)}
\eta_{\mu\nu}dx^\mu dx^\nu+ dy^2
\;.\label{xc}
\end{equation}
Einstein equations of this system are
$$
{\cal R}_{MN}-\frac{1}{2}g_{MN}
(
{\cal R}-\frac{2}{M^3}\Lambda)
=\frac{1}{4M^3}\sqrt{-\bar g(x)}
{\bar g}_{\mu\nu}(x) \delta_{M}^\mu
\delta_N^\nu \delta(y)
\;,$$
where $\mathbb{R}$ is parametrized by $y\in (\infty,+\infty)$. 

Away from the origin, the above equations are just the 
standard 5-dimensional Einstein equations with a cosmological 
constant, and they reduce to 
$$
{\cal R}_{MN}=\frac{\Lambda}{(d+2)M^{d+2}}
g_{MN}\;.
$$

At the origin, the solutions for the equations
require proper matching at the location of the 
source. This will obviously impose a relation between
the tension of the brane $\lambda$ and the warp factor $\sigma$, 
and consequently the cosmological constant
$\Lambda$. The procedure of matching differs according to the model, 
however this feature still holds. 

The discontinuity at $y=0$ is 
characterized by the extrinsic curvature,
$K_{\mu\nu}$,
which contains information of how the hypersurface is 
embedded in 
the higher-dimensional space-time. 
The extrinsic curvature is defined as $$
K_{\mu\nu}=-\eta^{\rho\mu}  \nabla_\nu n_\rho
\;\;\;,\;\;\;\perp_{\mu\nu}=n_\nu
n_\nu\;,$$
where $n_\rho$ is a normal vector in the direction of 
the extra dimension(s), which 
defines the orthogonal complement components of the induced metric
on the brane $\perp_{\mu\nu}=g_{\mu\nu}(x,y)-{\bar g}_{\mu\nu}(x)$. 

\begin{figure}
\centerline{\leavevmode\epsfysize=2.5cm \epsfbox{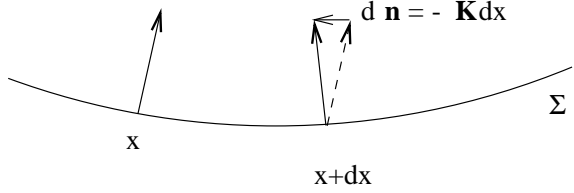}}
\vspace{10pt}
\caption{\small The extrinsic curvature measure the deformation 
of a figure (here the {\it brane})
lying in a spacelike hypersurface, $\Sigma$, 
that takes place when each point in the figure is
carried out forward  unit interval 
of proper time ``normal'' to the hypersurface out into 
the enveloping spacetime.}
\end{figure}

Assuming that the extrinsic curvature varies smoothly within 
the neighborhood $[-\epsilon,+\epsilon]$ 
of the constant hypersurface, the discontinuity 
as measured at scales larger than the thickness 
$2\epsilon$
is 
given
by the jump in $K_{\mu\nu}$ across
the discontinuity which is  
described by the quantity
$K_{\mu\nu}^+-K_{\mu\nu}^-$ or equally by 
\begin{equation}
[K_{\mu\nu}]\equiv\int_{-\epsilon}^{+\epsilon}
d\epsilon\;
n^\rho \partial_\rho K_{\mu\nu}
\;.
\label{branej}
\end{equation}
One can relate 
the normal derivatives of $K_{\mu\nu}$ to the
the usual (intrinsic) curvature within the 
thin brane limit (the transverse
derivatives are negligible compared to the normal ones)
\cite{batt} in the formula,\footnote{
See \cite{Bonjour:1999kz} for a 
generalization to a thick-wall.}
$$
{K'}_{\mu\nu}\equiv n^\rho\partial_\rho
K_{\mu\nu}=\eta_\mu^\gamma
\eta^\lambda_\nu 
{{{\cal R}_{
\gamma}}
}_\lambda
\;,
$$
from which one can deduce that the dominant components of the 
background Ricci tensor  
${\cal R}_{\mu\nu}$ will be given by the asymptotic formula
$$
{\cal R}_{\mu\nu} \backsimeq
{K'}_{\mu\nu}+K' \perp_{\mu\nu}
$$
where $K'= {\perp^\rho}_{\sigma}{{\cal R}^\sigma}_\rho$. Hence
$$
{\cal R}_{\mu\nu}-\frac{1}{2}g_{\mu\nu}{\cal R}
\backsimeq {K'}_{\mu\nu}-\eta_{\mu\nu} K'
\;.$$

Matching at $y=0$ of the equations of motion at $y=0$
requires 
\begin{equation}
[K_{\mu\nu}]
=\frac{1}{M^{d+2}}
\lim_{\epsilon \to 0}
\int_{-\epsilon}^{+\epsilon} d\epsilon
[(d+2)T_{\mu\nu}-g_{\mu\nu}g^{\rho\gamma}T_{\rho\gamma}]
\;.
\label{junc}
\end{equation}

The above condition is a higher-dimensional version of the so 
called Israel matching conditions \cite{is}. \footnote{ 
Further
discussion 
can be found in \cite{batt,mis,tk}.}

The matching condition at the origin of the solutions
to the equations of motion
coming from (\ref{act}) reads 
$$
3 \sigma' g_{\mu\nu}=4\frac{\lambda}{M^3}g_{\mu\nu}\;\;\;
\mbox{at}\;\;y=0 
\;,$$
where the prime refers to a derivative with respect to $y$.
The above equation yields to
\begin{equation}
\sigma(y)=\frac{\lambda}{12M^3}|y|
\;.
\label{11}
\end{equation}

The equations of motion are satisfied, in the 
case of a flat four-dimensional space, when   
\begin{equation}
{\sigma'}^2 =
-\frac{\Lambda}{12M^3}
\;,
\label{22}
\end{equation}
where the additive integration constant is omitted as it 
just amounts to an overall rescaling of $x^\mu$'s.
Clearly, $\Lambda$ has to be negative; which means that 
the five-dimensional space is an $AdS_5$.\footnote{The notation 
used here, as in \cite{Randjbar-Daemi:1983hi},
is such that $\Lambda <0$ indicates an anti-de Sitter space.}

The equations (\ref{11}) and (\ref{22})
imply that 
the tension of the brane and the cosmological 
constant are related by 
$$
\lambda^2=-12M^3\Lambda\;.
$$
In other words, the argument of the warp factor in (\ref{xc}) 
is \cite{Randall:1999ee,randal2} 
\begin{equation}
\sigma(y)=\sqrt{\frac{-\Lambda}{12 M^3}}|y|\;.
\label{rs3}
\end{equation}

This indicates a genuine singularity in the curvature, as 
can be seen by substituting the warp factor (\ref{rs3}) 
in (\ref{sing}). 
Therefore, this solution should be embedded in a context where the
singularity is avoided (like string theory or by
considering a thick wall, as in
\cite{sh2},
where the thickness of the wall would act as regulator)
for it to make sense.

Apart from the singularity, proving an ansatz to be a solution
for the equations of motion is not at all enough, it should also be
proved stable at least classically. 

The stability can be insured if the brane is a BPS state 
with exact supersymmetry in the full space-time, and/or 
carries a conserved charge thorough its coupling to a 
four form. In fact, the stability of the 
brane-models have been questioned in some 
recent papers, as
\cite{Binetruy:2000wn}-\cite{Ida:2001qw}, and it is 
far from being obvious.

\subsection{Randall-Sundrum scenario}
\label{brane}

The Randall-Sundrum model \cite{randal2} 
where the hierarchy problem is addressed is an extension of 
the above simple example. Here is how it works. 
Let us add another brane to 
(\ref{action}) which has a tension $\lambda'$ and consider
the warp ansatz (\ref{xc}) of $M_4\times S^1/Z_2$, where
$S^1/Z_2$ has a length $r_c$ and parametrized by the 
angular coordinate $ \phi\in [-\pi,\pi]$ (related to the 
old coordinate by $y=r_c\phi$). 

The background metric (\ref{xc}) would
satisfy the equations of motion 
of the action with the two branes
$$
\frac{{\sigma '}^2}{r_c^2}=\frac{-\Lambda}{12M^3} 
\;\;\;,\;\;\;
\frac{\sigma''}{r_c}= 
\frac{\lambda}{12M^3}\;\delta(\phi)-
\frac{\lambda'}{12M^3}
\delta(\phi-\pi)\;,\nonumber 
$$
if similar conditions to the previously discussed are fulfilled. 
There is a different way 
in obtaining the matching conditions 
here, which is coordinate dependent, and it is done
by making use of the $Z_2$ symmetry of the internal 
space.  
The consistency of the first equation 
with the orbifold symmetry $\phi\rightarrow 
-\phi$ 
implies (\ref{rs3}) and this 
makes sense only if $\Lambda <0$; indicating again
that
the 5-dimensional space-time
between the two branes is an anti-de Sitter space $AdS_5$.

\begin{figure}
\centerline{\leavevmode\epsfysize=5cm \epsfbox{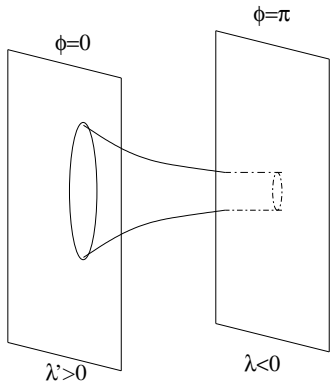}}
\vspace{10pt}
\caption{\small A sketch of Randall-Sundrum
two-brane model with a damping warp factor towards the 
brane with a negative tension.}
\end{figure}

The second equation of motion makes sense when 
\begin{equation}
\lambda' =-\lambda= 
\sqrt{3M^3\Lambda}=6M^3k\;\;\;;\;
\;\;k\equiv 
\sqrt{\frac{-\Lambda}{12M^3}}\;.
\label{eq}
\end{equation}


The five-dimensional gravity scale $M$ is again linked to 
$M_P$ through the ``volume'' of the internal space
\begin{equation}
M_P^2=M^3r_c \int_{-\pi}^{\pi} d\phi\; \mbox{e}^{-2kr_c|\phi|} 
=\frac{M^3}{k}(1-\mbox{e}^{-2kr_c\pi}  )\;.
\label{hier}
\end{equation}
Assuming the bulk curvature $\Lambda$ to be less than $M^5$,
and $kr_c\gg 1$, 
one deduces that $M_P$ depends weekly on $r_c$, unlike 
the case in Kaluza-Klein type models (\ref{mass1}). 

If we live on the brane located at $\phi=\pi$, then 
mass scales in the 4-dimensional theory will be 
lowered generally by the value of 
the metric at the location which is 
a factor of $\mbox{e}^{-kr_c\pi}$. This will remove any
fine tuning in generating the hierarchy between 
the Planck and electroweak scales 
since $M_P\sim \mbox{e}^{k\pi r_c} M_{EW}$.

The masses of the KK excitations are quantized
with gaps $ \Delta m\sim k\mbox{e}^{-kr_c}$ 
\cite{randal2,rubn,ran3}.
From the point of view of an observer on the brane with 
$\lambda<0$,
the KK gravitons will appear to have physical masses of 
order $\sim k$, {\it i.e.} of order TeV, while their
dimensionful couplings to the above 
matter will be characterized
by mass scales of order $(M^3/k)^{1/2}$, which is roughly 
the weak scale. 


Solving the hierarchy problem by living on the 
brane with negative tension does not seem to be 
satisfactory.
The reason is 
that
the weak energy principle
characterized by the energy momentum tensor 
$T_{\mu\nu}\zeta^\mu\zeta^\nu \geq 0$
(for an observer with $4$-velocity $\zeta^\mu$) 
will 
be violated \cite{witten55}
and hence one may expect unphysical modes to appear 
in the spectrum, like modes with arbitrary large negative 
energies 
\cite{Gregory:2000jc}, which puts the stability  
of the model under question.

There are in the literature however stable vacua with negative
cosmological constant, mainly within a 
supersymmetry context,
as then one of Maldacena \cite{mald} for instance which is not meant 
to describe the world as it is, however did 
shed a light on a deeper understanding of the relation
between general relativity and conformal field theory 
beyond the 
gauge principle. 

The above model \cite{Randall:1999ee}  
resembles in its general setup the one in \cite{horava}
of the eleven-dimensional supergravity compactified
on a manifold with boundaries, this configuration
was shown to be relevant to the strong coupling limit of 
the $E_8\times E_8$ heterotic
string theory \cite{horava}.       
In five dimensions, the effective strongly coupled 
heterotic string theory is shown to be 
a gauged version of 
$N=1$ five-dimensional supergravity with four-dimensional 
boundaries which are identified with a pair
of $3$-brane world-volumes \cite{lukas}. 
These branes couple to 
a $4$-form field which is the dual 
of the cosmological constant.
There have been various attempts to provide a supersymmetric 
realization to the Randall-Sundrum  
\cite{Randall:1999ee} scenario, \footnote{See for example 
see \cite{Bergshoeff:2001ii,rand5} as well as other references in  
this section.} however 
no explicit link with string theory has
yet been established.


Living on the brane with positive tension has the advantage of
having a 4-dimensional localized gravity \cite{randal2}, 
and the hierarchy can be 
generated when 
the electroweak or supersymmetry breaking
is transfered to us by some mechanism on the hidden 
brane of negative tension \cite{Gherghetta:2001kr}.
Further discussion regarding anti-de Sitter/Conformal 
field theory correspondence in the above
two brane model is present in \cite{Arkani-Hamed:2001ds}.
Aother phenomenological and cosmological aspects of 
brane-world models can be found in 
\cite{Goldberger:2000un}-\cite{Arkani-Hamed:2001ds}.

\subsection{
General solutions with Yang-Mills fields
}

Here we present a generalization of the general 
solutions for spontaneous compactification on symmetric
spaces \cite{percacci} to warp compactification developed in 
\cite{Randjbar-Daemi:2000ft}. 
As seen in section \ref{KK}, the coupling to Yang-Mills fields
may lead to interesting physics and
could also 
be necessary for 
the consistency of the theory. The most relevant point about the 
presence of a gauge fields background configuration is that 
it is crucial in obtaining chiral fermions, as well as 
localizing them on a brane. 
The case for pure gravity 
was worked out in \cite{wett2}, 

Consider
the background ansatz, 
\begin{equation}
ds^2=\mbox{e}^{A(r)}\eta_{\mu\nu}dx^\mu dx^\nu
+\mbox{e}^{B(r)}g_{mn}dy^mdy^n+dr^2\;,
\label{mod}
\end{equation}
for an Einstein-Yang-Mills system in $D=d_1+d_2+1$ dimensions,
where we are thought to live on a 
$d_1$-dimensional world-volume of a flat 
brane
with a $d_2$-dimensional compact internal space, $K$, and
another one-dimensional non-compact space, 
$\mathbb{R}^+$. 

This ansatz is fairly general since
several models can be regarded as special cases 
of (\ref{mod}) by 
choosing specific forms of the functions $A$ and $B$.
\footnote{
For instance 
the case for which $d_2=0$ and 
$A=c|r|$ ($c<0$) is the Randall-Sundrum
model \cite{randal2};  $d_1$ and $A=B=c|r|$
($c<0$) is  
\cite{Gherghetta:2000qi}; $d_2=1$, $A=c|r|$ ($c<0$), 
as well as other higher-dimensional 
solutions which localize gravity 
\cite{Gherghetta:2000jf,Olasagasti:2000gx}.}
Let us also assume that $K=G/H$ and that the Yang-Mills gauge
group contains $H$ as explained in section \ref{KK}. 
On such spaces, one can construct the solutions 
\cite{percacci,Randjbar-Daemi:2000ft}
$$
F_{mp}^a F_{n}^{ak}=\frac{\mbox{e}^B}{d_2} F^2 g_{mn}
\;,$$
where $F^2=g^{mk}g^{nq}F^a_{mn}F^a_{kq}$ must be 
constant by virtue of $G$-invariance. 

The starting action is 
$$
S= \int d^D x \sqrt{-G} \left(\frac{1}{\kappa^2}
{\cal R} -\frac{1}{2g^2}\mbox{Tr}F^2 +\Lambda \right)
+\lambda \int d^{d_1}x 
\sqrt{\bar g(x)}
$$
where the last term represents a Dirac $\delta$-function like 
simple {\it brane}-source
which depends only on 
$r$.  
${\bar g(x)}_{\mu\nu}$ is the induced metric,
and
$\lambda$ is the tension of the $d_1$-dimensional 
world-volume brane. The {\it brane} is chosen 
to be located at 
$r=0$. 

Around the origin, the solutions for the 
equations of motion will
require relating the brane tension, $\lambda$, 
together with the values of $A'$ and 
$B'$ in a similar manner as presented in 
sections \ref{junc3} and \ref{brane}.
More about such treatment can be found in 
\cite{Gherghetta:2000qi,Olasagasti:2000gx}. 

Away from $r=0$ the equations of motion are 
\cite{Randjbar-Daemi:2000ft}
$$
A''+\frac{d_1}{2}{A'}^2
+\frac{d_2}{2}A'B'=\frac{4\kappa^2}{D-2}\left(
-\Lambda+\frac{F^2}{4g^2}\mbox{e}^{-2B}
\right)\;,
$$
$$
B''+\frac{d_2}{2}{B'}^2
+\frac{d_1}{2}A'B'=\frac{4\kappa^2}{D-2}\left(
-\Lambda-\frac{2D-d_2-4}{d_2} \frac{F^2}{4g^2}\mbox{e}^{-2B}
\right)+a \mbox{e}^{-B}
\;,
$$
$$
d_1A''+d_2B''+\frac{d_1}{2}{A'}^2
+\frac{d_2}{2}{B'}^2
=\frac{4\kappa^2}{D-2}\left(
-\Lambda+\frac{F^2}{4g^2}\mbox{e}^{-2B}
\right)\;,
$$
where $a$ is the scalar curvature of $K$. 
The above equations
are valid everywhere apart from the region 
$r=0$ where the brane is positioned
in $\mathbb{R}^+$. 

When $K$ is Ricci flat $a=0$ ({e.g.} a torus or a Calabi-Yau), 
the solutions for $F^2=0$ are
$$
A(r)=f_- \mbox{log}[z'(r)+f_+\mbox{z(r)}]\;\;,\;\;
A(r)= s_-\mbox{log}[z'(r)+s_+\mbox{z(r)}]
\;,
$$
where 
$$
f_{\mp}=\frac{2}{D-1}[1\mp
\sqrt{
\frac{d_2(D-2)}{d_1}
}
]\;\;
,\;\;
s_{\mp}=\frac{2}{D-2}[
1\mp\sqrt{\frac{d_1(D-2)}{d_2}}]\;,$$
and
$$
z=\mbox{Re}(\alpha \mbox{e}^{\gamma r}+
\beta \mbox{e}^{-\gamma r}
)\;\;,\;\;\gamma=\sqrt{\frac{-(D-1)\kappa^2\Lambda}{2(D-2)}
}\;.$$

These solutions are generalizations of the ones in 
\cite{Rubakov:1983bz,wett2,Gherghetta:2000jf}.
For $\Lambda<0$ a simple solution is 
$$
A=B=-\sqrt{-\frac{8\kappa^2\Lambda}{(D-1)(D-2)}}\;r\;.
$$

One can also extend the fifth direction to 
$r\in(-\infty,+\infty)$ for which the solution
is 
$A=B=-c|r|$,
and the behavior of the 
solutions is similar to the warp behavior in 
\cite{randal2}.
For $a\neq 0$, solutions can be 
be $AdS_{d_1+1}\times G/H$ ($\Lambda <0$) 
\cite{Randjbar-Daemi:1983qa,Randjbar-Daemi:1983hi}
with $A=-cr$ (or $A=-c|r|$ for $r\in (-\infty,+\infty)$)
and $B=\mbox{const.}$ (as in 
\cite{Gherghetta:2000jf}).

The solutions of both cases $a=0$ and $a\neq 0$, are compatible with 
the symmetry of Einstein equations $r\rightarrow -r$ and lead 
to the localization of gravity. The presence of the brane
as a singular source is essential as a consequence of the 
junction condition (\ref{junc}).

In section \ref{localf} we will see how these types of solutions
can be used to localize chiral fermions on the brane.




\subsection{Localization of Matter}\label{local}

The conventional way to avoid long-range observable effects 
of a non-compact internal space 
is to localize 
the fields on a thin 3-dimensional wall (brane).
In this case, there will not be infinite images of the brane world, 
and in principle one can have only one single brane on which ordinary 
matter is localized.

A particle will be localized if 
its
wave function dependence on the extra dimensions 
has an amplitude resembling a distribution function sharply 
localized around a particular point in the internal space, 
and this requires a specific non-trivial 
dependence of this wave function
on the coordinates of the internal space.   

In the direct product compactification, the 
zero modes are usually constant functions over the extra space,
unless they are coupled to a background configuration, as in 
\cite{sh2}.\footnote{There are no explicitly worked out 
examples for localizing fermions on a hyperspace in 
Kaluza-Klein field-theoretic compactification. Attempts
can be found in \cite{add1}. }
However in the warped geometry they may have non-trivial dependence
on the transverse space through the warp factor. 
\footnote{
It is perhaps worth mentioning that the theory 
of fermions in a non-trivial background
in a non-compact space with direct product  
can resemble a theory with warped product, with the 
warp factor being the wave function of the normalized 
zero modes in the internal space, see section \ref{rb}.}

An equivalent way to read whether a particle is 
``trapped''
or not, is done by looking at its
propagator in the full space-time. If its pole is of a scale
much larger than the typical low energy scale on the brane, then 
the zero modes (light particles) will 
be bound to the brane and the bulk modes will not be excited 
at those low energies. 

The first example of localizing 
gravity on a domain wall was given in \cite{randal2}
fermions in \cite{sh2,Gibbons:1987wg}, 
and 
abelian
gauge fields in
\cite{Dvali:1997xe}. The following section will 
provide
examples to some of these mechanisms.


\subsubsection{Gravity}
An interesting example of the warped solutions is
to consider living on the 
positive tension brane explained in section \ref{brane}
and to push the one of negative tension to 
infinity by taking the limit $r_c\rightarrow \infty$. In 
this case, it is convenient to 
parameterize the fifth dimension
with $y=\phi r_c$. It is interesting precisely because
this configuration 
``traps'' the massless graviton on the brane. 
To see this, let us 
write 
the linearized Einstein equations for the graviton,
$h_{\mu\nu}(x,y)=h_{\mu\nu}(x) h(y)$,
using for convenience a change of variable  
$
z=\mbox{sgn}[y](\mbox{e}^{k|y|}-1)
$; the  
gauge $\partial^\mu h_{\mu\nu}=0,\;h_\mu^\mu=0$;
and a proper normalization for $h(y)$, 
\cite{randal2}
$$
\left[
-\frac{1}{2}\partial_z^2+\frac{15k^2}{8(k|z|+1)^2}
-\frac{3k}{2}\delta(z)\right]h(z) = m^2 h(z)
\;.$$
Looking at the above equation one expects
1) a single bound state supported by the $\delta$-function trapped
at $z=0$, and 2)  
a continuous 
spectrum due to the 
non-compactness of the space, no mass
gap, and   
asymptotic behavior as plane-waves. The amplitudes 
of the continuum modes
should be 
suppressed near $z=0$ due to the potential barrier.
Further details related to the linearized gravity in 
a brane scenario have been worked out in \cite{rangrav}.

The normalized zero mode of this operator, corresponding to $m^2=0$,
is \cite{randal2,ran3}
$$
h_0(z)=\frac{1}{k(|z|+1/k)^{3/2}}\;,$$ 
and is trapped in the potential, representing the brane. 
This zero mode, unlike in the KK picture, 
depends non-trivially on the extra coordinate, luckily 
with a decreasing dependence. 
This observation will be replicated in section \ref{local}
where the wave function of the localized fermion zero modes 
will have similar
features.

The continuum modes are
$$
h_m(z) =N_m(|z|+1/k)^{1/2}\left[
Y_2(m(|z|+1/k))+\frac{4k^2}{\pi m^2}J_2(m(|z|+1/k))
\right]
\;,$$
where $m$ is the mass of the mode, $Y_2$ and $J_2$ are Bessel functions,
and $N_m$ is a normalization constant.\footnote{
Further details related to the linearized gravity in 
a brane scenario have been worked out in \cite{rangrav}, 
and more about gravitational trapping solutions can be 
found in \cite{Karch:2001ct,locgrav}.
}
The potential generated by the exchange of the zero and 
continuum 
modes behaves as \cite{randal2} 
$$
V(r)=G_N\frac{m_1m_2}{r}\left(1+\frac{1}{r^2k^2}\right)
\;,$$
where $G_N$ is the four-dimensional Newton's constant. 
So, at scales lower than $k$, the scenario mimics a brane with 
pure $4$-dimensional gravity (it was shown, however, 
that the five-dimensional gravity will again be 
manifest at very large distances,  
and that the above potential will be a good approximation in 
a finite region only \cite{Gregory:2000jc}).  

Close to the brane, the wave function of the continum modes 
is suppressed by a factor $\sqrt{m/k}$, and hence their
coupling to matter on the brane is weak for small $m$ 
and their production will be  insignificant 
at low energies
\cite{Hebecker:2001nv}.  

At large $z$, 
the massive KK gravitaional modes have strong coupling away from 
the brane, as their wave function behaves in this limit 
as $\sim\mbox{e}^{imz/k}$. However, the overlap 
between the wave functions of the zero mode and the continum modes 
is small \cite{randal2}, and this will ensure 
that the zero mode exchange and their self-couplings 
are four-dimensional
and that there is no ultimate coupling of the 
KK excitations back to the matter on the brane. 
It was shown in \cite{Garriga:2000yh} in an explicit
way that the gravitational interaction of \cite{randal2}
correspond to 
4-dimensional general relativity.
It was shown in 
\cite{Binetruy:2000ut}-\cite{Cline:1999wi}
that the standard Friedmann expansion of the Universe
can be recovered under certain conditions.  

\subsubsection{Fermions}
\label{localf}

Localizing fermion fields together with gravity, 
in the manner explained above, 
turns out to be impossible 
without additional Yukawa type couplings
\cite{Randjbar-Daemi:2000cr}. Equivalently, a 
non-trivial vacuum configuration
is required.  
The essence of the localization here is that the kernel 
of the diagonalized 
Dirac operator along the fifth direction is 
modified in the presence of a
non-trivial background in the extra space, 
very much the same as  
explained in section \ref{chiralint}.
It seems that the presence of this background is 
equally essential in the warped geometry as in the 
direct one for obtaining chiral fermions on the brane. 
For example, the coupled Dirac operator to a 
background configuration 
in five dimensions
\cite{sh2}
admits 
two zero modes, one of them turns out to be normalizable
while the other is not. Hence, all modes which couple to 
the non-renormalizable zero-mode will decouple from the action.\footnote{
The fermion localization procedure is very similar 
to the one 
of lattice gauge theories.  
The story began with the realization that dimensional regularization
explicitly violates the chiral gauge symmetry as 
does the other regularization procedures as Pauli-Villars
and zeta function teachniques. A lattice regularization
was seeked, however the na\"\i ve approach suffered from 
a doubler problem \cite{kars} as was shown 
that it is impossible to formulate a gauge theory with 
continuous chiral symmetry on
a lattice without doubling the species of fermions. 
The concept of domain wall fermions was introduced
\cite{kap}
in order to solve this problem by 
simulating the behavior of chiral fermions in
an even dimensional space, $2n$, by considering a lattice theory 
of interacting massive fermions in $2n+1$ dimensions. 
The localization mechanism in the theories of 
extra dimensions is not very different. For comprehensive 
and analytical description 
of chiral fermions on lattice see 
\cite{Narayanan:1994sk,Randjbar-Daemi:1995sq} and for
a recent review \cite{Neuberger:2001nb}.}

Let us first see why a Yukawa coupling to fermions is needed,
in a localized gravity context, by reviewing
\cite{Randjbar-Daemi:2000cr}. 
Consider the relatively general metric
ansatz (\ref{mod}). The Dirac equation on the 
($D=d_1+d_2+1$)-dimensional 
space $M_4\otimes_f K\times \mathbb{R}^+$
coupled to a gauge field is 
\begin{equation}
\Gamma^A E^M_A(\partial_M-\Omega_M+A_M)\Psi(x,y,r)=0,
\;,
\label{ss10}
\end{equation}
where $E^M_A$ is the vielbeins, and 
$\Omega_M=\frac{1}{2}\Omega_{M[A,B]}\Sigma^{AB}$
is the spin connection, 
$\Sigma_{AB}=\frac{1}{4}[\Gamma_A,\Gamma_B]$, and
$A_M$ is the gauge fields of a gauge group $G$. The non-vanishing
components of the spin connection is 
\begin{eqnarray}
\Gamma_\mu&=&\frac{1}{4}A'
\mbox{e}^{A/2}
\delta_\mu^a\Gamma_r\Gamma_a\;,\\
\Gamma_m&=&\frac{1}{4}A'
\mbox{e}^{B/2}
\delta_m^{\underline{a}}\Gamma_r\Gamma_{\underline{a}}
+\omega_m
\;,
\end{eqnarray}
where $\Gamma$ matrices are the constant Dirac matrices, and 
$\omega_m=\frac{1}{8}\omega_{m[\underline{a},\underline{b}]}
[\Gamma_{\underline{a}},\Gamma_{\underline{b}}]
$ is the spin connection on $K$ derived from the metric
$g_{mn}(y)=
e^{\underline{a}}_m 
e^{\underline{b}}_n
\delta_{\underline{a}\underline{b}}$. 
Assuming a background gauge field configuration only in the internal 
space $K$ and the absence of other Yang-Mills couplings 
to fermions the 
Dirac equation (\ref{ss10}) becomes
\begin{equation}
\left\{
\mbox{e}^{A/2}\slash{\partial}_{M}+ \Gamma_r
\left(
\partial_r+\frac{d_1}{4}A'+\frac{d_2}{4}B'
\right)+\mbox{e}^{-B/2}\slash{D}_K
\right\}\Psi=0\;,
\end{equation}
where $\slash{D}_K$ is the Dirac operator on $K$ in the 
background gauge field $A_m$, which is assumed
to admit zero modes (see section \ref{chiralint}). 
Let the zero modes of $K$ be $\psi(y)$, then 
$$
\Psi(x,y,r)=\psi(y)\zeta(r)\chi(x)\;,
$$
where $\zeta$ and $\chi$ satisfy
$$
\slash{\partial}_{M}\chi(x)=0\;\;\;,\;\;\;
\zeta(r)=\mbox{exp}\left[-\frac{d_1}{4}A-\frac{d_2}{4}B
\right]\;.
$$
The effective action in $d_1$ dimensions is hence
\begin{equation}
S_{M}=
{\bar \chi}(x){\slash{\partial}}_{M} \chi(x)
\int dr dy \sqrt{g(y)}
\;\mbox{e}^{-A/2} \psi^\dagger(y)\psi(y)
\label{locf}\;.
\end{equation}
On the other hand, 
the $d_1$-dimensional Newton's constant is
\begin{equation}
G_{d_1}^{-1}=G_{D}^{-1}V_{d_2}\int dr \;\mbox{exp}
\left[\frac{d_1-2}{2}A+\frac{B}{2}
\right]\;,
\label{locgrav}
\end{equation}
where $V_{d_2}$ is the volume of the compact manifold
$K$. The localization of gravity requires a finite $G_{d_1}$, 
while localizing fermions demands that (\ref{locf}) should 
converges. However, 
the integrals (\ref{locgrav}) and (\ref{locf})
do not simultaneously converge in the case of an 
exponential warp factor $A,B\propto -|r|$ so far considered in the 
literature since the function $\zeta(r)$ diverges.

Now let us introduce a scalar field $\Phi$ to the model,
through the Yukawa coupling ${\bar \Psi}\Phi\Psi$. The new Dirac
equation is 
$$
\left\{
\mbox{e}^{A/2}\slash{\partial}_{M}+ \Gamma_r
\left(
\partial_r+\frac{d_1}{4}A'+\frac{d_2}{4}B'
\right)+
g\Phi+
\mbox{e}^{-B/2}\slash{D}_K
\right\}\Psi=0\;.
$$
For the purpose of localization, as will be seen from the 
example given below, the dynamics of $\Phi$ is irrelevant. 
What matters is that $\lim_{|r|\to\infty}
\Phi =|\phi|\epsilon (r)$ where $\epsilon(r)$ is the sign function
and $\phi\equiv\left<\Phi\right>$
({\it i.e.} 
behaves as 
a kink at $\pm \infty$). Assuming this together with 
imposing the chirality condition 
$\Gamma^r\Psi=+\Psi$ or 
$\Gamma^r=1$ 
(in the case when
$d_1+d_2$ is even) the Dirac equation away from the origin becomes
$$
\slash{\partial}_{M}\chi(x)=0\;\;\;,\;\;\;\slash{D}_K\psi(y)=0\;,
$$
$$
\left(
\partial_r+
\frac{d_1}{4}A'+\frac{d_2}{4}B'+g|\phi|\epsilon(r)
\right)\zeta(r)=0\;,
$$ 
which admits the solution
$$
\Psi(x,y,r)=\mbox{exp}\left[
-\left(\frac{d_1}{4}A+\frac{d_2}{4}B
\right)-g|\phi|\epsilon(r)
\right]\psi(y)\chi(x)\;.
$$
The condition for having localized fermions is hence
\begin{equation}
-\frac{A}{2}-2g|\phi|r\epsilon(r)\; <\;0\;.
\label{cond}
\end{equation}
This can be achieved for large enough values of $g|\phi|$.

The chirality of the localized (normalizable) zero modes will 
be determined by the solutions of $\slash{D}_K\psi(y)=0$ in the
presence of the background. $K$ will be even 
dimensional if $M=M_4$, and hence the solutions to the Dirac 
equation on $K$ will have a definite chirality. Moreover, 
in this case $\psi$ and $\chi$ will have the same chirality, 
and the number of chiral fermions will be equal 
to the index of $\slash{D}_K$ which is the difference
between the number of negative and positive chirality 
zero modes. 

\subsubsection*{Rubakov-Shaposhnikov mechanism}
\label{rb}
Now let us present an illustrative example
of fermion localization in the 
presence of a background \cite{sh2} in five dimensions. 
The idea of \cite{sh2} is a special case of the above
with $A=B=0$ and no compact space $K$. 
The condition (\ref{cond}) is automatically 
satisfied when $\Psi$ is in the positive region 
$\epsilon=+1$. 
As explained above, we necessarily need a 
scalar field $\Phi$ which acquires a non-zero , and 
varying,vacuum 
expectation value 
only along the extra space.
Obviously, $\phi(y)$ 
breaks the full translational invariance, as it is
needed to have a preferred direction orthogonal to the wall.

The vacuum expectation value of $\Phi$ can be regarded
as background gauge field, as in \cite{Kaplan:1996pe}, 
which 
has a domain-wall configuration, {\it e.g.} a kink, 
and this will provide an elegant dynamical origin 
for the spontaneous breaking of the 5-dimensional 
translation invariance. 

The fermionic field will 
localize, as will be explained below, 
where its total mass $${\hat m}_0
\equiv m_0 + g \phi(y)$$ 
vanishes ($m_0$ is
the bare fermionic mass in the five-dimensional theory), 
on a
wall with three spatial dimensions characterized by a particular 
position $y$
in the transverse direction.
\begin{figure}
\centerline{\leavevmode\epsfysize=3cm \epsfbox{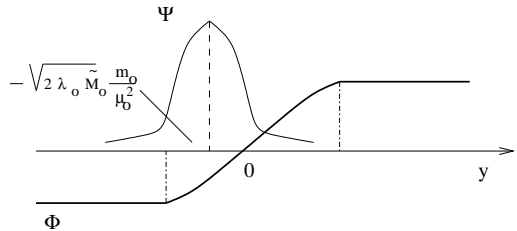}}
\vspace{10pt}
\caption{\small 
Profile of the scalar field $\Phi$ and of the fermionic field
$\Psi$ along the fifth dimension, in the presence of 
the kink approximated in (\ref{approx}). 
The fermionic field is localized where its total
mass
vanishes.}
\label{figkf}
\end{figure}

For definiteness, we consider the theory described by the action
\begin{equation}
S =\int d^4xdy\;
{\Psi}(x,y) 
\bigl[
i\slash{\partial}_4+ i 
\slash{\partial}_5 +
\frac{1}{{\tilde M}_0^{1/2}}  
\phi (y) + m_0 \bigr] \Psi(x,y) 
\label{lagpp}\;,
\end{equation}
where the subscript 
``$0$'' indicates the value of the parameter at zero
temperature (we will discuss finite temperature effects later on), 
and the fields and the parameters
have the following mass dimensions
$$
\left[ \phi \right] = \frac{3}{2} \;,\; \left[ \Psi \right] = 2 \;,\;
\left[ m_0 \right] = \left[ {\widetilde M}_0 \right] = 1 \;\;.
$$

For the time being
we will ignore the dynamics of 
the field $\Phi$, since the localization mechanism 
is concerned only with the background configuration. 
It is
possible, however, that the configuration can decay 
due to 
the dynamics of the field $\Phi$,
{\it e.g.} interaction with fermions, 
thermal, and other effects.
We will discuss thermal effects in section \ref{baryogm}, 
and shall 
assume the stability of the vacuum configuration 
regarding what follows. At this point the shape of the 
configuration does not matter, what matters is that
it is not constant in $y$, ${\phi'}(y)\neq 0$.

The coupled Dirac equation will hence be \footnote{We follow
here the notation of \cite{Arkani-Hamed:2000dc} concerning
Dirac matrices,...{\it etc}.}
\begin{equation}
\left( i \gamma^\mu \partial_\mu + 
\gamma^5 \partial_5 + {\hat m}_0 
\right) \psi = 0\;. 
\label{dirpsi}
\end{equation}
The spinor $\Psi$ can be decomposed, in a Lorentz non-invariant
way in 5 dimensions, 
into a left and right handed fermion with respect to the chirality 
matrix $\gamma_5$ in four dimensions as in (\ref{ferm}). The Fourier
expansion can be carried out \footnote{We perform a discrete expansion
here in order to illustrate the idea in an easier way, 
while we keep in mind that the length of the fifth dimension
is to be take to infinity at the end. One can alternatively 
keep the length big ``enough'', and not necessarily infinite,
however this will be at the cost of introducing an 
inexplicably
small mass scale to the theory.}
\begin{eqnarray}
\Psi (x,y)
&=& \sum_n 
L_n (y)P_L \psi_n(x)+
\sum_n R_n (y) P_R \psi_n (x) \;\;, \nonumber \\
{\bar \Psi} (x,y)
&=&\sum_n {\bar \psi}_n(x) 
P_R L_n^* (y)
+  \sum_n {\bar \psi}_n(x) 
P_L R_n^* (y)  \;\;,
\label{exppsi}
\end{eqnarray}
where $P_{L,R}=(1\pm i\gamma^5)/2\;$. Since the kernel 
of the Dirac is equal to the kernel of its square, let us 
look at the square of the equation (\ref{dirpsi}) in our 
search for the zero modes
\begin{equation}
\left( \slash{D}_4^2 - 
\partial_5^2 - i \gamma^5 
{{\hat m}'}_0 +
{\hat m}_0^2 
\right) \Psi = 0 \;\;,
\label{eqsq}
\end{equation}
where the prime again 
denotes a derivative 
with respect to $y$. 
The equation (\ref{eqsq}) can be rewritten using 
(\ref{exppsi}) as
\begin{eqnarray}
\left( - \partial_5^2 + 
{\hat m}_0^2 
- {{\hat m}'}_0
\right) L_n &=& \mu_{n,L}^2 L_n\;, 
\label{auto1}\\
\left(-\partial_5^2 + 
{\hat m}_0^2 
+{{\hat m}'}_0
\right) R_n &=& \mu_{n,R}^2 
R_n \;,\label{auto2}
\end{eqnarray}
where $\mu_{n,L}$ and $\mu_{n,R}$ are the eigenvalues of the
four-dimensional Dirac operator $\slash{D}_4$ related to the 
eigenfunctions $P_L\psi_n$ and $P_R\psi_R$ respectively.
 
Define
\begin{equation}
a \equiv \partial_5 +  {\hat m}_0
\;\;\;,\;\;\; a^\dagger 
\equiv - \partial_5 +{\hat m}_0.
\label{defop}
\end{equation}
Therefore
\begin{eqnarray}
a^\dagger a &=& 
-\partial_5^2 + 
{\hat m}_0^2
-  {{\hat m}'}_0
\;, \nonumber \\
aa^\dagger  
&=& - \partial_5^2 + 
{\hat m}_0^2
+{{\hat m}'}_0.
\end{eqnarray}
In this notation, the equation (\ref{auto1}) and (\ref{auto2}) 
can be written as
\begin{eqnarray}
a^\dagger a L_n &=& \mu_{n,L}^2  L_n , \label{susyqm1} \\
a a^\dagger R_n &=& \mu_{n,R}^2 R_n . \label{susyqm2}
\end{eqnarray}
The eigenfunctions 
$L_n$ and $R_n$ can be normalized to form two 
sets of orthonormal functions. Note that the operators
$aa^\dagger$ and $a^\dagger a$ commute only when 
$\phi(y)$ is constant in $y$. Also
multiplying (\ref{susyqm1}) by $a$ from the left shows that 
$a L_n$ is an eigenfunction of the operator 
$aa^\dagger$ with eigenvalue $\mu_{n,L}^2$ 
for $n$ different from zero. 
So, for $n\neq 0$ one can write 
$\mu_{n,L}=\mu_{n,R}\equiv \mu_{n}$, or in other words
\begin{equation}
R_n =\frac{1}{\mu_n} a L_n \;\;\;,\;\;\; 
L_n = \frac{1}{\mu_n} a^\dagger R_n .
\label{rellr}
\end{equation}
The above relation does not hold for the zero modes $L_0$ and $R_0$
corresponding to $\mu_0 = 0$.
The zero mode wave functions are found by integrating the two equations
$a L_0 = 0$ and $a^\dagger R_0=0$.
The solutions are
\begin{eqnarray}
L_0 &\sim& \mbox{exp} 
\left[- \int^{y} d s \;
{\hat m}_0(s) \right] \;\;, \nonumber \\
R_0 &\sim& {\mbox{exp}} \left[\int^{y} 
d s \,  {\hat m}_0
(s) \right]\;. \label{psish}
\end{eqnarray}
If, for example,  
$\phi(y)$ has a kink configuration as in the figure
\ref{figkf} and if the extra
dimension is infinite, only the left-handed mode 
$L_0$ is normalizable and
is
localized around the zero of its total 
mass 
$m_{\mbox{\tiny tot}}$. $R_0$ will not be normalizable 
and its coupling to other fields will always be suppressed
by the length of the fifth direction. 
The effective four-dimensional theory will hence contain a  
massless chiral fermionic field (in this case left-handed). 
The action (\ref{lagpp})
can now be expressed 
in terms of the 4-dimensional fields, using the 
orthogonality of 
$\left\{ L_n\right\}$ and $\left\{R_n\right\}$, as
\begin{equation}
S = \int d^4 x \left[
{\bar \psi}_{0,L} i \gamma^\mu \partial_\mu \psi_{0,L} + 
\sum_{n=1}^{\infty} {\bar \psi_n} \left(i\gamma^\mu 
\partial_\mu + \mu_n \right) \psi_n \right]\;.
\end{equation}
The first two terms correspond to 4-dimensional two-component massless
chiral fermions, and arise from the zero modes of equations
(\ref{susyqm1}) and
(\ref{susyqm2}). The third term describes an infinite tower of Dirac
fermions corresponding to the (Kaluza-Klein) modes with non-zero $\mu_n$
in the expansion (\ref{exppsi}). If the extra space is 
sufficiently large,
these modes decouple completely from the low energy theory.

\begin{figure}
\centerline{\leavevmode\epsfysize=5cm \epsfbox{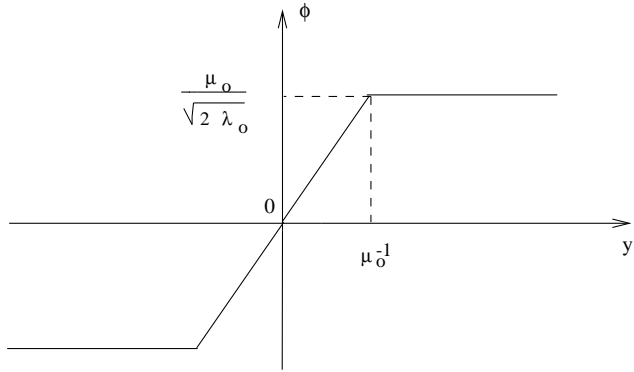}}
\caption{\small A linearized approximation to the kink 
solution of (\ref{kink}).}
\label{kink2}
\end{figure}

Let us now consider a specific shape for the configuration,
as this will serve for 
section \ref{baryogm}. Let the field $\Phi$ have the following 
Lagrangian 
\begin{equation}
{\cal L}_\Phi=\frac{1}{2}\partial_M\Phi \partial^M\Phi
-
(-\mu_0^2\Phi^2+\lambda_0\Phi^4)\delta(y)\;,
\label{lagk}
\end{equation}
where $\mu_0$ and $\lambda_0$ have mass dimensions
1 and -1 respectively. A solution of the equations of motion is
\begin{equation}  
\phi(y)=\frac{\mu_0}{\sqrt{2 \lambda_0}} 
\mbox{tanh} 
[\mu_0 \;y]\;,
\label{kink}
\end{equation}
which can be approximated with a 
straight line interpolating between the two
vacua (see figure \ref{kink2})
\begin{eqnarray} 
\phi (y) &\simeq& \frac{\mu_0^2}{\sqrt{2 \lambda_0}} 
\;y \;\;,\;\; |y| < \frac{1}{\mu_0}\nonumber \\
\phi( y) &\simeq& \pm \frac{\mu_0^2}{\sqrt{2 \lambda_0}}
\;\;,\;\; |y| > \frac{1}{\mu_0}\;. 
\label{approx}
\end{eqnarray}
The localization can occur only if
\begin{equation}
m_0 < \frac{\mu_0}{\sqrt{2\,\lambda_0\,{\widetilde M}_0}} \;\;,
\label{lcon}
\end{equation}
since otherwise the total fermion mass $\hat m$
never vanishes at any specific point $y$.

It can be shown that, from the four-dimensional point of view, a left handed
chiral massless fermionic field results from the localization mechanism, 
if the
above configuration (\ref{kink}) is assumed for the scalar $\phi$. 
The right
handed part remains instead de-localized in the whole space. This is not a
problem since it is customary to limit the Standard Model 
fermionic content only to left handed fields. 
The right handed fields can
also be localized  if a kink--antikink solution is assumed for the scalar
$\phi$. As a result, the left fields continue to be localized on the kink,
while the right ones are confined to the antikink. 
If the kink and the antikink
are sufficiently far apart, the left handed and right handed fermions however
do not interact and again the model reproducing our 
4-dimensional world must
be built by fermions of a definite chirality. 
The fermionic content of the full
dimensional theory is in this case doubled with respect to 
the usual one, and
observers on one of the two walls will refer to the other as 
to a ``mirror
world''. The presence of this kink--antikink 
configuration may be required by
stability consideration if thermal effects 
are considered. 

In order to give mass to the fermions, some other scalar field acting as a
Higgs in the four-dimensional theory must be considered. 
As it is shown in 
\cite{Arkani-Hamed:2000dc}, the 
mechanism described above could give an explanation to the
hierarchy among the Yukawa couplings responsible 
for the fermionic mass matrix.
If indeed one chooses different 
five-dimensional bare masses for the different
fermionic fields, 
their wave functions will only
partially overlap, as a consquence, 
and increasing the distance  
between them in the transverce direction to the brane
results in suppressing their mutual 
interaction.

\section{Living with Extra Dimensions}
\label{lwed}
\subsection{Introduction}

Imagining our Universe being extended to more than four dimensions
at energies around TeV may imply dramatic experimental consequences.
First of all, the force of
gravity is expected to become comparable to the other gauge 
forces at around TeV which would enable LHC and NLC \footnote{
LHC (CERN) has a center of mass energy around 14 TeV
and is scheduled to operate in 2005, 
while
NLC (SLAC), the 
anticipated $e^+e^-$ collider, 
has a center of mass energy 500-1500 GeV.
For reviews see \cite{Murayama:1996ec,Peskin:1999hj}.}
to probe the quantum structure of gravity. 
This could also be checked by the new experiments
measuring gravity at sub-millimeter 
distances \cite{price,Long:1999dk}. Secondly, since the masses
of the Kaluza-Klein excitations are typically of order of
inverse the compactification scale, they will be excited
by being exchanged or emitted
at such energies and will be expected to appear 
in collider experiments, Cosmological, and 
Astrophysical environments \cite{add2}. 
Further, a multi-dimensional Universe will 
generally evolve different from the 
usual Friedmann-Robertson-Walker expansion law. 
\footnote{See \cite{Kolb:2000se} for a breif summary
of
the present status of the standard Big Bang 
model, and 
\cite{Riess:1998cb,Perlmutter:1999np} for the 
recent
interesting
developments.} To fit the 
expected contribution of the infinitely new degrees
of freedom within the known experimental and observational 
data, it is necessary to bounds on the parameters of the
models with extra dimensions, both compact or non, 
mainly lower bounds on the fundamental scale of gravity whether 
this scale is set by the size of the compact dimensions, the 
characteristic scale which enters in the localization 
of matter transverse a non-compact space, or other relevant 
model dependent scales. 
In any case, the larger the number of the extra dimensions
is the looser the bounds become. 

In the following we review some of the rather 
established
bounds 
imposed
on models with large extra dimensions,
\footnote{Discussion about searches 
for extra dimensions in future colliders can be found 
in \cite{kap3} and \cite{Mirabelli:1999rt,Cheung:1999fj,Rizzo:2001sd}.} 
with more emphasis on 
constraints arising from compatibility with 
Cosmological observations. After having done so, we 
go on to present examples where the standard big bang
Cosmology can be recovered 
in Kaluza-Klein compactification
schemes and in Randall-Sundrum type scenarios. 
Then we move on to discuss briefly the 
issue of stabilizing the scale of compactification, 
proton stability in theories with extra
dimensions, and finally baryogenesis with low scale
gravity. We conclude this chapter by presenting a model 
for baryogenesis in 5 dimensions. 

\subsection{General bounds}

The most stringent bound 
\footnote{It was claimed racently 
nucleon-nucleon gravi-bremsstrahlung
in the early Universe could give a stronger bound 
that SN1987A provides
\cite{Fairbairn:2001ct}.}
on the size of 
the compact space comes from
the emission of the 
supernova 
SN1987A
core of 
large fluxes of Kaluza-Klein gravitons
which affects its energy release
\cite{Cullen:1999hc,Hall:1999mk} (see \cite{add2} 
for a review).
For $n=2$, these constrains turn out to be  
$R<0.9\times 10^{-4}$mm and for 
$n=3$, $R<1.9\times 10^{-7}$mm \cite{Hannestad:2001jv}. 
Collider and other bounds on the size of compact
extra dimensions can be found in
\cite{add2},
and a more recent analysis in
\cite{Hall:1999mk}-\cite{Giudice:1999ck}.
Higher-dimensional operators suppressed by small scales are
model dependent and they do not significantly 
modify the 
SM cross sections and precision observables 
for a gravity scale greater than 1TeV \cite{add2}.
\footnote{A serious problem in Kaluza-Klein large radii compactification
is proton decay. For that, a TeV gravity scale 
will cause a serious problem \cite{Benakli:1999ur}, 
unless a specific symmetry is imposed. }
The bounds on non-compact warped, Randall-Sundrum type, 
scenarios are 
less elaborated and more model-dependent; 
such constraints can be found in  
\cite{Besancon:2001wp,Rizzo:2000cv,Davoudiasl:2000jd}.

\subsubsection*{KK gravitons}

The existence of 
light Kaluza-Klein gravitons is a general 
feature of compact large extra dimensions, although
their coupling properties to ordinary matter
is more model-dependent.\footnote{For the 
manifestation of 
these states in a string theory context see for example
\cite{Cullen:1999hc}.} 
Each KK graviton mode interacts with the ordinary matter
through the four-dimensional 
energy-momentum tensor and its coupling is 
gravitationally week, suppressed by $1/M_P$, 
as can be read from 
$$
S_n=M^{d+2}\int d^{d+4}z \;
\left[\partial h(x)\mbox{e}^{iq_ny}\right]^*
\left[\partial h(x)\mbox{e}^{iq_ny}\right] +\int d^4x\;
h(x)T(x)
$$
which shows that upon integrating over
$y$ the volume of the compact space will appear in
front
of the first term producing $M_P^{2}$. The proper rescaling
gives a coupling $M_P^{-1}$ with the energy momentum tensor
(ordinary matter). 

Despite the gravitationally weak coupling, 
there are a large multiplicity of those massive gravitons 
below an energy level, $E$, which is enormous $\sim (ER)^d$, 
and their combined effect is much stronger than the 
gravitational suppression. One of the 
typical processes for producing a KK graviton 
is 
$$
e^+e^-\rightarrow \gamma+\mbox{\small graviton}
\;.$$
The graviton produced here 
will have the form of a missing energy. 
The total cross section is of order $\alpha/M_{P}^2$ 
and is
$$
\sigma(
e^+e^-\rightarrow \gamma+\mbox{\small graviton}
)\sim \frac{\alpha}{M_P^2}(ER)^d\sim\frac{\alpha}{E^2}\left(
\frac{E}{M}
\right)^{d+2}
$$
which becomes comparable to the electromagnetic cross section
at energies $E\sim M$ (for further 
details 
of this and others, like hadronic processes, 
we refer the reader to 
\cite{Giudice:1999ck}). 

Beside their possible detection in colliders, through their
emission or exchange, 
they can be produced at high temperature sometime in the early
Universe, and if they decay after the Big Bang 
Nucleosynthesis (BBN) they will distort the Cosmic Microwave Background
Radiation. 
Not only that, their abundant number and 
large mass will change the expansion of the Universe, 
and this may be done during the BBN by slowing the expansion 
down since they red shift as matter, $\sim R^{-3}$
rather than radiation $\sim R^{-4}$. 
Furthermore, their energy density may over-close the Universe
if their abundance is comparable to the photon abundance at 
early times.
These obstacles are enough 
to impose a strong upper bound on the temperature of the Universe,
$T_*$, after which it should behave as the usual Friedmann-Robertson-Walker
(FRW) Universe \cite{add2}.  

We explain below how these bounds arise
(more or less along the lines of \cite{rubn}). 

At high enough temperature $T\gg R^{-1}$, the creation rate 
per unit time and volume of a 
KK graviton of mass $m_n\lesssim T$, based on dimensional 
analysis,\footnote{Strictly speaking, 
assuming a four-dimensional KK gravity production
at scales much higher than $R^{-1}$ makes little sense, since
at such high scales the Universe appears multi-dimensional
presuming the general coordinate invariance is restored
at high energies (the break down of the isometry of the 
full space, and the formation of a domain wall,  
is usually due to a spontaneous breaking characterized by $R^{-1}$). }
$
\Gamma\sim \frac{T^6}{M_P} 
$,
where the factor $M_P^{-1}$ is the strength of the coupling of 
the KK graviton to matter. The estimate for the 
total rate of KK graviton production would be 
$$
\frac{dn}{dt}\sim \frac{T^6}{M_P}(TR)^d
\sim T^4\left(\frac{T}{M}\right)^{2+d}
.$$
Assuming that after a certain temperature, $T_*$,
after which 
the Universe evolves in the standard way (presumably during the
reheating after 
inflation) $H=T_*^2/M_P^{*  }$, 
where $M_P^{*2}=M_P/(1.66g_*^{1/2})\sim 10^{18}\mbox{GeV}$, and
$g_*$ is the effective number of relativistic degrees of 
freedom, one finds that the number of KK gravitons created in a Hubble
time per relativistic species (photons) is
$$
\frac{n_{KK}(T_*)}{n_\gamma}
\sim  M^*_P\frac{T_*^{1+d}}{M^{2+d}}
,$$
which is fairly large, and requires $n_{KK}\ll n_\gamma$. 

A stronger upper bound on the temperature $T_*$ comes from 
slowing the expansion of the Universe
at temperature below $T_*$, since KK gravitons
produced after this temperature red-shift as non-relativistic particles.
At the time of BBN their mass density per energy density 
of photons is of order
$$
\frac{\rho_{KK}(1\mbox{MeV})}{\rho_\gamma}
\sim
\frac{T_*}{\mbox{1MeV}}\times \frac{T_*^{d+1}M_P^*}{M^{d+2}}\;.
$$
Requiring $\rho_{KK}\ll \rho_\gamma$ yields to the 
following upper bound on the reheating temperature (or {\it 
normalcy} temperature) \cite{add2}
$$
T_*\lesssim 10^{\frac{6d-9}{d+2}}
\;\mbox{MeV}\; \frac{M}{1\mbox{TeV}}\;.
$$
The upper bound reads for 
$M=1$TeV in two dimensions $10$MeV which it too severe. The 
bound relaxes somehow for $d=6$ and is $T_*\lesssim 1$GeV.

Finally we come to the most constraining Cosmological 
bound which is the risk of overclosing the Universe.
The energy density of the massive KK gravitons should
not exceed the critical energy density today which 
correspond to $\sim 3\times 10^{-9}$GeV.
The life time of a graviton with an energy $E$ is
$$
\tau(E)=\frac{M_P^2}{E^3}
\sim 10^{10}\mbox{yr.}\left(
\frac{10^2\mbox{MeV}}{E}
\right)^3\;.
$$

Gravitons produced at temperatures below $10^{2}$MeV
have lifetime at least as the age of the Universe. 
The energy density stored in the gravitons produced at a 
temperature $T_*$ is
$$
\frac{\rho_{KK}
(T_*)}{\rho_\gamma}
\sim T_* n_{KK}
\simeq
\frac{T_*^{d+1}M_P}{M^{d+2}}
$$
which gives for $\rho_{KK}\ll \rho_\gamma$
$$
T_*\lesssim 10^{\frac{6d-15}{d+2}}\;\mbox{MeV}\; \frac{M}{1\mbox{TeV}}
$$
and this constrains $T_*$ severely from above by 
a value of $1.7$MeV, $0.3$GeV, and $2$GeV for $2$, $4$, 
and $6$ extra dimensions with fundamental 
gravity scale $M\simeq 10$TeV!. This bound 
suggest that $M> 10$TeV for this scenario to be 
compatible with BBN in the case of $2$ extra dimensions. 
\begin{figure}
\centerline{\leavevmode\epsfysize=6cm \epsfbox{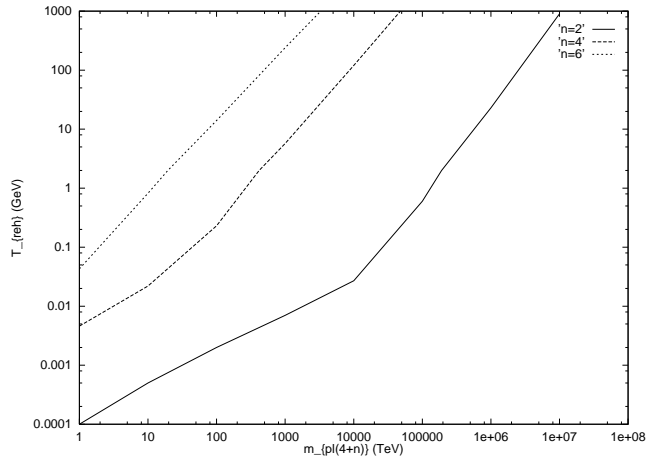}}
\vspace{10pt}
\caption{\small The maximum allowed normalcy temperature
$T_*$ as a function of the fundamental gravity scale $M$
for various number of extra dimensions (taken from 
\cite{Benakli:1999ur}).}
\end{figure}

\subsection{Kaluza-Klein Cosmology}\label{kkcosmology}
It is obvious that the evolution of  
the Universe as a system in 
more than four dimensions will 
be different than the standard Friedmann-Robertson-Walker 
scenario.

The question to pause is whether or not recovering the 
standard Cosmological expansion is possible and under 
which conditions. The answer to this question 
is yes, it is possible for the four-dimensional 
Universe to evolve according to the usual Freedmann
expansion law if the internal space is 
small enough compared to the size of the Universe
\cite{Chodos:1980vk}-\cite{Randjbar-Daemi:1984jz}. 
Other Cosmological consequences were
also  
discussed as massive relics \cite{Kolb:1984fm}, inflation
\cite{Shafi:1983hj}, and other finite temperature effects 
in \cite{Alvarez:1983kt,Maeda:1984fq}.
In the few past years the some of the above issues 
were re-examined in addition to some new cosmological 
and phenomenological  
aspects \cite{Krori:1990gw}-\cite{Starkman:2001xu}. 
 
Let us look at what happens to a Universe living 
in $D$ dimensions with the following metric
\begin{equation}
g_{MN}=\left(\begin{array}{ccc}
-1 &&\\
&R(t)^2{\tilde g}_{\mu\nu}(x)&\\
&&a(t)^2{\tilde g}_{mn}(y)
\end{array}\right)\;,
\label{kkmetric}
\end{equation}
where ${\tilde g}_{\mu\nu}$ and ${\tilde g}_{mn}$ are the 
metrics on the $d_1$-dimensional space $M$ ($d_1=4$ eventually) 
and the $d_2$-dimensional compact one $K$ respectively. 
Further, we assume the Universe to have the 
perfect fluid description in $D$ dimensions, {\it i.e.}
\begin{equation}
T_{MN}=\left(\begin{array}{ccc}
-\rho &&\\
&p{\tilde g}_{\mu\nu}&\\
&&p'{\tilde g}_{mn}
\end{array}\right)\;,
\label{etensor}
\end{equation}
where the energy density $\rho$ and the pressures $p,p'$
may depend on time, but not the space coordinates $x^\mu, y^m$.
They are also assumed to satisfy 
the equation of state $\rho=(D-1)p$,\footnote{This 
equation of state is not well justified. For a generic
treatment see \cite{Randjbar-Daemi:1984jz}.} $p=p'$.

The Einstein equations for the scale factor $R$ and $a$
in terms of the entries of the energy-momentum tensor can 
be derived from the action 
\begin{equation}
S= \int d^{d_1} x d^{d_2}y \;\sqrt{-g} \left(\frac{1}{\kappa^2}
{\cal R}+\Lambda \right)\;.
\label{action5}
\end{equation}
For $\Lambda=0$ they are
$$
(d_1-1)\frac{\ddot R}{R}
+d_2\frac{{\ddot a}}{a}=
-\kappa^2\rho\;,
$$
$$
\frac{{\cal R}_M}{R^2}+
\frac{d}{dt}\left(\frac{\dot R}{R}\right)
+\left((d_1-1)\frac{\dot R }{R}
+d_2\frac{\dot a}{a}
\right)\frac{\dot R}{R}=\frac{\kappa^2}{D-1}\rho
\;,
$$
$$
\frac{{\cal R}_K}{a^2}+
\frac{d}{dt}\left(\frac{\dot a}{a}\right)
+\left((d_1-1)\frac{\dot R}{R}
+d_2\frac{\dot R}{R}
\right)\frac{\dot a}{a}=\frac{\kappa^2}{D-1}\rho
\;,
$$
where ${\cal R}_M$ and ${\cal R}_K$ are the curvature scalars
of $M$ and $K$ respectively. 
Looking at the equations of motion (setting $\Lambda=0$)
it was shown in \cite{Sahdev:1984fp}
that, for a flat $M$, the typical distance $R(t)$
between two point on $M$ increases monotonically from $0$
to $\infty$ at every value of $R_0$, while $a $, 
the radius 
of $K$, increases from zero to a maximum then drops back to zero.
A similar varying behavior
of the compactification scale has been noted
in \cite{Rubin:1983ap} \footnote{There is a major difference 
in the two results as in \cite{Rubin:1983ap}
the compact space tends to increase after a certain critical length, 
in contrast with \cite{Sahdev:1984fp}.} 
by taking into account combined thermal and quantum mechanical
fluctuations 
effects, 
and the instability feature remains to be accounted for.
Therefore a stabilization mechanism is needed to set 
$a$ at the desired value whatever it is chosen to be
in a short enough period, 
not only for keeping the fundamental scale of gravity low enough, 
but also in order not to cause any conflict with
the variations of the four-dimensional 
Newton's constant $\kappa^2$ \cite{Steigman:1979kw} and
big bang nucleosynthesis \cite{Kolb:1986sj}.  

Assuming that such a localization mechanism exists and 
that its net effect is to induce a constant effective 
pressure $p_c=\frac{1}{a_c^2}$
the independent four-dimensional 
Einstein equations to be solved at low energies become
\cite{Sahdev:1984fp} 
$$
\frac{d}{dt}\left(\frac{\dot R}{R}\right)
+(d_1-1)\frac{{\dot R}^2 }{R^2}
={\kappa^2}\rho\;\;\;,\;\;\;
(d_1-1)\frac{\ddot R}{R}=\kappa^2\rho
$$
which clearly resemble the evolution of
FRW Universe. 

As pointed out in \cite{Sahdev:1984fp}, the 
temperature profile of the early Universe is radically
changed due to the presence of the extra dimensions. 
If all the dimensions were expanding, the Universe would 
have only cooled down by time. However, if the extra space 
undergoes a period of expansion and contraction, 
the $D$-dimensional Universe will start
off very hot, cools down to a certain limit, 
afterwhich it reheats until it becomes effectively 
the observed four-dimensional Universe. 

The energy momentum tensor (\ref{etensor}) was shown in 
\cite{Randjbar-Daemi:1984jz} to be derived from a free energy function
of free bosonic gas in thermal equilibrium. It was shown that
at energies below the compactification mass scale, the 
Universe expands as usually observed in four dimensions 
while the radius of compactification
remains constant.\footnote{For alternative expansion, 
{\it e.g.} inflation, see for instance 
\cite{Shafi:1983hj,Arkani-Hamed:1999kq,Arkani-Hamed:2000gq}.}  
This is done by adding a source term 
to (\ref{action5}) 
$$
\Gamma =\int d^{4}xd^dy\sqrt{-g} 
\left(\frac{1}{\kappa^2}
{\cal R}+\Lambda \right)+\Gamma_1\;,
$$
where $\Gamma_1$ represents the contribution of matter
fields, vacuum fluctuations,...{\it etc}. Its variations
defines the energy-momentum tensor
$$
\delta \Gamma_1
=\int d^{4}xd^dy\sqrt{-g} \frac{1}{2}\delta g^{MN}
T_{MN}\;.
$$
The stress tensor $T_{MN}$ is as usual 
expected to have the same invariances as the metric. 
In \cite{Randjbar-Daemi:1984jz} the 
full manifold is considered $R^1\times S^3\times S^d$, where 
the 3-dimensional space is taken to be closed. The case 
of flat Universe can be obtained by taking $R\rightarrow \infty$
after a careful transition from the discrete to continuum 
limit. The symmetry of $T_{MN}$
in this case is $O(1,3)\times O(d+1)$.
The action with this metric ansatz is hence
\begin{equation}
\Gamma=\int dt \;\frac{\Omega_3\Omega_d}{\kappa^2}\left[
\frac{6{\dot R}^2}{R^2}
+6d \frac{\dot R}{R}\frac{\dot a }{a}
+d(d-1)\frac{{\dot a}^2}{a^2}
\right]+U\;,
\label{action7}
\end{equation}
where $\Omega_3$
and $\Omega_d$ denote the spatial volumes
$$
\Omega_3=2\pi^2R^3\;\;\;,\;\;\;
\Omega_d=(2\pi)^{\frac{d+1}{2}}\frac{a^d}{\Gamma(\frac{d+1}{2})}
\;,
$$
and $U$ is the effective ``potential'' which incorporates a classical
gravity terms arising from ${\cal R}+\Lambda$,
from the one-loop quantum part $\Gamma_1$, and from 
the thermal part depending on the entropy $S$,
\begin{equation}
U=\frac{\Omega_3\Omega_d}{\kappa^2}
\left[
-\frac{6}{R^2}-d\frac{(d-1)}{a^2}
+\Lambda
\right]+ 
U_{1-{\tiny loop}}
\;.
\label{en5}
\end{equation}

The Free energy to be computed is obtained by
Legendre transform of $U$
\begin{equation}
U_{1-{\tiny loop}}=
F+TS=\Omega_3\Omega_d\rho(R,a,S)\;,
\label{energy6}
\end{equation}
where the temperature is given by $T=\partial U/{\partial S}$.
Let the pressures $p$ and $p'$ entries in (\ref{etensor})
be defined by the thermodynamic identity
$$
d\left[
\Omega_3\Omega_d\rho(R,a,S)
\right]=TdS-\Omega_dpd\Omega_3-\Omega_3p'd\Omega_d\;,$$
then the conservation of energy $\nabla_MT^{MN}=0$
implies the conservation of entropy.\footnote{Recall that
in the usual treatment \cite{weinbc}
the energy density and pressure 
must satisfy the integrability condition
$$\frac{\partial^2S(V,T)}{\partial T\partial V}=
\frac{\partial^2S(V,T)}{\partial V\partial T}\;,
$$
where $dS(V,T)\equiv dQ/T=[\rho(T)dV+p(T)dV+Vd\rho]/T$
which imply $dp/dT=(\rho(T)+p(T))/T$. Since the 
equations of motion are $d/dt[R^3(\rho+p)/T]=0$, the entropy 
will be constant in time.}
Since we have no a priori idea of what could the equation 
of state for Kaluza-Klein Cosmology be, let us suppose
that $\rho$, $p$, and $p'$ are due to a gas of non-interacting
bosons in thermal equilibrium. The free energy 
of this system can be written as follows
\begin{eqnarray}
\beta F&=&\frac{1}{2}\mbox{ln$\;$det}
\left[-\Delta+\mu^2\right]\nonumber
\\
&=&
\sum_{r,m,n=0}^{\infty}D_{mn}
\mbox{ln}
\left[
\left(\frac{2\pi r}{\beta}\right)^2+
m\frac{m+2}{R^2}+
n\frac{n+d-1}{a^2}
+\mu^2
\right]
\label{div}
\end{eqnarray}
where $\mu$ is a mass parameter, $\beta=(kT)^{-1}$,
and $\Delta$ denotes the 
Laplacian on the full compact manifold $S^1\times 
S^3\times S^d$ and the three first terms in (\ref{div})
represent its eigenvalues. $D_{mn}$
is the multiplicity factor related to each of the eigenvalues
and is equal to the dimension of 
$O(1,3)\times O(d+1)$ representation
$$
D_{mn}=(m+1)^2\;(2n+d-1)\frac{(n+d-2)!}{(d-1)!n!}
\;.
$$

The divergent sum (\ref{div})
can be regularized by making use of the identity
$$\mbox{ln}X=\frac{d}{ds}X^s|_{s=0}=\frac{d}{ds}\left[
\frac{1}{\Gamma(-s)}\int_0^\infty dt t^{-s-1}
\mbox{e}^{-tX}
\right]_{s=0}$$
The free energy can thus be rearranged as 
$$
\beta F=
\frac{d}{ds}\left[
\frac{1}{\Gamma(-s)}\int_0^\infty dt\; t^{-s-1}
\mbox{e}^{-t\mu^2}
\sigma_1(4\pi^2t/\beta^2)
\sigma_3(t/R^2)
\sigma_d(t/a^2)
\right]_{s=0}\;,
$$
where the functions $\sigma_{1,3,d}$ are defined by 
\begin{eqnarray}
\sigma_1(4\pi^2t/\beta^2)&=&2\sum_{r=0}^\infty
\;\mbox{e}^{-16\pi^2 r^2t/\beta^2}
\nonumber\\
\sigma_3(t/R^2)&=&
\sum_{m=0}^\infty \;(m+1)^2
\mbox{e}^{-m(m+2)t/R^2}
\label{summs12}
\\
\sigma_d(t/a^2)
&=&
\sum_{n=0}^\infty \;(2n+d-1)\frac{(n+d-2)!}{(d-1)!n!}
\mbox{e}^{-n(n+d-1)t/a^2}\;.
\nonumber
\end{eqnarray}

Note that each of the $\sigma$ function can be written in terms
of the Jacobi Theta function
$$
\theta_3(0,iu/\pi)\equiv
\sum_{n=-\infty}^\infty\;
\mbox{e}^{-\tau n^2}
=\sqrt{\frac{\pi}{\tau}}\;
\mbox{e}^{-\pi^2 n^2/\tau}
$$
which has away from $\tau=0$ the modular property
$$
\theta_3(z/\tau,-\tau^{-1})=\sqrt{\frac{\tau}{i}}
\;\mbox{e}^{i\pi z^2/\tau}
\theta_3(z,\tau)\;.
$$

The sums (\ref{summs12}) hence 
converge and define analytic functions
on the half plane $\mbox{Re}t>0$ and are singular
at $t=0$
$$
\sigma_1\sim t^{-1/2}\; ,\;
\sigma_3\sim t^{-3/2}\; ,\;\sigma_d\sim t^{-d/2}\;.
$$

For $\sigma_1$, the zero temperature limit corresponds to 
$\sigma_1\simeq \sqrt{4\pi t/\beta^2}$, and the flat four-dimensional
space limit of the energy density 
$F/(2\pi^2R)$
corresponds to $\sigma_3\simeq\frac{1}{4}\sqrt{\pi}
R^3t^{-3/2}$.

Now, in a regime in which 
$R\gg T^{-1}> a$ and a relativistic 
gas of particles,
$\mu <T$, the formula for the 
free energy converges at $t=0$ (see 
\cite{Randjbar-Daemi:1984jz} for the technical details) 
and the ``s'' regularization
can be removed, and the approximate expression
becomes 
$$
F\simeq\Omega_3\left(
\sigma_d a^{-4}-\frac{1}{90}\pi^2\beta^{-4}+...
\right)\;,
$$ 
where
$$
\sigma_d=-\frac{1}{32\pi^2}\int_0^\infty du\;u^{-3}
\sigma_{d}(u)\;.
$$

The 1-loop contribution to the potential $U$ related
to $F$ by (\ref{energy6}) therefore is
$$
U_{1-\mbox{\tiny loop}}
\simeq \Omega_3\frac{\sigma_d}{a^4}+ S^{4/3}
\frac{\tau}{R}\;\;\;,\;\;\;\tau=\frac{3}{4}\left(\frac{45}{4\pi^4}
\right)^{1/3}
$$
The full potential (\ref{en5}) up to one loop 
is therefore
$$
U=\frac{\Omega_3\Omega_d}{\kappa^2}
\left[
-\frac{6}{R^2}-d\frac{(d-1)}{a^2}
+\Lambda
\right]+ \sigma_d \frac{\Omega_3}{a^4}+\tau S^{4/3}
R+\zeta m\;,
$$ 
where the last term $\zeta m$ is identified by the 
energy for a constant $a$,
\begin{equation}
E=-\frac{\Omega_3\Omega_d}{\kappa^2}\frac{{\dot R}^2}{R^2}+U
\;.
\label{en3}
\end{equation}

The time evolution of the $D$-dimensional Universe
is governed by the Euler-Lagrange equations
derived from (\ref{action7}). Is is shown 
that for an appropriate
choice of $\Lambda$ and the parameters in $U$, it is possible to obtain 
a solution where the 
internal radius is constant while
the large radius evolves as in the standard Cosmology 
\cite{Randjbar-Daemi:1984jz}. 
For $\dot a=0$ the equations
of motion become 
\begin{equation}
\partial_t(R\dot R)-\frac{1}{2}{{\dot R}^2}=\frac{1}{12}
\frac{\kappa^2}{\Omega_3\Omega_d}R\frac{\partial U}{\partial R}
\label{eqm2}
\end{equation}
\begin{equation}
R^{-1}\partial_t(R^2\dot R)-{{\dot R}^2}
=
\frac{\kappa^2}{6d}\frac{R^2}{\Omega_3\Omega_d}
a\frac{\partial U}{\partial a}
\label{eqm3}
\end{equation}
The compatibility of (\ref{en3}), (\ref{eqm2}), and (\ref{eqm3})
requires $$
R\frac{\partial U}{\partial R}
-\frac{2}{d}
a\frac{\partial U}{\partial a}=E-U
\;,$$
and this constraint allows us to determines
$\Lambda$ and $a$ in terms of $\kappa^2$ which upon
substituting back into (\ref{eqm3}) gives
\begin{equation}
\partial_t(R\dot R)=-1
\;\;\;\mbox{or} \;\;\;
R^2=t\left(t_0-t\right)\;,
\label{scale}
\end{equation}
where $t_0$ is computed from (\ref{en3}) to be $t_0^2=(\tau/3\pi^2)
\kappa^2S^{4/3}/\Omega_d$. 

As a summary, there are 
solutions for Einstein equations
in the regime 
$R\gg T^{-1}> a$ 
where the size of the internal space remain constant, and 
the 4-dimensional Universe evolves according to the 
conventional Friedmann-Robertson-Walker
Cosmology, as (\ref{scale}) clearly indicates,
for a closed Universe in a radiation dominated era.
It remains to check the classical stability, at least, 
of the value of $a$
by solving the equations for small perturbations around the 
desired value.

\subsection{Brane Cosmology}
\label{rscosmology}
Here we review an example \cite{Binetruy:2000hy} 
of the evolution of 
a brane-world embedded in five dimensions, and under 
which condition the FRW expansion can be recovered. 
The time evolution of the scale factor, is worked out 
explicitly for  
the a specific
choice of the equation   
of state $p=\omega \rho$ ($\omega=\mbox{const.}$), 
and it is shown that the standard cosmological evolution 
can be obtained 
\cite{Binetruy:2000hy}. 

Since we are interested in a FRW metric for the four-dimensional 
Universe, we start from the following maximally symmetric
three-dimensional metric ansatz
$$
ds^2=-n^2(\tau,y)d\tau^2+a^2 (\tau,y){\tilde g}_{ij}dx^idx^j+
b^2(\tau,y)dy^2
\;,
$$
and we assume the brane as a hypersurface in five 
dimensions
defined by $y=0$.

Upon substituting this ansatz into the five-dimensional 
Einstein equations, the non-vanishing components of the 
Einstein tensor\\ 
$G_{MN}\equiv {\cal R}_{MN}-\frac{1}{2}{\cal R}\;g_{MN}$
are \cite{Binetruy:2000hy} 
\begin{eqnarray}
G_{00}&=&
3\left\{\frac{\dot a}{a}\left(
\frac{\dot a }{a}+\frac{\dot b }{b}
\right)-\frac{n^2}{b^2}
\left[
\frac{a'' }{a}+\frac{ a' }{a}\left(
\frac{a' }{a}-\frac{ b' }{b}
\right)
\right]+\kappa\frac{n^2}{a^2}
\right\}\;,\label{g00}\\
G_{ij}&=&
\frac{a^2}{b^2}\gamma_{ij}\left\{\frac{a'}{a}\left(
\frac{a' }{a}+2\frac{n' }{n}
\right)-\frac{b'}{b}
\left(
\frac{n' }{n}+2\frac{ a' }{a}\right)
+2
\frac{a'' }{a}+\frac{ n'' }{n}
\right\}\nonumber\\
&&+\frac{a^2}{n^2}{\tilde g}_{ij}
\left\{\frac{\dot a}{a}\left(-
\frac{\dot a }{a}+2\frac{\dot n }{n}
\right)-2\frac{\ddot a}{a}
+\frac{\dot b}{b}
\left(-2
\frac{\dot a }{a}+\frac{ \dot n }{n}\right)
-
\frac{\ddot b }{b}
\right\}-\kappa 
{\tilde g}_{ij}\;,
\;\;\;\;\;\;\;
\;\;\;\;\;\;\;\;
\;\;\;\; \label{gjj}\\
G_{05}&=& 3\left(\frac{n'}{n}\frac{\dot a}{a}+
\frac{a'}{a}\frac{\dot b}{b}-\frac{{\dot a}'}{a}
\right),\\
G_{55}&=&
3\left\{\frac{a'}{a}\left(
\frac{a' }{a}+\frac{n' }{n}
\right)-\frac{b^2}{n^2}
\left[
\frac{\dot a }{a}\left(\frac{\dot a }{a}-\frac{\dot n }{n}
\right)
+\frac{\ddot a }{a}\right]
-\kappa\frac{b^2}{a^2}
\right\}\;,
\label{g55}
\end{eqnarray}
where a prime stands for a derivative with respect
to $y$ while a dot denotes the derivative with respect to $\tau$. 

The energy-momentum tensor can be decomposed into two parts; the 
bulk matter source and the brane matter one
$$
{{T}^A}_B= {{\check T}^A}_B|_{\mbox{\tiny bulk}}+
{{\tilde T}^A}_B|_{\mbox{\tiny brane}}\;,
$$
where ${{\check T}^A}_B|_{\mbox{\tiny bulk}}$ is the energy-momentum
tensor of the bulk matter which we assume to have the form
$$
{{\check T}^A}_B|_{\mbox{\tiny bulk}}+
\mbox{diag}(-\rho_B,p_B,p_B,p_B,p_T)\;,
$$
where the energy density $\rho_B$ and the pressures $p_B$
and $p_T$ are independent of the coordinate $y$. Later we will 
be interested in the special case of a Cosmological constant
$-\rho_B=p_B=p_T$.

The second term ${{\bar T}^A}_B|_{\mbox{\tiny brane}}$ 
corresponds to the matter content on the brane, {\t i.e}
at $y=0$. According to our metric ansatz,
the matter description on the brane is the one of 
a perfect fluid in four dimensions
$$
{{\tilde T}^A}_B|_{\mbox{\tiny brane}}=\frac{\delta y}{b}
\mbox{diag}(-\rho_b,p_b,p_b,p_b,0)\;,
$$
where $\rho_b$ and $p_b$ are, respectively, 
the energy density and pressure on the brane and are functions of time
only. The last entry being zero means physically that there is no 
flow of ordinary matter along the fifth direction, and consequently 
the vanishing of $G_{05}$ and this leads in turn that 
(\ref{g00}) and (\ref{g55}) become
\begin{eqnarray}
F'=2\frac{a'a^3}{3}\kappa^2{{\check T}^0}_0\label{fp}\\
{\dot F}=2\frac{{\dot a}a^3}{3}\kappa^2{{\check T}^5}_5
\;,
\label{fq}
\end{eqnarray} 
where $F$
is a function of $\tau$ and $y$ defined by
\begin{equation}
F(\tau,y)=\frac{{a'}^2a^2}{b^2}-\frac{{\dot a}^2a^2}{n^2}
\kappa a^2
\;.
\label{ftau}
\end{equation}

Integrating (\ref{fp}) over $y$, keeping in mind that ${\rho_B}'=0$,
we obtain
$$
F+\frac{\kappa^2}{6}a^4\rho_B+C=0\;,
$$
where $C$ is a constant of integration. 

Now, by taking the time derivative of (\ref{fp}), the
$y$ derivative of (\ref{fq}), and
assuming
$-\rho_B=p_T$ we see that 
$$
(a'-{\dot a})a^3 \frac{d}{dt}\rho_B=0
$$
which indicates that 
$\rho_B$ is independent of time for
$a'-{\dot a}\neq 0$,
the thing we will assume from now on. 
\footnote{If $a$ is, for example, an exponential function
of $y$ and $\tau$ as $e^{g y\tau}$ this assumption
is no more valid.} 
This will indicate that ${\dot C}=0$. 

Coming back to (\ref{gjj}), $G_{ij}$ can be written using 
the Bianchi identity $\nabla_AG^{A0}=0$ as 
$$
\partial_\tau
\left(
\frac{F'}{a'}\right)=\frac{2}{3}{\dot a}a^2{\tilde g}_{ij}G^j_i
$$
which is identically satisfied when $-\rho_B=p_B$
as can be seen using (\ref{fp}).

Hence, when the bulk source is a Cosmological constant, any set
of functions $a$, $n$, and $b$ satisfying (\ref{ftau})
or more explicitly
\begin{equation}
\left(\frac{\dot a}{na}\right)^2=\frac{1}{6}\kappa^2\rho_B+
\left(\frac{a'}{ba}\right)^2-\frac{\kappa}{a^2}+\frac{C}{a^4}
\label{set}
\end{equation}
together with $G_{05}=0$ will locally be solution of Einstein equations 
in five dimensions, away from $y=0$. 

At the position of the brane, $y=0$, the junctions condition should
be taken into account (see section \ref{junc3}) which here 
take the form
\begin{eqnarray}
\frac{[a']}{a(0)b(0)}&=&-\frac{\kappa^2}{3}{\rho_b}\label{junction1}\\
\frac{[n']}{n(0)b(0)}&=&-\frac{\kappa^2}{3}(3p_b+{\rho_b})
\;,
\end{eqnarray}
where $[Q]=Q(0^+)-Q(0^-)$ defines the jump of the 
function $Q$ across $y=0$.

Assuming the symmetry $y\rightarrow -y$ (for simplicity, which is 
not necessarily imply the presence of another brane), the junction
condition (\ref{junction1}) can be used to compute $a'$ at
the two sides of the brane, and by continuity, 
when $y\rightarrow 0$, one sees that 
the equation (\ref{set})
implies the generalization of the first Friedmann equation (after
setting $n(0)=1$ by a suitable change of time):
\begin{equation}
\frac{{\dot a}(0)^2}{{\dot a}(0)^2}=
\frac{\kappa^2}{6}\rho_B+\frac{\kappa^4}{36}\rho_b^2+\frac{C}{a(0)^2}
-\frac{\kappa}{a(0)^2}\;.
\label{evolution7}
\end{equation}

Note from the above equation that the bulk energy density enters linearly
while the brane energy density enters quadratically, and
the Cosmological evolution depends on the constant
$C$, an effective radiation term, 
which is determined by the initial conditions can be constrained 
by BBN \cite{Binetruy:2000hy}.  

Equation (\ref{evolution7}) is enough to study the cosmological evolution
on the brane whatever the metric outside looks like and whether
or not $\dot b=0$.

However, let us assume $\dot b=0$ and set it to $b=1$.
Using the equation $G_{05}=0$ one can express $n$ in terms of 
$a$ according to 
$$
\frac{\dot a}{n}\equiv \alpha(t)\;, 
$$
where $\alpha$ depends solely on time. Inserting this 
into (\ref{fp}) leads to 
$$
\alpha^2+\kappa-(aa')'=\frac{\kappa^2}{3}
a^2\rho_B
$$  
which applies in the bulk at both sides of the brane. Integrating 
it over $y$ gives
$$
a^2=A\;\mbox{cosh}(\sigma y)+B\;\mbox{sinh}(\sigma y)\;,
+C
$$
with 
$$
\sigma =\sqrt{-\frac{2\kappa^2}{3}\rho_B}
\;,$$
for $\rho_B<0$.

For $\rho_B>0$, the solution is

$$
a^2=A\;\mbox{cos}(\sigma y)+B\;\mbox{sin}(\sigma y)
+C\;,
$$
with 
$$
\sigma =\sqrt{\frac{2\kappa^2}{3}\rho_B}
\;.$$

For $a=0$ it is 
$$
a^2=(\alpha^2+\kappa)y^2+Sy+E
\;.$$
where 
the functions $A,B,C,S$, and $E$
are functions of time (or constants) 
and can be determined
in terms of the input parameters of the theory.
 
In order to see more explicitly the Cosmological evolution on the 
brane, let us decompose the energy density on the brane 
into two parts:
$$
\rho_b=\rho+\rho_\Lambda
\;,$$
where $\rho_\Lambda$ is a constant, and $\rho$ stands for the 
contribution of matter on the brane. Substituting in 
(\ref{evolution7}) we get
$$
\frac{{\dot a}(0)^2}{{a}(0)^2}=
\frac{\kappa^2}{6}\rho_B+\frac{\kappa^4}{36}\rho_{\Lambda}^2
+\frac{\kappa^4}{18}\rho_{\Lambda}\rho
+\frac{\kappa^4}{36}\rho^2
\frac{C}{a(0)^2}
-\frac{\kappa}{a(0)^2}\;.
\label{evolution}
$$

If we chose 
\begin{equation}
\frac{\kappa^2}{6}\rho_B+\frac{\kappa^4}{36}\rho_{\Lambda}^2=0
\;,
\label{cosmo}
\end{equation}
we see that for $\rho\ll \rho_\Lambda$ 
the standard Cosmology is recovered 
\cite{Binetruy:2000ut,Binetruy:2000hy} (the solution for
the case $\rho=0$ is also in \cite{Kaloper:1999sm}).

Assuming the equations of state to be $\rho=\omega p$, the
energy density and pressure become
$$
\rho=\rho_c\left(\frac{a(0)}{a_c}\right)^{-q}\;\;\;,
\;\;\;q=3(1+\omega)\;,
$$
where $\rho_c$ and $a_c$ are constants. The scale factor
describing the time evolution on the brane, 
when (\ref{cosmo}) is satisfied, behaves as 
\cite{Binetruy:2000hy}
$$
a(0,t)=a_c \kappa^{2/q}\rho_c^{1/2}\left(
\frac{q^2}{72}\kappa^2\rho_\Lambda t^2+\frac{q}{6}t
\right)^{1/q}\;,
$$
and therefore
the evolution of the Universe will be non-conventional 
$a(t)\sim t^{1/q}$ at
early times                                                                    while at late times it is described 
by the standard Cosmology $a(t)\sim t^{2/q}$ (further
discussion can be found in \cite{Binetruy:2000hy}).

\subsection{Stability of the scale of compactification}
\label{stab}

So far, the examples explained in sections \ref{kkcosmology} and 
\ref{rscosmology} 
represent equilibrium configurations, what remains
is to check their 
stability against the perturbations of the metric
and other background fields. 

The procedure of checking 
classical perturbative 
stability of a solution to Einstein equations
was explained in section
\ref{com} under\ref{stabspec}. There, it was argued that one way to 
see whether the solution is stable or not is by looking at the 
spectrum and making sure it contains no tachyons or 
ghosts. 
The absence of 
such modes
would indicate that the radius of compactification
would only oscillate around the background value and 
that the configuration, 
say $M\times K$, will not acquire a shape different from the 
initial ansatz. \footnote{It was 
argued in \cite{Candelas:1984ae}
that even a manifold which is stable against all deformations
will become unstable if the temperature is raised
to a critical value, which is independent of the shape of 
the manifold, as on rather general grounds the free
energy could become increasingly negative above the critical 
temperature. The phase transition due to this phenomena
indicates the necessity of having time-dependence 
for the field associated with the evolution of the compact
space.}

For example, the theory provided 
in section \ref{com} 
(of reference\cite{Randjbar-Daemi:1983hi})
is free of 
both tachyons and negative norm fields, 
in addition to having a fixed value of the radius in 
of $S^2$ in terms of the other parameters in the theory
(equation (\ref{radius})),\footnote{There is no analogue for this
relation in the original Kaluza-Klein compactification
performed on a circle since $S^1$ is flat.}
and therefore the radius of compactification is expeted to 
be oscilating around the value $(8M^{4}g^2/n^2)^{-1}$.
It is may also be helpful to note that in this model 
there is no massless mode in the gravity sector
to associate with the {\it radion}, 
the radius of compactification. 

In general, this not the case. The radion
is usually massless and it has no potential (or 
a flat one).
Therefore, there is no reason why the scale of its
vacuum expectation
value should be
fixed at a specific point which is moreover close to 
the electroweak scale. When put in a cosmological 
context, the value of the radius should not 
vary much as this may have strong impact on the 
Universe as observed today.
Therefore, the ground state state of the radion 
field has to have a determind value for at least two 
reasons. The first is that there is no experimental evidence for 
a variation of the fundamental constants \cite{Kolb:1986sj}; 
according to observations the internal space(s) should
be static or nearly static at least from the time of 
nucleosynthesis. 
The second is that we wish to have a lower gravity 
scale, preferably as low as few TeV, 
as a cutoff for the standard model of particle physics. 
Therefore, a good stabilization mechanism for the 
radion should be a part 
of any realistic model.

In superstring theories, the fundamental physical constants
are related to the vacuum expectation values of the moduli fields
which are defined by the shape and size of internal space(s) 
of compactification. In these theories, some stabilization mechanisms
are proposed, for example in 
\cite{Krasnikov:1987jj,Taylor:1990wr}, and recently it was shown that 
the stabilization can be enhanced due to the coupling of the
dilaton, the string coupling constant $\mbox{e}^{\left<\phi\right>}$ 
to the kinetic energy of ordinary matter fields
\cite{Huey:2000jx}. In the context of Kaluza-Klein 
theory this issue was, and is still, subject to 
numerous investigations (see for instance 
\cite{Kikkawa:1985qc}-\cite{Huey:2000jx} for tensor product 
compactification, and 
\cite{Goldberger:1999uk}-\cite{Chacko:2001em} for warped).

Another way to stabilize the 
radion, than the one provided in \cite{Randjbar-Daemi:1983hi}, 
is by generating a potential for the 
radion, through an {\it ad hoc}
coupling to a scalar field for example (as in \cite{Goldberger:1999uk}), 
and checking whether this potential has a minimum
which could correspond to the desired value of the radion. 
\footnote{In \cite{Goldberger:1999uk} the back reaction 
of the scalar field into the metric was not taken into account, 
for a discussion about this point see 
\cite{Gibbons:2001tf}.} 
In fact, the background solution for the scalar field in 
\cite{Goldberger:1999uk} is, in a sense, 
similar to the Yang-Mills background (\ref{back1}), 
and the relation determening the value of the radion
in \cite{Goldberger:1999uk} in terms of the values of the minima
of scalar potential and other paramters of the model 
is somehow analogous to equation (\ref{radius}). 

In \cite{Arkani-Hamed:2001kx} the issue 
of stabilization was discussed within 
a cosmological time evolution context, which is 
essentially the same idea. Again a
potential was introduced for this purpose, 
with the background 
metric similar to (\ref{kkmetric}), and it 
was argued that the fluctuations of the radion
were oscillatory 
with positive frequency,
around a value $r_0$, 
and this subsequently lead to the understanding that
$r_0$ is a minumum of the potential. 
Of course one has to check that {\it all} 
physical modes of the model have the feature.

We believe that the method in section \ref{com} is nice since
it does not 
involve {\it ad hoc} potentials to stabilize the radion
as usually alternative stability checks require. 
It may even be that higher order corrections 
result in a potential \cite{Randjbar-Daemi:1985fs}
which could lead to a 
better understanding to the origin of stabilization potentials, 
which may also shed some light on the way how supersymmetry 
is broken in superstring theories, and to possible
relation   
of the value of the radion at the minimum with the rest
of parameters in the theory. This problem 
remains challenging.

\subsection{Proton stability}
\label{prot}
Another problematic issue arising from models with 
large extra (and also infinite) dimensions
is the potentially too fast proton decay. Generally, 
the proton decay rate due to gravitational effects of,
{\it e.g.} 
dimension $5$ operators,
is given by 
$$
\tau_p^{-1}\sim \frac{m_p^5}{M_P^4}.
$$
which we know from the bounds on proton life time it 
should satisfy $\tau_p\lesssim 10^{33}$ years.
This implies a bound, too well known in 
Grand Unified Theories, that the smallest mass
which could substitute $M_P$ and does not 
lead to an unacceptably fast 
proton decay is $10^{15}-10^{16}$GeV. Therefore
one does generally expect a problem with proton 
decay at low fundamental scales $\sim$TeV. 

If the particle mediating proton decay, say via
the dimension five non-renormalizable operator 
$qqql/m_X$,
has a mass
$m_X \sim$TeV range, it should 
be incredibly weakly coupled in order not to lead to 
proton decay.
A way to suppress such an operator could be by 
imposing certain symmetries 
\cite{Benakli:1999ur}
(see also \cite{Feruglio:2001nf}-\cite{Huber:2001ie}. 
Since global symmetries are are expected to be 
broken by quantum effects, 
those symmetries better be 
gauged. 
Moreover, the procedure of orbifold compactification
with fixed point 
(which is used by some in order to achieve chiral fermions,
project out unwanted states,...{\it etc})
is allowed only in a string theory context, and therefore a
neater way to achive acceptable proton decay rate could be 
by choosing an appropriate Yang-Mills gauge group in 
$D$-dimensions together with an appropriate compact space.
It is possible to construct examples of such 
compactification where operators leading to proton decay 
are forbidden because the quantum numbers of quarks and 
leptons can not form invariant couplings under the 
unbroken symmetry (could be the unbroken part of the gauge
group, the isometry group, or both)
as can be checked with the $U(6)$ model in chapter \ref{ssb}.
This possibility will be discussed elsewhere.  

We review here a simple 
mechanism for proton stability 
suggested in 
\cite{Arkani-Hamed:2000dc}, as this will 
serve in explaining chapter \ref{baryogm}.  
\footnote{See also \cite{add3,Benakli:1999ur,Shiu:1998pa} 
for alternative suggestions.} 
The same idea of fermion localization \cite{sh2} 
explained in section \ref{localf}
can be adopted to guarantee proton
stability \cite{Arkani-Hamed:2000dc}:
quarks and leptons are localized at two slightly 
displaced 
point along the fifth direction, and this suppressed their
mutual interaction. To see how, 
let us give leptons
and baryons the following 5-dimensional masses
\begin{equation}
( m_0)_l = 0 \;\;,\;\; (m_0)_b = m_0\;,
\label{ourch}
\end{equation}
which correspond to the localizations \footnote{The last inequality in
the next expression comes from (\ref{lcon}). 
We assume quarks of different
generations to be located in the same $y$ position 
in order to avoid
dangerous FCNC mediated by the Kaluza-Klein modes of the gluons
\cite{Delgado:2000sv}.}
\begin{equation}
y_l = 0 \;\;,\;\; y_b = \frac{m_0 \sqrt{2 \lambda_0
{\widetilde M}_0}}{\mu_0^2} < \frac{1}{\mu_0} \;\;.
\end{equation}

The shape of the fermion wave functions along the 
fifth dimension is the one in (\ref{psish})
up to a normalization. 
It can be cast in an explicit and simple form if we
consider the limit $y_b \ll 1/\mu_0$, in which the 
effect of the plateau
for $y>1/\mu_0$ can be neglected: \footnote{This is also the limit in
which the approximation (\ref{approx}) is valid.}
\begin{eqnarray}
f_l ( y ) &=& \left( \frac{\mu_0^2}{\sqrt{2 \lambda_0
{\widetilde M}_0} \pi} \right)^{1/4}  \exp \left \{ {-
\frac{\mu_0^2y^2}
{2\sqrt{2\lambda_0\,{\widetilde M}_0}}}\right \} \nonumber\\
f_b ( y ) &=& \left(\frac{\mu_0^2}{\sqrt{2\lambda_0
{\widetilde M}_0} \pi} \right)^{1/4}  
\exp \left \{{-\frac{\mu_0^2\,\left(y-y_b\right)^2}{2 
\sqrt{2\lambda_0{\widetilde M}_0}}}\right \} \;\;.\nonumber
\end{eqnarray}
\begin{figure}
\centerline{\leavevmode\epsfysize=3cm \epsfbox{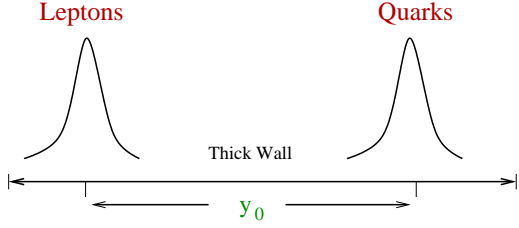}}
\vspace{10pt}
\caption{\small The wave functions of quarks and leptons 
in the transverse space to the brane world.}
\end{figure}

We assume the Standard Model to be embedded in some theory which, in
general, contains some additional bosons $X$ whose interactions violate
baryon number conservation. If it is the case, the four fermion
interaction $qq \longleftrightarrow ql$ can be effectively described by
\begin{equation}
\int d^4 x \, d y \:\frac{q\,q\,q\,l}{\Lambda \, m_X^2}\;\;,
\label{sca}
\end{equation}
where $m_X$ is the mass of the intermediate boson $X$ and $\Lambda$ is a
parameter of mass dimension one related to the five-dimensional coupling
of the $X$-particle to quarks and leptons.
This scattering is thus suppressed by \footnote{From the approximation (\ref{approx}), only the squared difference of the five-dimensional masses affects the suppression factor. For this reason, the above choice $\left( m_0 \right)_l = 0$ was only done in order to simplify notation and does not have any physical meaning.}
\begin{eqnarray} 
I &=& \frac{1}{\Lambda m_X^2} \int dy \frac{\mu_0^2}{\pi \sqrt{2\lambda_0
{\widetilde M}_0} }\exp \left \{ -\frac{\mu_0^2 /2}{\sqrt{2\lambda_0
{\widetilde M}_0}} \left[ y^2 + 3 \left( y - y_b \right)^2 \right]\right
\}\nonumber \\
&=& \frac{\mu_0}{\Lambda  m_X^2  \sqrt{2\,\pi} \left(
2\lambda_0{\widetilde M}_0 \right)^{1/4}}  \exp \left \{{-\frac{3
\left( 2\lambda_0\,{\widetilde M}_0 \right)^{1/2}}{8}\frac{m_0^2}
{\mu_0^2}}\right \}\;\;.
\label{supp}
\end{eqnarray}

Current proton stability
\cite{Shiozawa:1998si} requires $I \sim
\left(10^{16} \,\mbox{GeV} \right)^{-\,2}\;,$ that is
\begin{equation}
\frac{m_0}{\mu_0} \gtrsim 
\frac{\sqrt{200-6\,\mbox{Log}_{10}\left(\frac{\Lambda
m_X^2}{\mu_0}\Big / 
\mbox{GeV}^2\right )}}{\left( 2\lambda_0
{\tilde M}_0 \right)^{1/4}}\;\;.
\label{con2}
\end{equation}

The numerator in the last equation is quite insensitive to the mass scales
of the model, and -- due to the logarithmic mild dependence -- can be
safely assumed to be of order $10\,$. For definiteness, we will thus fix
it at the value of $10$ in the rest of this chapter.
Conditions (\ref{lcon}) and (\ref{con2}) give altogether
\begin{equation}
\frac{10\,\mu_0}{\left( 2\,\lambda_0\,{\widetilde M}_0 \right)^{1/4}}
\lesssim m_0 \lesssim \frac{\mu_0}{\left( 2\,\lambda_0\,{\widetilde M}_0
\right)^{1/2}} \;\;,
\end{equation}
that we can rewrite
\begin{eqnarray}
2\lambda_0 {\tilde M}_0 &\lesssim& 10^{-4} \nonumber\\
\frac{m_0}{\mu_0} &\gtrsim& 10^2 \;\;\;\;.
\label{limits}
\end{eqnarray}

Notice that the last limit in equations
(\ref{limits}) is stronger than the one
given in \cite{Arkani-Hamed:2000dc} where proton stability is
achieved if the ratio of the massive scales of the model is of order
$10\,$. However, in ref.~\cite{Arkani-Hamed:2000dc} the field $\Phi$
simply scales linearly as a function of $y\,$, while we expect that
whenever a specific model is assumed, conditions analogous to
our~(\ref{lcon}) and~(\ref{limits}) should be imposed
\cite{Masiero:2000fr}.

\subsection{Baryogenesis with Low Scale Gravity}
\label{baryogm}
In this section we discuss a mechanism for generating baryon
number violation in a model with one extra dimensions. 
As explained in section \ref{prot}, the Rubakov-Shaposhnikov
localization mechanism \cite{sh2} was used in 
\cite{Arkani-Hamed:2000dc} in order to insure proton 
stability at zero temperature.\footnote{
Beside providing an way around fast proton decay in
theories with low fundamental scales, this idea has
further applications 
as one may reproduce the entries of the
Cabibbo-Kobayashi-Maskawa matrix just placing quarks of 
different
generations at different and appropriately chosen positions (this requires
the presence of at least two extra dimensions, as shown
in \cite{Mirabelli:2000ks})
without assuming any hierarchy in the original
Yukawa couplings of the fermions to the Higgs field.
Discussion for collider signature is in \cite{Arkani-Hamed:2000za}.}
The baryon number is almost conserved at zero
temperature due to the slight overlap of the wave functions
of quarks and lepton in the internal space due to 
their different localization along the fifth 
direction. The suppression
decreases due to finite temperature effects leading
to bayon number violation at earlier times in the history
of the Universe.

Here, we review
\cite{Masiero:2000fr}
where the idea
of \cite{Arkani-Hamed:2000dc}
is adopted 
for generating operators leading 
to baryon number violation at finite temperature, 
while such operators remain appropriately suppressed at zero 
temperature. 
 
The starting point is to assume that
quarks and leptons are now ({\it i.e.} 
at zero temperature) sufficiently apart
in the extra space. However, we wonder if finite temperature effects could
change this picture increasing the interactions between quarks and leptons
at early times. This would render proton stability now compatible
with baryogenesis in the early Universe. An exact
computation of the corrections on a solitonic background presents some
technical difficulties. Moreover, relying on a perturbative analysis at a
scale close to the cut-off of the theory may be unsafe. For this reason,
we do not give a precise final value for the temperature necessary to
achieve the observed baryon asymmetry. We limit ourselves to a
perturbative analysis made on dimensional arguments, which anyhow
indicates that finite temperature effects should indeed increase the
interactions between baryons and leptons.

As it is well known, baryon number violation alone is not sufficient for
baryogenesis. To present a more complete analysis
we discuss a particular model reminiscent of GUT
baryogenesis. In doing so, we meet another problem typical of these brane
scenarios. Due to the low energy densities involved, the expansion rate of
the Universe is always very small. If baryogenesis originates from the
decay of a boson $X$, the out of equilibrium condition requires $m_X$
much higher than the physical cut-off of brane models. This problem can be
overcome for example if the temperature of the Universe never exceeds
$m_X$ and if these bosons are created non-thermally (for instance at
preheating). This and some other options are briefly discussed in the last
section.

\subsubsection{Thermal correction to the coefficients} \label{ctre}
Once the localization mechanism is incorporated in a low energy effective
theory -- as the system described above may be considered --, one can
legitimately ask if thermal effects could play any significant role. We
are mainly interested in any possible change in the argument of the
exponential in (\ref{supp}), that will be the most relevant for the
purpose of baryogenesis. For this reason, we introduce the dimensionless
quantity
\begin{equation}
a(T)=\frac {m(T)^2}{\mu(T)^2}\sqrt{2\,\lambda (T) {\widetilde M}(T)}\;\;\;.
\label{defa}
\end{equation}

From (\ref{con2}) and (\ref{limits}), we can set $a(0)
\simeq 
100$
at zero temperature. Thermal effects will modify this value. There are
however some obstacles that one meets in evaluating the finite temperature
result. Apart from some technical difficulties arising from the fact that
the scalar background is not constant, the main problem is that
nonperturbative effects may play a very relevant role at high temperature.
As it is customary in theories with extra dimensions, 
the model described
by (\ref{lagpp}) and (\ref{lagk}) 
is nonrenormalizable and one
expects that there is a cut-off (generally related to the fundamental
scale of gravity) above which it stops holding. Our considerations will
thus be valid only for low temperature effects, and may be assumed only as
an indication for what is expected to happen at higher temperature.

Being aware of these problems, by looking  at the dominant
finite-temperature one-loop effects, we estimate the first corrections to
the relevant parameters to be
\begin{eqnarray}
\lambda \left( T \right) &=& \lambda_0 + c_\lambda \, \frac{T}{{\widetilde
M}_0^2} \nonumber\\
{\widetilde M} 
\left( T \right) &=& {\widetilde M}_0 + c_{{\widetilde M}} \, T \nonumber\\
m \left( T \right) &=& m_0 + c_m \, \frac{T^2}{{\widetilde M}_0} \\
\mu^2 \left( T \right) &=& \mu_0^2 + c_{\mu} \, \frac{T^3}{{\widetilde M}_0} \;\;,
\label{thermal}
\end{eqnarray}
where the $c$'s are dimensionless coefficients whose values are related to the exact particle content of the theory.

In writing the above equations, the first of conditions (\ref{limits}) has
also been taken into account. For example, both a scalar and a fermionic
loop contribute to the thermal correction to the parameter $\lambda_0\,$.
While the contribution from the former is of order $\lambda_0^2 \, T\,$,
the one of the latter is of order $T / {{\widetilde M}_0}^2$ and thus
dominates. \footnote{Notice also that with our choice~(\ref{ourch}) loops
with internal leptons dominate over loops with internal quarks, since the
former have vanishing 5-dimensional bare mass and thus are not
Boltzmann suppressed. However, although this choice is the simplest one,
one may equally consider the most general case where all the fermions have
a nonvanishing five-dimensional mass.}

Substituting equations (\ref{thermal}) into eq.~(\ref{defa}), we get, in the
limit of low temperature,
\begin{equation}
a \left( T \right) \simeq a\left(0\right) \cdot \left[ 1 +
\frac{T}{{\widetilde M}_0} \: \left(
\frac{c_\lambda}{2\,\lambda_0\,{\widetilde M}_0} + \frac{c_{\widetilde
M}}{2} + \frac{2\,c_m\,T}{m_0} - \frac{c_{\mu}\,T^2}{\mu_0^2}
\right) \right] \;\;.
\end{equation}

From the smallness of the quantity $\lambda_0 \, {\widetilde M}_0\;$
(see (\ref{limits})) we can safely assume (apart from high hierarchy
between the $c$'s coefficients that we do not expect to hold) that the
dominant contribution in the above expression comes from the term
proportional to $c_\lambda\,$.

We thus simply have
\begin{equation} \label{acorr}
a \left( T \right) \simeq a\left(0\right) \left( 1 + c_{\lambda} \,
\frac{T}{2\,\lambda_0\,{\widetilde M}_0^2} \right) \:\,.
\end{equation}

We notice that the parameter $c_\lambda$, being related to the thermal
corrections to the $\phi^4$ coefficient due to a fermion loop, is expected
to be {\it negative} 
\cite{Dienes:1999hx}: the first thermal effect is to
decrease the value of the parameter $a(T)$, making hence the baryon number
violating reactions more efficient at finite rather than at zero
temperature.

There is another effect which may be very crucial at finite temperature,
linked to the stability of the $Z_2$ symmetry. When a temperature is
turned on, we generally expect the formation of a fermion--antifermion
condensate $\langle {\bar \psi} \: \psi \rangle \neq 0\,$. If it is the
case, the Yukawa coupling $\Phi\,{\bar \psi}\,\psi$ in the
Lagrangian (\ref{lagpp}) renders one of the two vacua unstable. While this
leads to an instantaneous decay of the kink configuration, a
kink--antikink system could have a sufficiently long lifetime provided the
two objects are enough far apart.

Let us conclude this section with an important remark. While in this
discussion we have considered only the localization of fermions along one
extra dimension, almost everything we have said can be generalized if two
or more additional dimensions are present~\footnote{This is mandatory in
the scenario \cite{add1}, since the presence of only one
large extra dimension is phenomenologically excluded.}. 
The index theorem
guarantees 
\cite{Jackiw:1976fn, Weinberg:1981eu} 
indeed the possibility of
localizing fermions on a topological defect of an arbitrary dimension.
Just to give an example, let us consider the localization on a
Nielsen--Olesen
\cite{Nielsen:1973cs} vortex in the case of two
extra--dimensions, as it is discussed in \cite{add1}.
Also in this case, one can localize quarks and leptons at two different
positions (actually along different circles about the center of the
vortex). Once again, proton stability requires conditions completely
analogous to conditions (\ref{limits}) here discussed. Of course the
calculation of thermal corrections gives different results, since the
dimension of the couplings of the model changes according to the number of
spatial dimensions. However, also in the case of the two-dimensional
vortex, the qualitative result turns out to be identical to what has been
derived in the one-dimensional case: the most significant effect comes
from the variation of the coefficient $\lambda$ of the $\phi^4$
interaction, and it is in the direction of enhancing the quark--lepton
interaction with increasing temperature.

\subsubsection{A model for baryogenesis} \label{cquattro}
We saw in the previous section that thermal effects may increase the rate
of baryon number violating interactions of the system. This is very
welcome, since a theory which never violates baryon number cannot lead to
baryogenesis and thus can hardly reproduce the observed Universe. Anyhow
baryon number violation is only one of the ingredients for baryogenesis,
and the aim of this section is to investigate how the above mechanism can
be embedded in a more general context.

A particular scheme which may be adopted is baryogenesis through the decay
of massive bosons $X$. \footnote{We may think of these bosons as the
intermediate particles which mediate the four fermion interaction
described by the term (\ref{sca}).} This scheme closely resembles GUT
baryogenesis, but there are some important peculiarities due to the
different scales of energy involved. In GUT baryogenesis the massive boson
$X\;$, coupled to matter by the interaction $g\,X\,\psi\,{\bar \psi}\,$,
has the decay rate
\begin{equation}
\Gamma \simeq \alpha \, m_x \;\;,\;\; \alpha = \frac{g^2}{4\,\pi} \;\;. 
\end{equation}

An important condition is 
that the $X$ boson decays when 
the temperature of the Universe is below its mass (out of equilibrium decay), in order to avoid thermal regeneration. 
From the standard equation for the expansion of the Universe,
\begin{equation}
H \simeq g_*^{1/2}\,\frac{T^2}{M_{\mathrm P}}
\end{equation}
\nd
(where $g_*$ is the number of relativistic degrees of freedom at the
temperature $T$), this condition rewrites
\begin{equation}
m_X 
\gtrsim
g_*^{-1/2}\,\alpha \, M_{\mathrm P} \;\;.
\label{out}
\end{equation}

If $X$ is a Higgs particle, $\alpha$ can be as low as $10^{-\,6}\,$. Even
in this case however the $X$ boson must be very massive. In principle this
may be problematic in the theories with extra dimensions we are interested
in, which have the main goal of having a very low fundamental scale. 

There are some possibilities to overcome this problem. One is related to a
possible deviation of the expansion of the system from the standard
Friedmann law. This is a concrete possibility, since the exact expansion
law is very dependent on the particular brane model one is considering and
on the fact that the size of the compact dimension is or is not
stabilized. For example, we will show in the next chapter that the
Randall-Sundrum model
\cite{Randall:1999ee} with a stabilized radion can
have (depending on the energy density on the zero brane) an expansion rate
higher than the standard one for temperatures close to the cut-off scale
of the system (that is TeV). This  accelerated expansion could in
principle favor the out of equilibrium condition for the $X$ bosons.

However this issue is very dependent on the specific cosmological scenario
adopted, and one may be interested in more general solutions for the out
of equilibrium problem. One very natural possibility is to create the $X$
particles non thermally and to require the temperature of the Universe to
have been always smaller than their mass $m_X\,$. In this way, one
kinematically forbids regeneration of the $X$ particles after their decay.
In addition, although interactions among these bosons can bring them to
thermal equilibrium, chemical equilibrium cannot be achieved.

Nonthermal creation of matter has raised a considerable interest in the
last years. In particular, the mechanism of preheating has proven quite
successful, as we have discuss in the first part of this work. The
efficiency of preheating has been exploited in the work
\cite{Kolb:1996jt}
to revive GUT baryogenesis in the context of standard four-dimensional
theories. Here, we will not go into the details of the processes that
could have lead to the production of the $X$ bosons. Rather, we will
simply assume that, after inflation, their number density is $n_X\,$. To
simplify our computations, we will also suppose that their energy density
dominates over the thermal bath produced by the perturbative decay of the
inflaton field.\footnote{An alternative way to overcome the
bound (\ref{out}) relies on the fact that, as observed in the
works
\cite{Chung:1999rq, Giudice:2001ex}, the maximal temperature reached
by the thermal bath during reheating can indeed be much higher than the
final reheating temperature. In this case, even if $T_{\mathrm {rh}}$ is
considerably lower than $m_X$, $X$ particles can be produced in a
significant amount, and the out of equilibrium condition is easily
achieved. However, the treatment of this mechanism is in our case somewhat
different from the one given in \cite{Chung:1999rq}: due to the
slowness of the expansion of the Universe, the $X$ bosons will decay
before the freeze out of their production. The final baryon asymmetry
cannot be estimated with the use of the formulae of \cite{Chung:1999rq},
which are valid only if the decay of the $X$ particles occurs well after
their freeze out.}

Just for definiteness, let us consider a very simple model 
where there are two species of $X$ boson 
which can decay into quarks and leptons, 
according to the 4-dimensional effective interactions 
\footnote{Here one can not really adopt a GUT regime, since
in unified theories the quarks and leptons are in the 
same multiplets which is not consistent with 
the idea of separation in the extra space.}
\begin{equation}
g\,X\,{\bar q}\,{\bar q} \;\;\;,\;\;\; g\,e^{-\,a/4}\,X\, l\,q \;\;,
\label{deca}
\end{equation} 
where (remember the suppression given by the different localization of
quarks and leptons) the quantity $a$ is defined in eq. (\ref{defa}). Again
for definiteness we will consider the minimal model where no extra
fermionic degrees of freedom are added to the ones present in the Standard
Model. Moreover we will assume $B - L$ to be conserved, even though the
extension to a more general scheme can be easily performed.

The decay of the $X$ bosons will reheat the 
Universe to a temperature that can be evaluated to be
\begin{equation}
T_{\mathrm {rh}}\simeq
\left(\frac{30}{\pi^2}\,\frac{m_X\,n_X}{g_*}\right)^{1/4}\,\,.
\end{equation}

Since we do not want the $X$ particles to be thermally regenerated after
their decay, we require $T_{\mathrm {rh}}
\lesssim  
m_X$, that can be
rewritten as an upper bound on $n_X$
\begin{equation}
n_X
\lesssim
30\,\left(\frac{g_*}{100}\right)\,m_X^3\,\,.
\end{equation}

Another limit comes from the necessity to forbid the $B$ violating four
fermion interaction~(\ref{sca}) to erase the $B$ asymmetry that has been
just created by the decay of the $X$ bosons. We thus require the
interaction~(\ref{sca}) to be out of equilibrium at temperatures lower
than $T_{\mathrm {rh}}$. From eq.~(\ref{supp}) we see that we can
parameterize the four fermion interaction with a coupling
$g^2\,e^{-3\,a/8}/m_X^2$. Hence, the out of equilibrium condition reads
\begin{equation}
g^4\,e^{-3\,a/4}
\lesssim
g_*\, \frac{m_X}{M_{\mathrm P}} \, \left( \frac{m_X}{T_{\mathrm {rh}}} \right)^3 \;\;.
\label{outfour}
\end{equation}

One more upper bound on the reheating temperature comes from the out of
equilibrium condition for the sphalerons. This requirement is necessary
only if one chooses the theory to be $B - L$ invariant, while it does not
hold for $B - L$ violating schemes. We can approximately consider the
sphalerons to be in thermal equilibrium at temperatures above the
electroweak scale. Thus, if $B - L$ is a conserved quantity, we will
require the reheat temperature to be smaller than about $100\, {\mathrm
{GeV}}$. 

If one neglects the presence of the thermal bath prior to the decay of the
$X$ bosons, the very first decays will be only into couples of quarks,
since the  channel into one quark and one lepton is strongly suppressed by
the $e^{-a\left(T=0\right)}$ factor due to the fact that the kink is not
modified by any thermal correction. However, the decay process is not an
instantaneous event. It is shown in \cite{Chung:1999rq} that the
particles produced in the very first decays are generally expected to
thermalize very rapidly, so to create a thermal bath even when most of the
energy density is still stored in the decaying particles. \footnote{As
shown in \cite{Chung:1999rq}, what is called the reheating
temperature is indeed the temperature of the thermal bath when it starts
to dominate. After the first decays, the temperature of the light degrees
of freedom can be even much higher than $T_{\mathrm {rh}}$.} The
temperature of this bath can even be considerably higher than the final
reheating temperature. The presence of the heat bath modifies in turn the
shape of the kink, as shown in the previous section, and we can naturally
expect that this modification enhances the $B$ violating interactions.

If the energy density of the Universe is dominated by the $X$ bosons before they decay, one has 
\begin{equation}
\eta_B\simeq\,0.1 \,\left(N_X\, T_{\mathrm {rh}}/m_X\right)\,\,\langle r-{\bar r}\rangle\;\;,
\label{etabar}
\end{equation}
where $N_X$ is the number of degrees of freedom associated to the $X$
particles and $\langle r-{\bar r}\rangle$ is the difference between the
rates of the decays $X\rightarrow q\, l$ and $\bar X\rightarrow\bar q\,
\bar l$.

We denote with $X_1$ and $X_2$ the two species of bosons whose
interactions~(\ref{deca}) lead to baryon number violation, and parameterize
by $\epsilon$ the strength of CP-violation in these interactions.
Considering that $e^{-2a}$ is always much smaller than one, we get
\cite{Nanopoulos:1979gx}

\begin{equation}
\langle r-{\bar r}\rangle\sim 3\, g^2\, e^{-a/2}\,\epsilon\; {\mathrm
Im}\,\mbox{I}_{SS}\left(M_{X_1}/M_{X_2}\right)\,\:,
\end{equation}
where the function ${\mathrm Im}\,\mbox {I}_{SS}
(\rho)=[\,\rho^2\;{\mathrm Log}(1+1/\rho^2)-1\,]\,/\,(16\,\pi)$ can be
estimated to be of order $10^{-3} - 10^{-2}$. It is also reasonable to
assume $\epsilon \sim 10^{-2} - 1$.

Collecting all the above estimates, and assuming $N_X$ to be of order $10$, we get 
\begin{equation}
\eta_B\simeq \left(10^{-5}\,-\,10^{-2}\right)\, g^2 \, \frac{T_{\mathrm {rh}}}{m_X}\,e^{-a\left(T_{\mathrm {rh}}\right)/2}\;\;.
\label{etabaryon}
\end{equation}

From the requirement $T_{\mathrm{rh}}
\lesssim 
m_X$ we get an upper limit on the baryon asymmetry
\begin{equation}
\eta_B
\lesssim 
\left(10^{-5}\,-\,10^{-2}\right)\,g^2\,e^{-a/2}\,\,,
\label{limit1}
\end{equation}
where the factor $a\left( T\right)$ has to be calculated for a value of $T$ of the order of the reheating temperature. 

We get a different limit on $\eta_B$ from the bound~(\ref{outfour}): assuming $m_X\sim \mbox{TeV}$ and $g_*\sim 100$ indeed one obtains
\begin{equation}
\eta_B
\lesssim
\left(10^{-6}\,-\,10^{-10}\right)\,g^{2/3}\,e^{-a/4}\,\,.
\label{limit2}
\end{equation}

Since the observed amount of baryon asymmetry is of order $10^{-10}$, even
in the case of maximum efficiency of the process (that is, assuming
maximal $CP$ violation and $g\sim 1$), the bounds~(\ref{limit1}) and
(\ref{limit2}) imply that $a\left(T_{\mathrm {rh}} \right) \sim 40\,$.
Unfortunately, the temperature at which the condition $a \left( T \right)
\sim 40\:$ occurs cannot be evaluated by means of the expansion of
eq.~(\ref{acorr}), that have been obtained under the assumption $\left |
a\left ( T\right )-a\left ( 0 \right )\right | \ll a\left ( 0 \right )$.
On the other hand, it is remarkable that our mechanism may work with a
ratio $a\left ( T_d\right ) / a\left ( 0 \right )$ of order one. We thus
expect that a successful baryogenesis may be realized for a range of the
parameters of this model which -- although not possible to 
evaluate through a
perturbative analysis -- should be quite wide and reasonable.

As we have discussed in the previous chapter, in scenarios with large
extra dimensions and low scale gravity, the maximal temperature reached by
the Universe after inflation is strongly bounded from above in order to
avoid overproducing Kaluza-Klein graviton modes, which may eventually
contradict cosmological observations
\cite{add2}. For
instance, in models with two large extra dimensions the reheating
temperature cannot exceed much $1$MeV (unless the fundamental scale $M$
is unnaturally high. 

This value is too low for the
scenario we are describing since $\eta_B$ is proportional to the ratio
$T_{\mathrm {rh}}/m_X$, and hence the observed amount of baryons would be
reproduced at the price of an unnaturally small value of $a\left(
T_{\mathrm {rh}}\right)$. However, other schemes with extra dimensions
exist where the bounds on $T_{\mathrm {rh}}$ are less severe. For example,
in the proposals \cite{Randall:1999ee, Kaloper:2000jb} the mass of the
first graviton KK mode is expected to be of order TeV. The reheating
temperature can thus safely be taken to be of order $10-100 \,
\mbox{GeV}$.

There are of course several possible baryogenesis schemes alternative to
the one just presented (see for instance
\cite{Mazumdar:2001nw,Ibanez:1999it}). 
A possible option which also requires a minimal
extension to the Standard Model could be to achieve the baryon asymmetry
directly through the $4$ fermions interactions $q + q \leftrightarrow q +
l$ in the thermal primordial bath. The out of equilibrium condition may be
provided by the change of the kink as the temperature of the bath
decreases. \footnote{This condition may be easily achieved due of the
exponential dependence of the rate of this process on the temperature, see
equations
(\ref{supp}) and (\ref{acorr}).} What may be problematic is the
source of $CP$ violation which may lead the creation of the baryon
asymmetry. A possibility in this regard may be provided by considering a
second Higgs doublet, but the whole mechanism certainly deserves a deep
analysis by itself.

\section{Gauge Hierarchy, Electroweak Spontaneous Symmetry\\
Breaking, \& Fermion Chirality Linked}\label{ssb}
\label{hssbch}

\subsection{Introduction}
In the SM,
the hierarchy of mass scales is present at both
classical and quantum levels.
At the tree level, there is a huge difference between the scales
associated with the electroweak and the gravitational interactions,
$
{M_W}/{M_p}
\sim { 10^{-17}}$.

If quantum corrections were not to significantly alter the value of the
Higgs (mass)$^2$ computed classically, one could simply
consider the number $10^{-17}$ above         
as one of the many extreme ratios
existing in nature
(like the mass ratio of a
feather and of an elephant). However,
there are huge
quadratic corrections to the Higgs
(mass)$^2$ at the quantum level
due to the fact that the Higgs particle is described by
a fundamental scalar field. These corrections change the
classical value of the Higgs (mass)$^2$ by many orders of magnitude, and
adjusting the value back to its classical one
requires a fine-tuning of order
${10^{-34}}$ in the case of a gravitational cutoff, and
another of order ${10^{-26}}$ in GUTs.

Supersymmetry is a very good example where the problem         
is nicely solved at the quantum level, in fact once the SUSY breaking 
scale
is set classically, usually taken to be around few TeV, only logarithmic 
corrections 
will alter this value. The key principle for the absence of
quadratic divergences in this theory is that the Higgs mass is protected 
by
the
symmetry above the cutoff. Being in a multiplet with
chiral
fermions, the Higgs
is deemed to be massless, due to
chiral symmetry, as long as SUSY is unbroken.     

The idea of large extra dimensions explains {\it perturbative
stability} of the weak scale versus the Planck 
scale,\footnote{Provided
that 
a convincing stabilization 
mechanism for the radius of compactification
is found.}, 
by lowering
the cut-off of the theory. Nevertheless, it does not address the
issue of {\it sensitivity} of the Higgs mass to the ultraviolet
cut-off.
This is an  important issue from the point of view of
the low energy calculability, and is the central point to be
addressed in \cite{Dvali:2001qr}.

In the chapter we present the 
work 
\cite{Dvali:2001qr}, where 
the above key principle is employed, 
though 
without supersymmetry, to explicit models where the Higgs mass is protected 
by a gauge symmetry in $4+d$ dimensions.
It is argued that 
identifying
the electroweak Higgs particle with the extra
components of the gauge field in $4+d$ dimensions provides a solution 
to the hierarchy problem.\footnote{An alternative approach was
discussed in 
\cite{barbieri}, 
where the Higgs mass is controlled
by a higher-dimensional extended supersymmetry, spontaneously
broken globally by Scherk-Schwarz mechanism.
An alternative to the Higgs mechanism was suggested in \cite{st}.
The idea can also be 
thought of as performing a compactification 
on a manifold with 
the first betti number $b_1(Y)=0$. 
}  

The absence of ultraviolate quadratic divergences is due to 
to the manifistation of exact gauge symmetry 
at energies beyond the cutoff.\footnote{Soon after the proposal in 
\cite{Dvali:2001qr}, several models
appeared, \cite{arkani,Hall:2001zb,Antoniadis:2001cv}, 
using the same idea
in explaining the finiteness of the 
Higgs mass.}
The higher-dimensional 
gauge symmetry plays the role of the ``protector''
in this case 
forbidding the Higgs (mass)$^2$ 
from receiving cut off dependent 
(and hence local) corrections. 
This symmetry is spontaneously
broken by compactification.

The idea is implemented within explicit models 
which also provide a link between fermion chirality
and electroweak symmetry breaking 
in four dimensions.

\subsection{The proposal}

After
compactifying extra dimensions on a monopole background,
via mechanism of \cite{sss1}, some of the extra components of the
gauge fields become tachyonic and spontaneously break the
electroweak symmetry. Their quantum numbers are identical to those
of SM Higgs doublet. Notice that the monopole background is
essential for generating the families of chiral fermions in
four-dimensions, and therefore is doing a double job. If the
tachyonic mass is a tree level effect the natural scale of the
symmetry breaking is $\sim 1/a$, the inverse radius of extra
compact space, since the only source of spontaneous breaking of
the higher-dimensional gauge invariance is the compactification
itself. In  other words, since in the infinite volume limit $a
\rightarrow \infty$ the full higher-dimensional gauge invariance
must be recovered, the weak scale must go as $M_W \sim 1/a$. Thus
in this case the size of extra dimensions to which gauge fields
can propagate should be $a \sim 1/$TeV (as in \cite{antoniadis}).
However, it is important to stress that in order for the theory
not to become infinitely strongly coupled above the
compactification scale, the cut-off $M$ must be lowered as in
\cite{add1}, possibly via increase of the volume of some
additional dimensions to which only gravity can spread. This
issue will not be discussed here.

The models constructed in \cite{Dvali:2001qr} were in the 
direction
of answering the following question: 
{\it ``How close can one get 
the to the Standard Model, in four dimensions, by Kaluza-Klein 
compactifying large extra dimensions of an Einstein Yang-Mills
theory coupled to fermions in $4+d$ dimensions?''}. 

Eventually, the effective action of a ``good'' theory should provide 
us in four dimensions with:
\begin{itemize}
\item[-]{ chiral fermions in 4 dimensions.} 
\item[-]{standard model gauge group (or a group containing it), 
spontaneously broken to $U(1)_{em}$.}
\item[-]{fermions 
and Higgs particles in the correct 
representations of the standard model gauge group.} 
\item[-]{a good solution to the 
hierarchy problem.} 
\item[-]{
correct quark and lepton masses.} 
\item[-]{suppressed proton decay.}
\end{itemize}
In the following, 
examples will explicitly given where the first four points 
mentioned above 
can be realized, with some hints to achieving the last two. 

In a an attempt to approaching the answer of the above question, 
two principal ideas were proposed: Identifying the electroweak 
Higgs with the 
extra components of the Yang-Mills
field as a solution for the hierarchy problem, and 
linking fermion chirality to the spontaneous electroweak 
symmetry breaking. \\

\nd
{\bf Approaching the hierarchy problem:}\\

\nd
The suggestion to solving the hierarchy 
problem is that the electroweak Higgs particle is identified with 
the extra components of the gauge field in $4+d$ dimensions. 

Note that this identification will not be meaningful, 
from the point of view
of explaining the absence of quadratic divergences to the Higgs mass, 
unless it is proven that those extra components (which are a scalar
in four dimensions) possess a $\phi^4$ potential with a negative
bilinear term so to derive the usual spontaneous symmetry breaking
described by the SM. Not only that, this scalar filed should
also
have the appropriate Yukawa couplings with the fermions in the
right SM group representations.

In the examples constructed in \cite{Dvali:2001qr}, a Higgs-type
potential is produced at the tree-level and therefore the 
electroweak breaking scale will be set by the compactification scale. 
A more appealing regime to implement 
the above idea is when the extra components of the gauge field
of interest have a vanishing tree-level (mass)$^2$ and acquire
a negative bilinear term via quantum corrections, at one loop
say, as can happen in SM \cite{Coleman:1973jx} (the procedure is 
outlined in section \ref{massless}). 
This will render the electroweak scale to be, as desired,
an order of magnitude or two less than the scale
of compactification, $1/a$, which is taken usually to be few TeV. 

Since the Higgs phenomena happens only
in four dimensions and at  
low energies, the spontaneous symmetry breaking will be dealt 
with exactly as in the standard model.
This description will be valid at low energies, below $1/a$,
as the 
four-dimensional observer will only see the effective 
Lagrangian 
of the standard model however with an extra advantage that 
most of its parameters 
are fixed (the magic hand is the higher-dimensional theory): the 
parameters in the Higgs potential; Yukawa couplings; and the 
gauge couplings are all determined by the scale of compactification
and the original Yang-Mills coupling in the higher-dimensional 
theory.


Suppose that we start from an Einstein Yang-Mills theory coupled to fermions
on a $4+d$ dimensional manifold, $W$, $W=M_4\times K$. The field content
to start with  
consists of a graviton, $g_{MN}(x,y)$, 
a gauge field $A^a_M(x,y)$ in the algebra of a Lie group $G$, 
and a fermion $\psi(x,y)$. Where $M,N=0,1,2,...,d+3$, $a=1,...\mbox{dim}G$, 
$x$ and $y$ are the coordinates on $M_4$ and $K$ respectively. We take 
$Y$ to be a compact manifold with a typical volume of order TeV$^d$, and 
$\times$ to indicate a tensor product.   
Arguing that the Higgs particle is  
$${H (x)\;\equiv \;A_\alpha(x)}\;\;\;\;\;\;\alpha\in K$$
implies a solution for the quantum instability of 
the electroweak scale
as was first pointed out by 
\cite{Hatanaka:1998yp,hata2}.\footnote{An earlier attempt for 
obtaining spontaneous symmetry breaking in 
six dimensions is in \cite{manton}.} 

Classically, the Planck scale is related to the fundamental 
$4+d$ dimensional gravity scale, as we 
mentioned previously, by the relation
$$
M_P= a^{\frac{d}{2}}\;M^{\frac{d}{2}+1}
$$
When $K=K_1\times K_2\times...\times K_n$, the $a^{d}$ 
should be replaced by the product of the volumes of each manifold.
As an example, we take 
$$
W=M_4\times S^2\times {\mathbb{C}}P^2
$$
as in the model we discussed in \cite{Dvali:2001qr} 
for quarks and leptons. In this example 
we have the gravity scale $M\sim 10^4$TeV, and hence new physics is 
expected to show off at around $1/a \sim 1$TeV.
Whether one considers the cutoff to be $M$, $1/a$, or the scale at 
which the gauge coupling in $10$ dimensions becomes strong 
(in our case this happens at around $1$TeV as well), the hierarchy 
between the weak scale and the cutoff, $\Lambda$, is much milder than 
the one in the ordinary gravity or grand unified theories: 
$$
\frac{m_H}{\Lambda} \sim 10^{-5} -- 10^{-1}
$$

Quantum mechanically, the absence of quadratic, or large, 
divergences as the ones present in the standard model
\footnote{$
\delta m^2_H=\frac{1}{8\pi^2}(\lambda^2_H-\lambda^2_t)\Lambda^2
+\mbox{log.div. + finite term}$, 
where $\lambda^2_H$ and $\lambda^2_t$ are the self coupling
of the Higgs and its coupling to the top quark respectively.}
can be understood via the argument of symmetry, as in the 
case of SUSY, however with no fundamental scalar to start with. 
The gauge symmetry in this case is spontaneously broken due 
to the presence of 
a topologically non-trivial background,\footnote{
In \cite{bachas, b2}, similar
ideas have been used to study dynamical breaking of supersymmetry
in the context of type I string theory. In 
\cite{Sakamoto:2001gn} the idea of achieving 
spontaneous breaking of the C, P, and rotational 
symmetries by topological defects in the internal space was
discussed.}
as will be explained  later on, 
however,
at energies larger than the compactification scale, 
it is recovered and the Higgs field is massless being
a component of the massless gauge field.
In other words the Higgs mass$^2$, $m_H^2$, can not be larger 
than $1/a$  
$$m_H^2\;=\; \frac{1}{a^2}\;f(Ea)$$
\nd
where $E$ is the common energy scale, and $a$ is the typical radius 
of compactification. We conjecture that 
$\displaystyle
\lim_{E\to \infty}
f(Ea)=0$. In fact, it was shown in 
\cite{hata2,hata3,host} that 
the function $f(Ea)$ is exponentially damping at energies higher 
than $1/a$. Finding the explicit form of $f$ in our 
case is technically more complicated due to the presence of a  
monopole background, however we believe that the 
$\displaystyle
\lim_{E\to \infty}
f(Ea)$ will always be finite. \\ 

\nd 
{\bf Linking chirality and SSB:}\\

As discussed in an earlier chapter, 
the only way to get chiral fermions in Kaluza-Klein 
type field theories  
couple them to a topologically non-trivial background 
\cite{Randjbar-Daemi:1983hi,Witten:ed.ux}. This 
in principle changes both the index and the kernel of  
$\slash{D}$ and hence allows for achieving a chiral theory, 
as the standard model, in four dimensions. Examples will 
be shown in details in the following sections.


\subsection{The background solution\label{2}}


As discussed earlier, the background solutions should satisfy 
the Einstein and Yang-Mills classical equations of motion. 
Although the background we are going to use will solve the field equations
of any generally covariant and gauge invariant action 
containing the metric
and the Yang-Mills fields only, for the sake of simplicity we
start from  Einstein-Yang-Mills system in
$D$-dimensions. The action is given by
$$
S= \int d^D x \sqrt{-G} \left(\frac{1}{\kappa^2}
{\cal R} -\frac{1}{2g^2}\mbox{Tr}F^2 +\lambda +
{\bar \psi } i \slash{\nabla} \psi \right )
$$
\nd
where $\psi$ is in some representation of the gauge
group $G$. This action can be the low energy string
field theory action with the $\lambda$-term induced
by some mechanism. The presence of $\lambda$ in our
discussion is required if we insist on having product
spaces like $M_1\times M_2\times...$ as a solution of
the classical bosonic field equations, where one of
the factors in the product is flat, e.g. the flat
$4$-dimensional Minkowski space (as discussed earlier). 
Our argument about
chirality is not sensitive to the flatness of any of
the factors in the product. The presence of tachyons,
however, depends on the definition of a mass
operator. This is different for example in AdS$^d$
and (Minkowski)$^d$. 

The bosonic field equations are (\ref{bosonic1}) and (\ref{bosonic2}).
In this paper we shall consider solutions of the form
$M_4\times K$, where $M_4$ is the flat $4$-dimensional
Minkowski space and $K$ is a compact manifold. Therefore
the equations of motion (\ref{ink}) will be used. 

The internal space $K$ will 
be mostly taken to be either $S^2$ or
$S^2\times {\mathbb{C}}P^2$. Furthermore we shall
assume that the gauge field configuration $A$ will be
non-vanishing only on $K$. One can of course think of
many other choices for $K$.


For $K={\mathbb{C}}P^1\times {\mathbb{C}}P^2$ the
metric is given by
\begin{equation}
ds^2=a_1^2\left(d\theta^2+\mbox{sin}^2\theta d
\varphi^2 \right) +\frac{4a_2^2}{1+\zeta^{\dag}\zeta}
d{\bar \zeta}^a \left(\delta^{ab}-
\frac{\zeta^a{\bar\zeta}^b}{1+\zeta^{\dag} \zeta}
\right)d\zeta^b
\label{fub}
\end{equation}
\nd
where $a_1$ and $a_2$ are the radii of
${\mathbb{C}}P^1$ and ${\mathbb{C}}P^2$ respectively,
and $\zeta=(\zeta^1,\zeta^2)$ is a pair of local complex
coordinates in ${\mathbb{C}}P^2 $. The ${\mathbb{C}}P^2$
metric is the standard {\it Fubini-Study} metric. There
are two facts about ${\mathbb{C}}P^2 $ which are of
importance for our present discussion. The first is the
isometry group $SU(3)$ of ${\mathbb{C}}P^2 $. Together
with the invariance group $SU(2)$ of the metric of $S^2$,
$SU(3)$ will form part of the gauge group in $M_4$.
$SU(3)$ will be identified with the strong interaction
color gauge group. The low energy $4$-dimensional gauge
group will be ${\tilde G}\times SU(2)\times SU(3)$, where
$\tilde G$ is the subgroup of the $D$-dimensional gauge
group $G$ which leaves the background solution invariant.
Note that even with $G=U(1)$ we can obtain a
$4$-dimensional gauge theory with a gauge group $U(1)
\times SU(2)\times SU(3)$. Although such a solution can
produce chiral fermions in a non-trivial representation
of $U(1)\times SU(2)\times SU(3)$, it is not possible,
however,  to obtain the correct Standard Model spectrum
of leptons, quarks, and the Higgs fields. For this we
need a bigger $G$. We shall discuss this point in a
greater detail in a later section. 

The second important fact about ${\mathbb{C}}P^2 $ is
that in the absence of a background $U(1)$ gauge field
it is not possible to have globally well defined spinor
field on it. This is principally due to the fact that
the complex coordinates $\zeta$ do not cover ${\mathbb{C}}
P^2 $ globally. We need at least three patches $(U,\zeta)$,
$(U',\zeta')$, and $(U{''},\zeta{''})$, where in $U\bigcap
U'$ we have the transition rule $\zeta_1'=
\frac{1}{\zeta_1}$ and $\zeta_2'=\frac{\zeta_2}{\zeta_1}$.
It needs some work to show that the
two chiral spinors of the
tangent space $O(4)$ of ${\mathbb{C}}P^2 $ can not be
patched consistently on the overlap. We shall give some more
details of this later on.

To write the solution of the Yang-Mills equations on
$K={\mathbb{C}}P^1\times {\mathbb{C}}P^2$ we first work out
the spin connection on $K$. It is given by
\begin{equation}
\Omega = -(\mbox{cos}\theta -1)d\varphi\frac{\tau^3}{2}+
\left(\begin{array}{cc}\frac{1}{2}\omega^i\sigma^i& 0\\
0& -\frac{3}{2}\omega \sigma_3\end{array}\right)
\label{X}
\end{equation}
\nd
where the first factor refers to ${\mathbb{C}}P^1$and the
second, which is a $4\times 4$ matrix, refers to
${\mathbb{C}}P^2$. Here $\tau^3$ as well as $\sigma^i$ and
$\sigma_3$ are Pauli matrices. Also the expressions are
valid on the upper hemisphere on ${\mathbb{C}}P^1$ and the
local patch $(U,\zeta)$ on ${\mathbb{C}}P^2$. The expressions
for $\omega^i$ and $\omega$ can be read from the Fubini-Study
metric (\ref{fub}) on ${\mathbb{C}}P^2$. We shall not need
the explicit expression for $\omega^i$. The one for $\omega$
is given by
\begin{equation}
\omega(\zeta, {\bar \zeta})= \frac{1}{2\left(1+\zeta^{\dag}
\zeta\right)}\left(\zeta^\dag d\zeta - d\zeta^\dag \zeta
\right)
\label{a}
\end{equation}
\nd
Note that $d\omega$ is the self dual K\"{a}hler form on
${\mathbb{C}} P^2$. It is thus an instanton type solution of
the Yang-Mills equation in ${\mathbb{C}} P^2$.

It is important note from (\ref{X}) that
the
${\mathbb{C}} P^2$ spin-connection takes its values in the
subgroup $SU(2)
\times U(1)$ of the tangent space $SO(4)$. Furthermore, under
$ SO(4)\rightarrow SU(2)\times U(1) $ the two chiral spinors
of $O(4)$ decompose according to
\begin{eqnarray}
2_+&=& 2_0 \\
2_-&=&1_{-\frac{3}{2}}+1_{\frac{3}{2}}
\end{eqnarray}
\nd
where the subscripts indicate the $U(1)$-charges. Using this
fact one can understand why spinors are not globally well
defined on ${\mathbb{C}}P^2$. The point is that in the overlap
of two patches $(U,\zeta)$ and $(U',\zeta')$ we have
\begin{equation}
\omega(\zeta')= \omega(\zeta) -id\varphi
\end{equation}
\nd
where $\varphi$ is defined by $\zeta_1=|\zeta_1|
\mbox{e}^{i\varphi}$. For $2_-$ to be globally well defined
$1_{\pm 3/2}$ should patch according to the rule $\psi'
(\zeta')=\mbox{e}^{\pm \frac{3}{2}i\varphi} \psi(\zeta)$. We
thus obtain transition functions which are anti-periodic under
$\varphi\rightarrow \varphi + 2\pi$. Coupling a background
gauge field proportional to $\omega$ can change this. With a
little more work one can show that a similar obstruction also
prevents $2_+=2_0$ from being well defined.

Now we are in a position to write our solution of the
Yang-Mills equation on  ${\mathbb{C}}P^1\times {\mathbb{C}}
P^2$. It is easy to show that the ansatz
\begin{equation}
A=\frac{n}{2}(\mbox{cos}\theta
-1)d\varphi + qi\omega
\label{back}
\end{equation}
\nd
where $n=\mbox{diag}(n_1, n_2,...)$ and $q=\mbox{diag}(q_1,
q_2,...) $ are matrices in the Cartan-subalgebra of $G$. The
consistent patching of spinors requires that $n_1, n_2,...$
be integers and $q_1,q_2,...$ be one half of an odd integer.
Note that the substitution of the above ansatz in the Einstein
equations will require that the radii $a_1$ and $a_2$ of
${\mathbb{C}}P^1$ and ${\mathbb{C}}P^2$ are quantized.

As mentioned in the beginning
of this section our ansatz for the background
configuration solves the field equations derived from any
generally covariant and gauge invariant Lagrangian in $D=10$, which contains
the metric and the Yang-Mills potentials only. Such an effective Lagrangian
will contain infinite number of parameters and therefore the relationship
between the radii and other parameters will be more involved.


\subsection{Chiral fermions\label{section3}}


It is a well known fact that in order to obtain chiral
fermions in $D=4$ we need topologically non-trivial
background
gauge fields on
${\mathbb{C}}P^1\times{\mathbb{C}}P^2$. Our solution for the
Yang-Mills equations consist of magnetic monopole on $S^2$
and the potential for the K\"{a}hler form on ${\mathbb{C}}P^2$.
The  K\"{a}hler form defines a topologically non trivial line bundle on
${\mathbb{C}}P^2$ . 

Consider the $D=10$ fermion Lagrangian
\begin{equation}
{\cal L}={\bar \psi}i \slash{\nabla}\psi
\end{equation}
\nd
where
\begin{equation}
\nabla_{\hat M}\psi = (\partial_{\hat M}+\omega_{\hat M}
-i A_{\hat M}
)\psi\;\;, \;\; {\hat M}=0,1,...,9
\end{equation}
\nd
$\omega_{\hat M}$ and $A_{\hat M}$ are, respectively, the
$SO(1,9)$ and the Lie algebra valued spin and gauge connections.
We analyze the fermion problem in two steps. In the first step
we write the manifold as $M_6\times  {\mathbb{C}}P^2 $.
Correspondingly we write the $D=10$ Dirac matrices as
\begin{eqnarray}
{\hat \Gamma}_a& =&\Gamma \times \gamma_a\;\;\;\;\;\;\;a=6,7,
8,9\nonumber\\
{\hat \Gamma}_A &=&\Gamma_A \times 1\;\;\;\;\;\;\;A=0,1,...,
5\nonumber
\end{eqnarray}
\nd
where $\gamma_a$ and $\Gamma_A$ are respectively $4\times 4$
and $8\times 8 $ Dirac matrices satisfying
\begin{eqnarray}
\{\gamma_a,\gamma_b \}&=&2\delta_{ab} \nonumber\\
\{\Gamma_A,\Gamma_B \}  &=&2\eta_{AB}\nonumber
\end{eqnarray}
\nd
and $\Gamma=\Gamma_0\Gamma_1...\Gamma_5$.

Substituting these $\Gamma$'s into ${\cal L}$ and recalling
that the geometry has factorized form we obtain
\begin{equation}
{\cal L}={\bar \psi}\Gamma i \slash{\nabla}_{ {\mathbb{C}}P^2
}\psi + {\bar \psi}i \slash{\nabla}_{M_6 }\psi
\end{equation}
\nd
The chiral fermions on $M_6$ will originate from those modes
for which
\begin{equation}
\slash{\nabla}_{ {\mathbb{C}}P^2
}\psi=0
\label{psi}
\end{equation}
\nd
Those $\psi$'s which are not annihilated by
$ \slash{\nabla}_{{\mathbb{C}}P^2}$ will give rise to massive
fermionic modes on $M_6$. The standard way to analyze
(\ref{psi}) is to operate one more time with
$ \slash{\nabla}_{{\mathbb{C}}P^2}$ on it. Using the background
connections (\ref{X}) and $(\ref{back})$ we obtain
\begin{equation}
(\nabla^2-\frac{3}{2})\;\psi_+=0
\label{psi+}
\end{equation}
\begin{equation}
\{ \nabla^2+(q\;\sigma_3-\frac{3}{2}) \}\;\psi_-=0
\label{-}
\end{equation}
\nd
where
\begin{equation}
\nabla \psi_+=(d+i\omega^r\frac{\sigma^r}{2}+\omega\;q)\;\psi_+
\end{equation}
\begin{equation}
\nabla \psi_-=\{d+\omega(q-\frac{3}{2}\sigma_3)\}\;\psi_-
\end{equation}
\nd
and
$$
\psi_\pm =\frac{1\pm {\hat \gamma}_5}{2}\;\psi\;\;\;\;,\;\;\;\;
{\hat \gamma}_5 =\gamma_6\gamma_7\gamma_8\gamma_9$$
\nd
The K\"{a}hler instanton $\omega$ is given by equation
(\ref{a}). Since
$\nabla^2\leq 0$ (\ref{psi+}) will have no non-zero solutions.
Thus fermions of $\psi_+$ type will all be non-chiral and massive.
Equation (\ref{-}), on the other hand, can have solutions. Their
existence depends on the eigenvalues of $q$. Clearly for $q=3/2$
we have only one solution with $\sigma_3=+1$. For $q=+5/2$ we
obtain $3$ solutions with $\sigma_3=+1$. They form a triplet of
the isometry group $SU(3)$ of ${\mathbb{C}}P^2$. For $q=-5/2$ and
$\sigma_3=-1$ we obtain a $3^*$ of $SU(3)$. These are the only type
of solutions we need to consider. 

Next we study the $M_6$ Dirac Lagrangian
\begin{equation}
{\cal L}={\bar \psi}i \slash{\nabla}_{M_6 }\psi
\label{6}
\end{equation}
\nd
where $\psi$ is assumed to be a solution of (\ref{-}). We shall
assume that the $D=10$ spinor is chiral and has positive
chirality. Then the spinor of $\psi_-$ type will have
negative
$D=6$ chirality. We choose the $D=6$ $\Gamma$ matrices to be
\begin{eqnarray}
\Gamma_\alpha& =&\Gamma_\alpha \times \tau_1\;\;\;
\;\;\;\;\alpha=0,1,2,3\nonumber\\
\Gamma_4 &=&\Gamma_5 \times \tau_1\;\;\;\;\;\;\;
\gamma_5=i\gamma_0\gamma_1\gamma_2\gamma_3\nonumber\\
\Gamma_5&=&1\times \tau_2
\end{eqnarray}
\nd
and $\tau_1= \left(\begin{array}{cc}
0&1 \\
1& 0\end{array}\right)$,  $\tau_2= \left(\begin{array}{cc}
0&-i \\
i& 0\end{array}\right)$.

Inserting the $\Gamma_A$'s in (\ref{6}) we obtain
\begin{equation}
{\cal L}={\bar \psi}i \slash{\nabla}_{M_6 }\psi
+\frac{i}{\sqrt{2}} \left\{{\bar \psi}(\gamma_5+1)D_-
\psi+ {\bar \psi}(\gamma_5-1)D_+\psi
\right \}
\end{equation}
\nd
where
\begin{equation}
D_\pm \psi=e_{\pm}^m\left(\partial_m +\frac{i}{2}\omega_m
(n-\gamma_5)
\right)\psi
\end{equation}
\nd
$e_{\pm}^m$ are the $U(1)$ components of an orthonormal frame
on $S^2$ and $\omega_m$ is the corresponding spin connection
($\omega_\theta =0$, $\omega_\varphi=-\mbox{cos}\theta+1$
in the upper hemisphere and
$\omega_\varphi=-\mbox{cos}\theta-1$ in the lower hemisphere
).\ Decomposing
$\psi=\psi_L+\psi_R$, where
$\displaystyle\psi_L=
\frac{1-\gamma_5}{2} \psi$,\ we obtain the analogue of
(\ref{psi+}, \ref{-}) for the Dirac operator on ${\mathbb{C}}
P^1$
$$
\left\{\nabla^2-\frac{1}{2}(1-n)  \right\}\psi_R=0
$$
$$
\left\{\nabla^2-\frac{1}{2}(1+n)  \right\}\psi_L=0
$$
\nd
$n=1$ produces one $\psi_R$ while $n=-2$ gives rise to two
$\psi_L$ which form a doublet of the Kaluza-Klein $SU(2)$.


\subsection{General rules for Higgs-type tachyons}\label{section4}


To obtain the spectrum of the effective theory in
$4$-dimensions we need to expand the functions about our
background solution in harmonics on ${\mathbb{C}}P^1\times
{\mathbb{C}}P^2$. These include fluctuations of the
gravitational, Yang-Mills, as well as fermionic fields. The
techniques of doing such analysis have been developed long
ago. In this paper we shall ignore the gravitational
fluctuations and consider only the Yang-Mills and fermionic
fields. The full set of linearized gravity Yang-Mills
equations can be found in \cite{sss1}. In the same paper it
was shown that there are tachyonic modes in the components of
the gauge field fluctuations tangent to $S^2$. Here, we would
like to show that the rule to identify the tachyonic modes
given in \cite{sss1} for $G\equiv SU(3)$ is in fact quite
general and applies to any gauge group $G$. It should be
emphasized that neglecting the gravitational fluctuations is
justified as they will not mix with the gauge field
fluctuations of interest for us. 

In general, we should write $A={\bar A}+V$ where $\bar A$ is
the background solution and $V$ depends on the coordinates of
$M_4$, $S^2$ and ${\mathbb{C}}P^2$. Our first interest is in
the fields which are tangent to $S^2$. It is these fields,
which if develop a tachyonic vacuum expectation value, can
break $SU(2)$, provided such modes are singlets of $SU(3)$
isometry of ${\mathbb{C}}P^2$.

We suppress the ${\mathbb{C}}P^2$ dependence of these fields
and denote by $V_{\underline 1}$ and $V_{\underline 2}$ their
components with respect to an orthonormal frame on $S^2$. It
is convenient to use the ``helicity'' basis on $S^2$ defined
by
$$
V_{\pm}=\frac{1}{\sqrt{2}}(V_{\underline 1} \mp i\;
V_{\underline 2})
$$
\nd
$V_{\pm}$ are matrices in the Lie algebra of $G$. What governs
their mode expansion on $S^2$ is their isohelicities. This is
basically the effective charge of $V_{\pm}$ under the
combination of $U(1)$ transformations which leave our background
configuration invariant. These charges can be evaluated in the
same way which was done in \cite{sss1}. For the sake of
simplicity, let us assumes $G=U(N)$ and assume that charge
matrices $n$ and $q$ introduced in (\ref{back}) are diagonal
$N\times N$ matrices. Then $V_\pm$ are $N\times N$ matrices with
elements ${V_{\pm i}}^j$, $i,j=1,...N$. Their isohelicities,
$\lambda({V_{\pm i}}^j)$, are given by
$$
\lambda({V_{\pm i}}^j)=\pm 1 + \frac{1}{2}(n_i-n_j)
$$
\nd
Note that there is a hermiticity relation
$$
{V_{+i}}^j = \left( {V_{-j}}^i \right)^*
$$
\nd
The harmonic expansion of ${V_{+i}}^j$ on $S^2$ will produce an
infinite number of Kaluza-Klein modes. These expansions are
defined by \begin{equation}
V_\pm (x,\theta,\varphi)= \sum_{l\geq |\lambda_\pm|}
\sqrt{\frac{2l+1}{4\pi}} \sum_{m\leq|l|} V_\pm^{lm} (x)\;
D_{\lambda_\pm,m}^l(\theta,\varphi)
\end{equation}
\nd
$ D_{\lambda_\pm,m}^l(\theta,\varphi)$ are $2l+1$-dimensional
unitary matrices.

The tachyonic modes are generally contained in the leading terms
with $l=|\lambda_\pm|$. The effective $4$-dimensional mass$^2$
of $V_\pm^{lm}(x)$ obtains contributions from the appropriate
Laplacian acting on $S^2$ and ${\mathbb{C}}P^2$. $V_\pm$ are
charged scalar fields on ${\mathbb{C}}P^2$. We shall analyze
their dependence on the ${\mathbb{C}}P^2$
coordinates in the next section.
Here we shall consider the $S^2$ contribution to their masses.
The condition for this contribution to be tachyonic is expressed
in the following simple rule
$$
M^2({V_{+i}}^j) < 0\;\;\;\;\mbox{if}\;\;\;\;\lambda( {V_{+i}}^j)
\leq 0$$
\nd
Likewise
$$
M^2({V_{-i}}^j) < 0\;\;\;\;\mbox{if}\;\;\;\;\lambda( {V_{-i}}^j)
\geq 0$$
\nd
To prove these claims let us make more detailed analysis. 

Since we are assuming $V_\pm$ are independent of the
${\mathbb{C}}P^2$ coordinates, their mass term comes from the
expansion of $\mbox{Tr} F_{mn}F^{mn}$, where $m,n$ indicate
indices tangent to $S^2$. The cubic and the quadratic parts in
$\mbox{Tr} F_{mn}F^{mn}$ will produce the interaction terms in
the Higgs potential. We have
\begin{eqnarray}
\mbox{Tr}
F_{mn}F^{mn}&=&  \mbox{Tr}{\bar F}_{mn}{\bar F}^{mn} +
\mbox{Tr}(D_+V_- - D_-V_+)^2\nonumber\\
&-&4\;i\;  \mbox{Tr}{\bar F}_{+-} [V_-,V_+]
-2\;i \;\mbox{Tr}(D_+V_- - D_-V_+) [V_-,V_+]\nonumber\\
&+&\mbox{Tr}[V_-,V_+]^2
\nonumber
\end{eqnarray}
\nd
where the covariant derivatives are defined by
$$
D_mV_n=\nabla_m V_n - i[{\bar A}_m,V_n]
$$
\nd
$\nabla_m$ denotes the ordinary Riemannian covariant
derivative on $S^2$. Now, since for $\lambda_+\leq 0$ \
($ \lambda_+\geq 0 $)
$D_- D_{\lambda_+,m}^{l=|\lambda_+|}=0  =  D_+
D_{\lambda_-,m}^{l=|\lambda_-|}   $
we see that such modes will be annihilated by $D_\pm$ and thus
the $S^2$ contribution to their $D=4$ action is given by
$$
S=-\frac{1}{2g^2}\int_{0}^{2\pi}d\varphi\int_{0}^{\pi}
\mbox{sin}\theta \;\mbox{Tr}\left\{ 4D_\mu V_+D^\mu V_- -4i
{\bar F}_{+-} [V_-,V_+]
+[V_-,V_+]^2 \right\}
$$
\nd
The mass terms hence come from $-4iTr{\bar F}_{+-}[V_-,V_+]$
term only. 

To proceed it is convenient to choose the Cartan-Weyl basis for
the Lie algebra of $G$. Let $Q_j$ denote the basis of the Cartan
subalgebra, $E_\alpha$ and $E_{-\alpha}=E_\alpha ^\dag $ the
generators outside the Cartan subalgebra. The only part of the
algebra needed for the evaluation of the mass terms is
$$
[Q_j,E_\alpha]= \alpha_j E_\alpha
$$
\nd
In this basis we can write
$$
V_\pm= V_\pm^\alpha E_\alpha+ (V_\mp^\alpha)^*E_{-\alpha}
+V_\pm^jQ_j
$$
\nd
It is easy to see that
\begin{equation}
\lambda(V_\pm^\alpha)=\pm 1+ p.\alpha
\end{equation}
\nd
where $p.\alpha=p^j\alpha_j $ and $p^j$ are defined by
$$
\frac{1}{2}n=p^jQ_j
$$
To
simplify the discussion consider the case when only one
$\lambda(V_+^\alpha)\leq 0$. Set the remaining modes to zero.
Of course this is not a loss of generality. In this case
$V_+=V_+^\alpha E_\alpha$ and
$V_-=(V_+^\alpha)^*E_{-\alpha}$. The mass term then becomes
\begin{eqnarray}
\mbox{Tr}\left(-4i {\bar F}_{+-}[V_-,V_+]\right)
&=&-4i
\mbox{Tr} V_+ [ {\bar F}_{+-} ,V_-]
\nonumber\\
&=& \frac{4}{a_1^2}\;p.\alpha\;|V_+^\alpha|^2
\mbox{Tr}E_\alpha E_{-\alpha}
\nonumber
\end{eqnarray}
\nd
where we inserted $\displaystyle {\bar F}_{+-}=-
\frac{i}{a^2}p^jQ_j$. The kinetic part of the action for
$V_+^\alpha$ thus becomes
$$
S_2= -\frac{2Tr( E_\alpha E_{-\alpha})}{g^2}
\int_{0}^{2\pi}d\varphi\int_{0}^{\pi} d\theta \;
\mbox{sin}\theta \left\{
\partial_\mu V_+^\alpha\partial^\mu {V_+^\alpha)}^*+
\frac{p.\alpha}{a_1^2}{|V_+^\alpha|^2}
\right\}
$$
Substituting $p.\alpha=\lambda(V_+^\alpha)-1$ we obtain the
mass of $V_+^\alpha$ in terms of its isohelicity as (recall
that our signature is $(-,+,+,...)$)
\begin{equation}
m^2=\frac{\lambda-1}{a_1^2}
\label{mass}
\end{equation}
\nd
which is negative for $\lambda\leq 0$.
Similar reasoning can be applied if for some $V_-^\alpha$
the corresponding isohelicity $\lambda(V_-^\alpha) $ is
non-negative.

This rule gives us an easy way of identifying possible
tachyonic modes which can act as Higgs scalars in the $D=4$
effective theory.


\subsection{Examples\label{examples}}


In this section we shall ignore the
${\mathbb{C}}P^2$ part and give some examples of a $D=6$
gravity Yang-Mills theories which produce standard model
type Higgs sectors upon compactification to $D=4$. Leptons
and quarks will be included in the next sections.
We basically need to choose the gauge group  $G$ and assign
magnetic charges $n$. 

The notation is always
\begin{equation}
{\bar A}=\frac{n}{2}(\mbox{cos}\theta\mp 1)d\varphi
\end{equation}
\nd
where $
n=\mbox{diag}(n_1,n_2,...)
$ is in the Lie algebra of $G$, $-(+)$ give the expression
for $\bar A$ in the upper (lower) hemispheres.


\subsubsection{Tachyons}


\subsubsection*{\mbox{\boldmath $G=SU(3)$}}

\begin{equation}
n=\mbox{diag}(n_1,n_2,-n_1-n_2),\;\;\;\;\;\;n_1,n_2\in {\mathbb{Z}}
\end{equation}
\nd
The isohelicities can be assembled in a $3\times 3$ matrix
\begin{equation}
\lambda(V_\pm)=
\left(\begin{array}{ccc}
\pm 1 & \pm 1+\frac{1}{2}( n_1-n_2)& \frac{1}{2}( 2n_1+n_2) \\
\pm 1- \frac{1}{2}(n_1-n_2) &\pm 1 &\pm 1 +\frac{1}{2}(n_1+2n_2) \\
\pm 1-\frac{1}{2}(2n_1+n_2) & \pm 1-\frac{1}{2} (n_1+2n_2)& \pm 1
\end{array}
\right)
\end{equation}

Using the results of
section
\ref{section3} we see that in order to obtain
left handed doublets and right handed singlets we had to
take $(n_1,n_2)=(1,1)$. With these values of $n_1$ and $n_2$,
${V_{-1}}^3$ and ${V_{-2}}^3$ will contain tachyonic modes
in the leading term of their expansion on $S^2$.\\

In this example the  $SU(2)\times U(1)$ subgroup of $SU(3)$
is unbroken and the tachyonic Higgs ${V_{-1}}^3$ and
${V_{-2}}^3$ form a doublet of $SU(2)$ with $U(1)$ charge
of $3/2$. We denote this doublet by $\phi$. Its isohelicity
is $+1/2$. Therefore it will also be a doublet of the
Kaluza-Klein isometry of $S^2$. One can integrate the
$(\theta,\varphi)$ dependence of $\phi$ on $S^2$ and work
out its $D=4$ effective action. The result is
$$
{\cal L}= -\frac{1}{2g^2}
\int_{0}^{2\pi}d\varphi\int_{0}^{\pi} d\theta \;
\mbox{sin}\theta \;\mbox{Tr} F_{MN}F^{MN}
\;\;\;\;\;\;\;\;\;\;\;
\;\;\;\;\;\;\;\;\;\;\;
\;\;\;\;\;\;\;\;\;\;\;$$
$$
\;\;\;\;\;
= -\frac{1}{4g_1^2} {F^8_{\mu\nu}}^2 -
\frac{1}{4g_2^2} {F^r_{\mu\nu}}^2
-\frac{1}{4e^2} {W^r_{\mu\nu}}^2
$$
\begin{equation}
\;\;\;\;\;\;\;\;\;\;\;\;
\;\;\;\;\;\;\;\;\;\;\;
\;\;\;\;\;\;\;\;\;\;\;
\;\;\;\;\;\;\;\;
-\mbox{Tr}\left\{
\nabla_\mu \phi^\dag \nabla^\mu \phi -\frac{3}{2a_1^2}
\phi^\dag\phi+2g_1^2 (\phi^\dag\phi)^2\right\}
\end{equation}
\nd
where we have regarded $\phi$ as a $2\times 2$ complex
matrix, and
$$
\nabla_\mu \phi=\partial_\mu\phi-\frac{3}{2}iV_\mu^8\phi
-iV_\mu^r \frac{\sigma^r}{2}\phi-i W_\mu^r \phi
\frac{\tau^r}{2}$$
\nd
where $V_\mu^8$, $V_\mu^r$, and $W_\mu^r$ are
respectively the $U(1)$, $SU(2)_L$, and the Kaluza-Klein
$SU(2)_R$ gauge fields. $g_1$, $g_2$, and $e$ are their
respective couplings. Some calculation show that
\begin{equation}
g_2=\frac{1}{2\sqrt{\pi}} \frac{g}{a_1}=\sqrt{3} g_1
\end{equation}
\nd
The Kaluza-Klein gauge coupling $e$ can also be expressed
in terms of the fundamental scales $g$ and $a_1$.\\

In the next section we shall work out the Yukawa couplings
for this model as well.

\subsubsection*{ \mbox{\boldmath $G=U(6)$} }

With $n=\mbox{diag}(n_1,..., n_5,n_6)$
we can again work
out the table of isohelicities for $V_\pm$. We shall see in
section \ref{5.2} that in order to obtain one family of leptons
and quarks we need to take $n=\mbox{diag}(-2,1, 1,-2,1,1)$.
Note that since the group is $U(6)$ rather than $SU(6)$, $n$
is not traceless. $\lambda(V_\pm)$ is given by
\begin{equation}
\displaystyle
\lambda(V_+)=
\left(\begin{array}{cccccc}
+1
\;&-\frac{1}{2}
\;&-\frac{1}{2}
\;&+ 1
\;&-\frac{1}{2}
\;&-\frac{1}{2}
\\
+\frac{5}{2}
\;&+1
\;&+1
\;&+\frac{5}{2}
\;&+1
\;&+ 1  \\
+\frac{5}{2}
\;&+1
\;&+1
\;&+\frac{5}{2}
\;&+1
\;&+1
\\
+1
\;&-\frac{1}{2}
\;&-\frac{1}{2}
\;&+1
\;&-\frac{1}{2}
\;&-\frac{1}{2}
\\
+\frac{5}{2}
\;&+1
\;&+ 1
\;&+\frac{5}{2}
\;&+ 1
\;&+ 1
\\
+\frac{5}{2}
\;&+1
\;&+1
\;&+\frac{5}{2}
\;&+1
\;&+1
\end{array}
\right)
\label{iso}
\end{equation}
\nd
Since
$V_-=V_+^\dag$,
therefore $\lambda({V_{-i}}^j)=-\lambda({V_{+j}}^i)$.\\

The tachyonic modes are contained in ${V_{+1}}^i$,
and
${V_{+4}}^i$ where $i=2,3,5,6$.
They will all be doublets
of the Kaluza-Klein $SU(2)$. They also transform under some
representation of the unbroken part of $U(6)$, which is $SU(2)
\times SU(2)
\times U(1)^3\times U(1)'$, which is generated by
$\displaystyle
\mbox{diag}(0,\frac{\sigma^i}{2}, 0,0,0)$, \\
$i=1,2,3$;
$\displaystyle\mbox{diag}(0,0,0,0,\frac{\sigma^i}{2})$;
$\mbox{diag}(-2,1,1,0,0,0)$;
$\mbox{diag}(0,0,0,-2,1,1)$;\\
$\mbox{diag}(1,1,1,-1,-1,-1)$;
and the $6\times 6$ unit matrix
$1_6$ which generates $U(1)'$. The tachyonic Higgs will be
neutral under this $U(1)'$, therefore  their tree level
vacuum expectation value will not break it. Under
$U(6)\rightarrow SU(2)\times SU(2)\times U(1)^3$
we have
\begin{equation}
\underline{6}= (1,1)_{(-2,0,1)}
+(2,1)_{(1,0,1)}+ (1,1)_{(0,-2,-1)}+
(1,2)_{(0,1,-1)}
\end{equation}
\nd
The quantum numbers of the relevant Higgs tachyons will be
\begin{eqnarray}
{V_{+1}}^i &\sim& (2,1)_{(-3,0,0)}\;\;\;i=2,3 \label{v+1}\\
{V_{+4}}^t &\sim& (1,2)_{(0,-3,0)}\label{67} \;\;\;t=5,6
\end{eqnarray}

As we said earlier, the tachyonic modes in all these fields
will be in the doublet representation of the Kaluza-Klein
$SU(2)$. The vacuum expectation value of
the fields ${V_{+1}}^i\sim (2,1)_{(-3,0,0)}$ and
${V_{+4}}^t \sim
(1,2)_{(0,-3,0)}$ will give masses to the quarks and leptons
respectively. In section \ref{section6} we shall show that
the leading term in their expansion on ${\mathbb{C}}P^2$
is a singlet of $SU(3)$ and therefore their masses
receive no contribution from the dependence on the
${\mathbb{C}}P^2$ coordinates. Thus they remain tachyonic.
The
other tachyonic fields,
namely ${V_{+1}}^i$, $i=5,6$; ${V_{+4}}^t$, $t=2,3$,
would induce Yukawa couplings between
quarks and leptons.
We shall show that in fact the leading term in
their harmonic expansion on  ${\mathbb{C}}P^2$ is
a triplet of
$SU(3)$. Thus the vacuum expectation value
of these fields can
break the color $SU(3)$.
We will determine the conditions to avoid this.


\subsubsection{Fermions\label{anomaly}}


We consider the two examples of the previous section.

\subsubsection*{
\mbox{\boldmath $G=SU(3)$} \label{anomaly1}}

Here we assume that $D=6$ and there is no ${\mathbb{C}}P^2$
factor. Let us take $\psi$ in $\underline{3}$ of $SU(3)$ and
$n=\mbox{diag} (1,1,-2)$. According to our rules this will
produce two right handed singlets of the Kaluza-Klein $SU(2)$
which we denote by $SU(2)_K$ and a left handed doublet. The
singlets will form a doublet of $SU(2)_G\subset SU(3)$ and the
doublet of $SU(2)_K$ will be a singlet of $SU(2)_G$. Thus under
$SU(2)_K\times SU(2)_G\times U(1)$ where $U(1)\subset SU(3)$ we
have $(1,2_R)_{1/2}+(2_L,1)_1$.
The $D=4$ Yukawa
and gauge
couplings can be easily worked out. The result is
\begin{eqnarray}
{\cal L}_F&=&
\int_{0}^{2\pi}d\varphi\int_{0}^{\pi}d\theta\;
\mbox{sin}\theta\; {\bar \psi}i \slash{\nabla}\psi
\;\;\;\;\;\;\;\;
\;\;\;\;\;\;\;\;
\;\;\;\;\;\;\;\;
\;\;\;\;\;\;\;\;
\;\;\;\;\;\;\;\;
\;\;\;\;\;\;\;\;
\nonumber\\
&=&{\bar \lambda}_L\; i \gamma^\mu\left(
\partial_\mu-ig_1V_\mu^8-ie W_\mu^i\frac{\tau^i}{2}
\right)\lambda_L
\nonumber\end{eqnarray}
$$
\;\;\;\;\;\;\;\;
+
{\bar \lambda}_R\; i \gamma^\mu\left(
\partial_\mu-i\frac{g_1}{2}V_\mu^8-ig_2 V_\mu^i
\frac{\sigma^i}{2}\right)\lambda_R
$$
$$\;\;\;\;\;\;\;\;
\;\;\;\;\;\;\;\;\;
\;\;\;
-2g_1
\left\{{\bar \lambda}_L\phi (i\sigma_2)\lambda_R
-{\bar \lambda}_R (i\sigma_2)\phi^\dag\lambda_L
\right\}
$$
\nd
where $\lambda_L=(2_L,1)_1$ and $\lambda_R=(1,2_R)_{1/2}$.

This expression together with the bosonic part given in
equation (\ref{6})
give the total effective $D=4$ action for the
$SU(3)$ example. Although this example leads to interesting
chiral and Higgs spectrum in $D=4$ can not be considered
satisfactory. It has both perturbative and global chiral
anomalies in $D=6$. The perturbative anomalies can be
eliminated with the standard Green Schwarz
mechanism\cite{gs}. To
apply this mechanism \cite{sss3} we need first to introduce an
antisymmetric rank two potential together with three right
handed
$D=6$ $SU(3)$ singlets to kill the pure
gravitational anomaly which is given by ${\cal R}^4$ term in the
anomaly $8-$ form. The remaining terms in the anomaly
$8$-form factorize appropriately in order to be canceled
by a judicious transformation of the antisymmetric
potential. This mechanism does not cancel the global
anomalies \cite{witten2}
whose presence is due to the fact that $\pi_6
(SU(3))=Z_6$ is non zero. To kill these ones we need to
introduce further $SU(3)$ multiplets or to change the gauge
group altogether and chose to a gauge  group
like
$E_6$ which has a trivial $\pi_6
(E_6)$.

\subsubsection*{ \mbox{\boldmath $G=U(6)$}\label{5.2}}

Now assume $D=10$ and choose $\psi$ to be in $\underline{6}$ of
$U(6)$ and $q=\mbox{diag}(5/2,5/2,5/2,\\
3/2,3/2,3/2)$. As before
$n$ will  be taken to be $n=\mbox{diag}(-2,1,1,-2,1,1)$.\
According to the  results of the previous section with
respect to the isometry  group $SU(2)\times SU(3)$ we have
the following chiral fermions
$$
(2_L,\underline{3})+
(1_R,\underline{3})+
(1_R,\underline{3})+
(2_L,1)+
(1_R,1)+(1_R,1)
$$
\nd
Clearly the first three triplets are candidates for $\left(
\begin{array}{c}
u\\ d
\end{array}\right)_L$, $u_R$ and $d_R$. The last two pieces can
be identified with the leptons $\left(\begin{array}{c}\nu_e\\ e
\end{array}\right)_L$ and $e_R$.\footnote{
We have an extra right handed singlet in the lepton sector. This
can be removed by choosing the last entry in $q$ to be for
instance $-1/2$ or the last entry in $n$ to be $0$. In this
way the unbroken subgroup of $U(6)$ will be
$SU(2)\times U(1) \times U(1)'$.}

These multiplets also transform in the following representation
of the unbroken $SU(2)\times SU(2)\times U(1)^3\subset G$
\begin{eqnarray}
(2_L,\underline{3})&\sim&(1,1)_{(-2,0,1)}\nonumber\\
(1_R,\underline{3})+(1_R,\underline{3})
&\sim& (2,1)_{(1,0,1)}\nonumber\end{eqnarray}
$$
(2_L,1)\sim  (1,1)_{(0,-2,-1)}\;\;\;\mbox{and}\;\;\;
(1_R,1)\sim  (1,2)_{(0,1,-1)}
$$
\nd
The Yukawa coupling between the quarks will be through the Higgs
field ${V_{+1}}^i$ given in (\ref{v+1}), while the electron will
get its mass through coupling to ${V_{+4}}^t$.  Thus our
construction leads to a multi Higgs theory in which the
quarks and leptons obtain their masses from their Yukawa
couplings to different Higgs scalars.  Note also that
there is no common $U(1)$ under which both Higgs
multiplets are charged. The hypercharge coupling in our
model is different from the standard electroweak theory.


\subsection{Higgs like Tachyons on \mbox{${\mathbb{C}}P^2
\times {\mathbb{C}}P^1$}\label{section6}}


If the total space-time dimension is $D=6$ the masses of the Higgs
like tachyons are given by (\ref{mass}). In the case of a $D=10$
theory we need to take into account the contribution of
${\mathbb{C}}P^2$ part as well. The fields $V_\pm$ are like scalar
fields on ${\mathbb{C}}P^2$ which are
charged with respect to the
${\mathbb{C}}P^2$ part of the background gauge field (\ref{back}),
viz, $iq\omega$. The
${\mathbb{C}}P^2$
contribution to the masses of $V_\pm$
come
from the commutator term in the ${\mathbb{C}}P^2$ covariant
derivative of $V_\pm$, i.e.
\begin{eqnarray}
DV_\pm&=&dV_\pm-i[i \omega q, V_\pm] \nonumber\\
&=&dV_\pm+\omega[q,V_\pm]\nonumber
\end{eqnarray}
\nd
To be specific let us consider the example of the $U(6)$ model
for which $q=\mbox{diag}(5/2,5/2,5/2;3/2,3/2,3/2)$. Write
\begin{equation}
V=
\left(\begin{array}{c|c}
\;\;v\;\;&\;\;u\;\;\\
\hline
\;\;{\tilde u}\;\;& \;\;{\tilde v}\;\;\end{array}
\right)
\label{V}
\end{equation}
\nd
where $v$, $\tilde v$, $u$, and $\tilde u$ each is a
$3\times 3$ matrix.
Then
$$
[q,V]=
\left(\begin{array}{c|c}
\;\;0\;\;&\;\;u\;\;\\
\hline
\;-{\tilde u}\;\;&\;\; 0\;\;\end{array}
\right)
$$
\nd
This indicates that out of the Higgs fields given in
equations (\ref{v+1}--\ref{67}) the ones which give masses to
quarks and leptons, namely, ${V_{+1}}^i$ and ${V_{+4}}^t$
(which lie respectively inside $v$ and $\tilde v$ in the above
notation), do not couple to the background $\omega$ field on
${\mathbb{C}}P^2$. The leading term in their harmonic expansion
on ${\mathbb{C}}P^2$ will be a constant (independent of the
coordinates of ${\mathbb{C}}P^2$). Their masses will be tachyonic
and will be given by (\ref{mass}) for $\displaystyle\lambda=-
\frac{1}{2}$, i.e.
$\displaystyle M^2=-\frac{3}{2}\frac{1}{a_1^2}$.

The remaining fields ${V_{+1}}^t$ and ${V_{-i}}^4$ on the other
hand are located inside $u$ and they couple to the background
$\omega$-field. Their masses will receive contribution from
${\mathbb{C}}P^2$ and in principle can become non-tachyonic.
To verify this we need to evaluate the eigenvalues of
$\nabla^2_{{\mathbb{C}}P^2}$ on these fields. Their covariant
derivatives are
$$D{V_{+1}}^t =d{V_{+1}}^t+ \omega {V_{+1}}^t$$
$$D{V_{-4}}^i =d{V_{-4}}^i+ \omega {V_{-4}}^i $$
\nd
Since they couple with the same strength to the $\omega$-field
they will receive the same contribution from $\nabla^2_{{\mathbb{C}}
P^2}$. It turns out that the leading term in the expansion of any of
these fields on ${\mathbb{C}}P^2$ is a triplet of $SU(3)$ and $D^2$
acting on it is $\displaystyle -\frac{1}{a_2^2}$. Thus the total
mass$^2$ of such modes will be
$$
-\frac{3}{2}\frac{1}{a_2^2}+\frac{1}{a_2^2}=\frac{1}{a_1^2}
\left( -\frac{3}{2}+\frac{a_1^2}{a_2^2} \right)
$$
\nd
If $a_1$ and $a_2$ were independent we could choose $\displaystyle
(\frac{a_1}{a_2})^2
\geq \frac{3}{2}$ and make these fields non-tachyonic. If we insist
on the validity of the background Einstein equations then the ratio
of $\displaystyle\frac{a_1}{a_2}$ will be fixed. Equation
(\ref{ink}) leads to $\displaystyle (\frac{a_1}{a_2})^2=
\frac{12}{17}$.
\footnote{To obtain $\displaystyle (\frac{a_1}{a_2})^2=
\frac{12}{17}$  we need to use the following results, which
can be obtained by straightforward calculation,\\
$\displaystyle{\cal R}(S^2)=
\frac{1}{a_1^2}
1_{2\times 2}$,\
$\displaystyle{\cal R}({\mathbb{C}}P^2)=\frac{3}{2}\frac{1}{a_1^2}
1_{4\times 4}$,\
$\displaystyle\mbox{Tr}F^2_{S^2}=6 \frac{1}{a_1^4}$,\
and
$\displaystyle\mbox{Tr}F^2_{
{\mathbb{C}}P^2
}=\frac{51}{2} \frac{1}{a_2^2}$.}
With this value
unfortunately the above mass$^2$ is still negative. The vacuum
expectation value of these fields will break the color $SU(3)$.

One way to change the ratio $\displaystyle\frac{a_1}{a_2}$ is to
couple a $U(1)$ gauge field to gravity in $D=10$. This $U(1)$ will
{\it not} couple to anything else. In particular the fermions will
be neutral under it, so the spectrum of the chiral fermions will be
unaltered. Its sole effect will be to add an extra term to the
right hand side of Einstein equations. In particular (\ref{ink})
will be replaced by
$$
{\cal R}_{{\hat m}{\hat n}}=\frac{\kappa^2}{g^2}\mbox{Tr}
F_{{\hat m}{\hat r}}{F_{\hat n}}^{\hat r}
+\frac{\kappa^2}{{g'}^2}\mbox{Tr}
{F'}_{{\hat m}{\hat r}}{{F'}_{\hat n}}^{\hat r}
$$
\nd
where $F'$ and $g'$ refer to the extra $U(1)$ system. Now if we set
$$
A'=\frac{n'}{2}(\mbox{cos}\theta
-1)d\varphi + q'i\omega
$$
\nd
where
$n'$ and $q'$ are real numbers, the ratio of $a_1/a_2$
will turns out to be
$$
\displaystyle
\frac{a_1^2}{a_2^2}=
\frac{36+3{n'}^2\frac{g^2}{{g'}^2}}{51+2{q'}^2\frac{g^2}{{g'}^2} }
$$
\nd
There is a big range of parameters for which  $a_1/a_2\geq 3/2$ .


\subsection{Other scalars \label{section7}}


The components of the gauge field fluctuations tangent to
${\mathbb{C}}P^2$ will also give rise to infinite tower of
Kaluza-Klein modes which will be scalars fields in $D=4$. These
modes will belong to unitary representations of $SU(2)\times
SU(3)$. If there is a tachyon Higgs among them they will break
$SU(3)$. We need to verify that this does not happen. To this end
we denote these fields by $V_a$, where $a$ is tangent to
${\mathbb{C}}P^2$, and write those terms in the bilinear part of
$\mbox{Tr}F_{MN}F^{MN}$ which contains $V_a$. In this section we
are considering only the $U(6)$ model. The $V_a$ are $5\times 5$
Hermitian matrices. After some manipulation and the imposition if
the $D=10$ background gauge condition $D_MV^M=0$, the bilinear
terms of interest to us can be written as
\begin{equation}
S_2= -\frac{1}{2g^2}\int d^{10}x \mbox{Tr}\{
2V_a(-\partial^2-D_m^2-D_{\hat m}^2+\frac{3}{2}\frac{1}{a_2^2})
V^a
+4iV^a[{\bar F}_{ab},V^b] \}
\label{s2}
\end{equation}
\nd
where $D_m$ and $D_{\hat m}$ are respectively the covariant
derivatives on $S^2$ and ${\mathbb{C}}P^2$ and
\begin{equation}
D_m V_a=\partial_m V_a-\frac{i}{2} \omega_m [n,V_a]
\end{equation}
\begin{equation}
D_{\hat m} V_a=\nabla_{\hat m}V_a-\frac{i}{2} \omega_{\hat m}[q,V_a]
\label{ru3}
\end{equation}
\nd
$\nabla_{\hat m}$ is the Riemann covariant derivative on
${\mathbb{C}}P^2$. The contribution of $D_m^2$ on each $SU(2)$
mode of $V_a$ will simply be $\displaystyle
\frac{1}{a_1^2}[l(l+1)-\lambda^2]$,
$l\geq |\lambda|$ where $\lambda$ represents the isohelicities of
various components of $V_a$, $\lambda({V_{ai}}^j)=\lambda (
{V_{+i}}^j)-1$, where $\lambda ({V_{+i}}^j)$ are given
in equation (\ref{iso}).

To work out the contributions of $D_{\hat m}^2$ and the commutator
term $[{\bar F}_{ab},V^b]$, we represent $V_a$ as in (\ref{V}),
i.e.
\begin{equation}
V_a=
\left(\begin{array}{c|c}
\;\;v_a\;\;&\;\;u_a\;\;\\
\hline
\;\;{\tilde u_a}\;\;& \;\;{\tilde v_a}\;\;\end{array}
\right)
\label{Va}
\end{equation}
\nd
where $v_a$, $\tilde v_a$,  $u_a$, $\tilde u_a$
each is a $3\times 3$ matrix.
Then
\begin{equation}
[q,V_a]=
\left(\begin{array}{c|c}
\;\;0\;\;&\;\;u_a\;\;\\
\hline
\;-{\tilde u_a}\;\;&\;\; 0\;\;\end{array}
\right)
\end{equation}
\nd
This indicates that the commutator terms in (\ref{s2}) and
(\ref{ru3}) do not contribute to $D_{\hat m}v_a$ and
$D_{\hat m}{\tilde v}_a$. Thus $D_{\hat m}$ acting on these fields
is just the Riemannian Laplacian acting on vectors and its
contribution to the masses of these fields will be non-tachyonic.\\

The only fields we need to be concerned about are those in
$u_a$. To  analyze the contribution of these terms we
introduce $2$ complex
$SU(2)$ vectors $u_\alpha$ and $u'_\alpha$ defined by
\begin{equation}
\left\{\begin{array}{l}
u_{\underline{1}}=\frac{1}{\sqrt{2}}(u_6+iu_7) \\
u_{\underline{2}}=\frac{1}{\sqrt{2}}(u_8+iu_9) \\
\end{array} \right.
\;\;\;\;\;\;\;\;\;\;\;\;\;\;
\left\{\begin{array}{l}
u'_{\underline{1}}=\frac{1}{\sqrt{2}}(u_6-iu_7) \\
u'_{\underline{2}}=\frac{1}{\sqrt{2}}(u_8-iu_9) \\
\end{array} \right.
\end{equation}
\nd
where $6$, $7$, $8$, and $9$ are directions tangent to
${\mathbb{C}}P^2$. In terms of these new fields the $u_a$ part
of (\ref{ru3}) can be rewritten as
\begin{equation}
D_{\hat m} u_\alpha=
(\partial_{\hat m}+i\omega_{\hat m}^i\frac{\sigma^i}{2}
-i\frac{5}{2} \omega_{\hat m}) u_\alpha
\label{x4}
\end{equation}
\begin{equation}
D_{\hat m} u'_\alpha=
(\partial_{\hat m}+i\omega_{\hat m}^i\frac{\sigma^i}{2}
-i\frac{5}{2} \omega_{\hat m}) u'_\alpha
\label{x5}
\end{equation}
\nd
The contribution of $D^2_{\hat m}$ on $u_\alpha$ and $u'_\alpha$
will again be positive. 

Finally we need to evaluate the contribution of $2i\mbox{Tr}V^a
[{\bar F}_{ab}, V^b]$ to the masses of $u_\alpha$ and $u'_\alpha$.
After some calculation this turns out to be
\begin{equation}
2i\mbox{Tr}V^a
[{\bar F}_{ab}, V^b]=
\frac{2}{a^2_2}\mbox{Tr}(u^\dag_
\alpha u_\alpha-{u'}^\dag_\alpha {u'}_\alpha)
\label{fi}
\end{equation}
\nd
It is seen that the contribution of this term to the $u_\alpha$
mass is non-tachyonic. However, it makes a negative contribution
to the mass$^2$ of $u'_\alpha$ field. Upon substitution of the
above in (\ref{ru3}) we find out that the negative contribution in
(\ref{fi}) is off-set by the $\displaystyle
\frac{3}{2}\frac{1}{a_2^2}$ term in
equation (\ref{s2}), with the result that $u'_\alpha$ is also
non-tachyonic. 

We thus conclude that all the tachyonic Higgs are singlets of
$SU(3)$ and doublets of $SU(2)$.


\subsection{Massless scalars and loop-induced hierarchy
\label{massless}}


So far we have been discussing tachyonic mass of the scalar
particles at the tree level of the effective $4$ dimensional
theory. The natural scale of this mass and therefore also of the
symmetry breaking is the compactification scale. This is few
order of magnitude above the electroweak symmetry breaking scale
of a 200 hundred GeV. It will be very desirable if we could find
a mechanism to lower the scale of the tachyonic mass. An obvious
idea is if the tree level mass of the scalars is zero and they
obtain their tachyonic value as a consequence of loop effects.
Our theory is of course a non renormalizable one, at least in
conventional sense. However, the Higgs mass is
controlled by
$1/a$ due to higher dimensional gauge invariance.
Our main point is that the sign of the one loop induced effective
mass will depend on the imbalance between the contribution of
fermions and bosons. By a judicious choice of the fermionic
degrees of freedom this sign can be made tachyonic. Any way
whatever the justification the first step in implementing this
idea is to find tree level massless scalars in the spectrum of the
effective four dimensional theory. Unlike the massless chiral
fermions whose presence is dictated by the topology of the gauge
field in compact subspace, to verify the existence of the massless
scalars in the spectrum requires more detailed analysis of the
mass spectrum and should be carried out separately for each case.
In this section we give an example of a model in $D=10$ in which a
monopole background on the $S^2 \times S^{'2}\times S^{''2}$
internal space leads to  massless scalars transforming non
trivially under the $SU(2)\times SU(2)\times SU(2)$ isometry group
of the internal space. This example which was  is only for
illustrative purpose
and is not going to be used for a realistic model building.

We start from a $U(N)$ gauge theory in $10$ dimensions and
consider a solution of equations (\ref{ink}) in which the internal space
is $S^2\times S^{'2}\times S^{"2}$. In the notation of previous
section we denote the magnetic charge matrices on the three
$S^{2}$'s by  $n$  $n'$ and $n"$. Denoting all the quantities on
$S{'2}$ with a prime our ansatz for the gauge field becomes
$$
A = {n\over 2}(cos \theta -1)d\phi + {n'\over 2}(cos \theta'
-1)d\phi' +{n''\over 2}(cos \theta'' -1)d\phi''
$$

The structure of the charge matrices will determine the unbroken
subgroup of $U(N)$. As before we shall take them to be $N\times
N$ diagonal real matrices.

The scalars of interest for us are those components of the
fluctuations of the vector potential which are tangent to
$S^2\times S^{'2}\times S^{"2}$ and are in the directions of
perpendicular to the Cartan subalgebra of $U(N)$. Consider the
field $V_{-i}^{j}$ tangent to $S^2$.

The masses of these fields can be calculated using the
appropriate modification of equation (\ref{s2}). The result is\\

$
\displaystyle
S_2= -\frac{1}{2g^2}\int d^{10}x  \Big\{
(V_{-i}^{j})^*(-\partial^2-D^2-D^{'2}- D^{''2}+{1\over a^2})
V_{-i}^{j}$
\begin{equation}
\;\;\;\;\;
\;\;\;\;\;\;
\;\;\;\;\;
\;\;\;\;\;\;\;\;\;\;\;
\;\;\;\;\;\;
\;\;\;\;\;
\;\;\;\;\;
\;\;\;\;\;
\;\;\;\;\;
\;\;\;\;\;
-{1\over a^2}(V_{-i}^{j})^*(n_{i}- n_{j})V_{-i}^{j}
\Big\}
\label{s22}
\end{equation}
\nd
where $D^2$, $D^{'2}$ and $D^{''2}$ are the appropriate
Laplacian on the three $S^2$'s. The eigenvalues of these
Laplacians are basically determined from the isohelicities of
$V_{-i}^{j}$  which are given by\\

$
\lambda(V_{-i}^{j}) = -1 +{1\over 2}(n_i-n_j),\;\;\;\;\;
\lambda^{'}(V_{-i}^{j}) = {1\over 2}(n_i^{'}-n_j^{'}),\;\;\;\;\;
\lambda^{''}(V_{-i}^{j}) ={1\over 2}(n_i^{''}-n_j^{''})$\\

Similar expressions can be written for the bilinear parts of the
fields tangent to $S^{'2}$ and $S^{''2}$.

For our illustrative example we consider an $n$ matrix which has
only the elements $n_1$ and $n_2$ different from zero and
such that $n_1 -n_2 \geq2$. Then $\lambda (V_{-1}^{2})\geq 0$ and
according to our general rule the leading mode in this field can
be tachyonic.  The question we would like to answer is if by an
appropriate choice of magnetic charges we can make the mass of
this field to vanish. It is not difficult to write down the
formula for the masses of the infinite tower of modes of
$V_1^2$. These are given by
$$a^{2}M^2= l(l+1) - \lambda^2 + {a^2\over a^{'2}}
(l'(l'+1) -\lambda^{'2})+ {a^2\over a^{''2}}
(l''(l''+1) -\lambda^{''2}) + 1-(n_1-n_2)$$

To verify the existence of a massless mode
first we employ the background
equations (\ref{ink}) to obtain the ratios
$$
{a^2\over a^{'2}}= {{\mbox{Tr}n^2}\over {\mbox{Tr} n^{'2}}}
,\;\;\;\mbox{and}\;\;\;
{a^2\over a^{''2}}= {{\mbox{Tr} n^2}\over
{\mbox{Tr} n^{''2}}}.$$
\nd
It is seen that for the choice of $n'_1 - n'_2 =
n_1- n_2$ , $\mbox{Tr} n^2= \mbox{Tr} n^{'2}$
and $ n_{1}^{''}- n_{2}^{''}=0 $
the leading mode is  indeed massless. For this choice there will
of course be a similar massless mode in the fluctuations $V_{-
1}^{'2}$ tangent to $S^{'2}$. The $SU(2)\times SU(2)\times SU(2)$
quantum numbers of these modes will be \\
$(l={1\over 2}(n_1 -n_2)-1,
l'={1\over 2}(n_1 -n_2), 0)$, and
$l=({1\over 2}(n_1 -n_2),
l'={1\over 2}(n_1 -n_2)-1, 0)$,
respectively.  We can make all
other modes to have positive masses by appropriate choices of the
remaining magnetic charges.

\section{Concluding Remark}

Considering the possibility of living in 
a higher-dimensional space-time 
seriously opens new gates towards 
understanding the present problems of particle
physics, mainly the unification of gravity with the 
other fundamental forces, and the gauge hierarchy 
problem.
So far, the dimensionality of our space-time is an assumption
based on our observational, technical,
and probably mental 
limitations. The new colliders, like LHC, 
will be able to test this possibility and detect a possible 
modification of Newton's low at energies around few TeV. 

The last three years has witnessed an enormous development in 
extra-dimensional model building. 
It is very desirable to significantly
increase the predictability of the new scenarios as
a compensation of having extra dimensions.
In particular, it would be nice if these
models are free of the same problems the standard model 
suffers from in four dimensions at low energies. 

The hope is that, eventually,
most of these models can be somehow 
a low energy effective field theoretical description 
of a supertring theory. 
Alternatively, one can be more modest and 
seek a consistent 
theory in a field theory context trying to 
make the best of this approach.

Interestingly enough, by going to higher dimensions
one can link some of the features of the 
standard model together. The fermion chirality 
and electroweak spontaneous symmetry breaking can 
have the same origin: a non-trivial Yang-Mills
background in the internal space. What is even 
more interesting is that the mechanism in which the
SM gauge group is spontaneously broken does not 
involve a fundamental scalar field, the thing
which leads to the absence of quadratic corrections
to the classical value of Higgs mass square.   

There are certainly many other interesting 
issues to discuss. Unfortunately,
it was impossible to incorporate all the 
work done in this subject as the recent
literature consists of more than 1,000 article written
from 1998 until now, touching various sides 
and implications of the new thrilling idea,
while the total number of 
papers, old and new, probably
exceeds 2,000. Therefore, only
examples could be provided.

As perhaps all new subjects in physics, 
the models with extra dimensions seem
to open a way for new problems and not to completely solve any.
It remains to check the classical and semiclassical 
stability for many of the scenarios with both
categories of warped and factorizable geometry.
The issue of semiclassical stability was not 
discussed here, and can be found in 
\cite{witten5}-\cite{ssez} for Kaluza-Klein compactification,
and for instance \cite{Ida:2001qw} for brane-worlds. 
Furthermore, the stabilization of the compactification scale 
close to the electroweak scale is an open problem although
some attempts exist 
\cite{Arkani-Hamed:2001kx,Goldberger:1999uk}.
Other aspects not discussed here include
unification
with large extra dimensions \cite{Dienes:1998vh},
brane-world in \cite{add1} scenario,
thick branes as 
\cite{Bonjour:1999kz}-\cite{Kobayashi:2001jd},
the cosmological constant problem
(see \cite{rubn} for a review),...{\it etc.},
in addition to many aspects of the 
older literature (for which the reader is refered 
to 
\cite{ssez} and \cite{kk}).

\subsection*{Acknowledgement}
I am grateful to Antonio Masiero and Seifallh Randjbar-Daemi for 
supervising my thesis, and to ICTP and SISSA for sponsoring me during
my studies.


\begin{thebibliography}{99}







\bibitem{history}
S.~ Dalley, 
``Myths from mesopotamia, creations, the flood, Gilgamesh 
and others'', a new translation, Oxford Paperbacks, 
Oxford Un. Press, ISBN 0-19-281789-2.


\bibitem{exp}
[LEP Collaborations],
``A combination of preliminary 
electroweak measurements and constraints  on the standard model,''
hep-ex/0103048.
For a review, see H.~Przysiezniak,
``Precision electroweak measurements,''
hep-ex/0101008.

\bibitem{Kajita:1999bw}
T.~Kajita  [Super-Kamiokande Collaboration],
Nucl.\ Phys.\ Proc.\ Suppl.\  {\bf 77} (1999) 123.

\bibitem{Brown:2001mg}
H.~N.~Brown {\it et al.}  [Muon g-2 Collaboration],
Phys.\ Rev.\ Lett.\  {\bf 86} (2001) 2227.

\bibitem{Carena:2000yx}
M.~Carena {\it et al.},
arXiv:hep-ph/0010338.

\bibitem{Lane:2000pa}
For a recent review see: K.~D.~Lane,
``Technicolor 2000,''
hep-ph/0007304.


\bibitem{Kounnas:1984cj}
C.~Kounnas, A.~Masiero, D.~V.~Nanopoulos and K.~A.~Olive,
{Singapore, Singapore: World Scientific 
(1984) 425 P. (International School For Advanced Studies Lecture Series, 2)}.

\bibitem{Nilles:1984ge}
H.~P.~Nilles,
Phys.\ Rept.\  {\bf 110} (1984) 1.

\bibitem{add1}
N.\ Arkani-Hamed, S.\ Dimopoulos, G.\ Dvali,
Phys. Lett. {\bf B429}\ (1998) 263.

\bibitem{add3}
I. Antoniadis, N. Arkani-Hamed,
S. Dimopoulos, G. Dvali, Phys. Lett. {\bf B436}\
(1998) 257.


\bibitem{sch}
J.~Scherk and J.~H.~Schwarz,
Phys.\ Lett.\ B {\bf 57} (1975) 463.

\bibitem{sh1}
K.~Akama,
Lect.\ Notes Phys.\  {\bf 176} (1982) 267.

\bibitem{sh2}
V.~A.~Rubakov and M.~E.~Shaposhnikov,
Phys.\ Lett.\ B {\bf 125} (1983) 136.

\bibitem{Rubakov:1983bz}
V.~A.~Rubakov and M.~E.~Shaposhnikov,
Phys.\ Lett.\ B {\bf 125} (1983) 139.

\bibitem{antoniadis}
I.~Antoniadis,
Phys.\ Lett.\ B {\bf 246} (1990) 377.



\bibitem{price}
 J.~C.~Price, in Proc. of Int. Symp. on Experimental gravitational 
Physics, ed. P.~F.~Michelson, Guangzhou, China
(World Sientific, Singapor, 1988).

\bibitem{Long:1999dk}
J.~C.~Long, H.~W.~Chan and J.~C.~Price,
Nucl.\ Phys.\ B {\bf 539} (1999) 23.


\bibitem{kap3}
J.~C.~Long,
``Laboratory Seach for Extra-Dimensional Effects in the 
Sub-millimeter Regime''; 
A.~Kapitulnik, ``Experimental Tests of Gravity Below
1 mm'', 
{\it Talks presented at the International Coference
on Physics Beyind Four Dimensions}, ICTP, Triest, Italy (2000).


\bibitem{kaluza}
Th.~ Kaluza, Sitzungsber. Preuss. Akad. Wiss. Phys. Math. Klasse 966 
(1921).

\bibitem{klein1}
O.~ Klein, Nature {\bf 188} (1926) 516.

\bibitem{Wetterich:1984uc}
C.~Wetterich,
Nucl.\ Phys.\ B {\bf 242} (1984) 473.

\bibitem{Randall:1999ee}
L.~Randall and R.~Sundrum,
Phys.\ Rev.\ Lett.\  {\bf 83} (1999) 3370.


\bibitem{Groom:2000in}
D.~E.~Groom {\it et al.}  [Particle Data Group Collaboration],
Eur.\ Phys.\ J.\ C {\bf 15} (2000) 1.

\bibitem{add2}
N.~Arkani-Hamed, S.~Dimopoulos and G.~R.~Dvali,
Phys.\ Rev.\ D {\bf 59} (1999) 086004.

\bibitem{nord}
G.~Nordstr\"{o}m, Phys. Zeitsch. {\bf 15} (1914) 504.

\bibitem{weyl}
H.~Weyl, Sitzunberger. Akademie der Wissenschaften
Berlin, (1918) 465.
An English presentation of this paper can be 
found in  
L.~O'Raifeartaigh and N.~Straumann,
hep-ph/9810524.

\bibitem{dewitt}
B.~S.~De Witt, in ``Relativity, Groups, and Topology'', 
Eds. C.~De Witt and B.~S.~De Witt, 
Gordon and Breach, New York (1964).

\bibitem{klein2}
O.~ Klein, Z. F. Physik {\bf 37} (1926) 895.

\bibitem{kk}
``Modern Kaluza-Klein Theories'', (Edited by T.~ Appelquist, 
A.~ Chodos, and P.~ Freund), Frontiers In Physics v. 65, Addison-Wesley 
Pub. Com. (1987).

\bibitem{kerner}
R.~Kerner,
Annales Poincare Phys.\ Theor.\  {\bf 9} (1968) 143.

\bibitem{Cremmer:1978km}
E.~Cremmer, B.~Julia and J.~Scherk,
Phys.\ Lett.\ B {\bf 76} (1978) 409.


\bibitem{Nahm:1978tg}
W.~Nahm,
Nucl.\ Phys.\ B {\bf 135} (1978) 149.

\bibitem{wit}
B.~de Wit and H.~Nicolai,
Phys.\ Lett.\ B {\bf 148} (1984) 60.

\bibitem{nie}
P.~van Nieuwenhuizen and N.~P.~Warner,
Commun.\ Math.\ Phys.\  {\bf 99} (1985) 141.

\bibitem{wett2}
S.~Randjbar-Daemi and C.~Wetterich,
Phys.\ Lett.\ B {\bf 166} (1986) 65.

\bibitem{wett3}
C.~Wetterich,
Nucl.\ Phys.\ B {\bf 222} (1983) 20;
Nucl.\ Phys.\ B {\bf 223} (1983) 109.

\bibitem{Witten:ed.ux}
E.~Witten,
``Fermion Quantum Numbers In Kaluza--Klein Theory'',
The Proc. of Second Shelter Island Meeting (1983) 227.

\bibitem{avijit}
A.~Mukherjee and R.~Tabbash,
Eur.\ Phys.\ J.\ C {\bf 20} (2001) 193.

\bibitem{strathdee}
Personal notes of S.~Randjbar-Daemi;
J.~Strathdee,
``Symmetry In Kaluza-Klein Theory,''
IC-82/228
{Based on lectures 
given at Summer School 
'Supergravity 82', Trieste, Italy, Sep 6-18, 1982}.

\bibitem{percacci}
S.~Randjbar-Daemi and R.~Percacci,
Phys.\ Lett.\ B {\bf 117} (1982) 41.

\bibitem{Percacci:1983yi}
R.~Percacci and S.~Randjbar-Daemi,
J.\ Math.\ Phys.\  {\bf 24} (1983) 807.


\bibitem{sch3}
J.~Scherk and J.~H.~Schwarz,
Nucl.\ Phys.\ B {\bf 153} (1979) 61.

\bibitem{salam}
A.~Salam and J.~Strathdee,
Annals Phys.\  {\bf 141} (1982) 316.

\bibitem{Randjbar-Daemi:1984ij}
S.~Randjbar-Daemi, A.~Salam and J.~Strathdee,
Nucl.\ Phys.\ B {\bf 242} (1984) 447.


\bibitem{tooft}
G.~'t Hooft,
Phys.\ Rev.\ D {\bf 14} (1976) 3432


\bibitem{luc}
J.~F.~Luciani,
Nucl.\ Phys.\ B {\bf 135} (1978) 111.


\bibitem{witten3}
E.~Witten,
Nucl.\ Phys.\ B {\bf 186} (1981) 412.


\bibitem{sch2}
E.~Cremmer and J.~Scherk,
Nucl.\ Phys.\ B {\bf 103} (1976) 399.



\bibitem{Horvath:1977st}
Z.~Horvath, L.~Palla, E.~Cremmer and J.~Scherk,
Nucl.\ Phys.\ B {\bf 127} (1977) 57.

\bibitem{palla1}
Z.~Horvath and L.~Palla,
Nucl.\ Phys.\ B {\bf 142} (1978) 327.


\bibitem{Omero:1980vx}
C.~Omero and R.~Percacci,
Nucl.\ Phys.\ B {\bf 165} (1980) 351.

\bibitem{Freund:1980xh}
P.~G.~Freund and M.~A.~Rubin,
Phys.\ Lett.\ B {\bf 97} (1980) 233.

\bibitem{sss1}
S.~Randjbar-Daemi, A.~Salam and J.~Strathdee,
Phys.\ Lett.\ B {\bf 124} (1983) 345.


\bibitem{Schellekens:1984dm}
A.~N.~Schellekens,
Nucl.\ Phys.\ B {\bf 248} (1984) 706;

\bibitem{Schellekens:1985ks}
A.~N.~Schellekens,
Nucl.\ Phys.\ B {\bf 262} (1985) 661;
``Boson Mass Spectrum Of Kaluza-Klein Theories On Hyperspheres,''


\bibitem{Randjbar-Daemi:1983hi}
S.~Randjbar-Daemi, A.~Salam and J.~Strathdee,
Nucl.\ Phys.\ B {\bf 214} (1983) 491.

\bibitem{lawson}H.~B. Lawson and M.~Michelsohn, 
``Spin Geometry'', Princeton University Press, (1989).


\bibitem{Randjbar-Daemi:1983qa}
S.~Randjbar-Daemi, A.~Salam and J.~Strathdee,
Phys.\ Lett.\ B {\bf 132} (1983) 56.


\bibitem{L} A.~Lichnerowicz, ``Spineurs harmonique'', 
C.R.
Acad.\ Sci.\ Paris S\'{e}r
.\ A {\bf 257} (1963) 7. 

\bibitem{palla}
L.~Palla, ``Spontaneous Compactification'', 
in Proc. of 1978 Tokyo Conf. on High 
Energy Physics, Eds. S.~Homma et al (Phys. Society 
of Japan, 1979) 629.


\bibitem{yau} P.~Li and S.-T.~Yau, ``Estimates
of Eigenvalues of a Compact Riemannian Manifold'', 
in ``Geometry of the
Laplace Operator'', Proc. Sym. Pure Maths. (AMS) Vol. 36, (1980) 205.


\bibitem{Kaloper:2000jb}
N.~Kaloper, J.~March-Russell, G.~D.~Starkman and M.~Trodden,
Phys.\ Rev.\ Lett.\  {\bf 85} (2000) 928.

\bibitem{rula}
R.~Tabbash,
in the Proc. 
of Cairo International Conference on High Energy Physics (2001) .

\bibitem{hijazi}
O.~Hijazi, Commun. Math. Phys. {\bf 104}, (1986) 151;
C.~B\"{a}r, Math. Ann. {\bf 293}, (1992) 39.

\bibitem{oneil}
R.~L.~Bishop and B.~O'Neill, 
``Manifolds of Negative Curvature'',Trans. Amer. Math. Soc. 
{\bf 145} (1969) 1. 

\bibitem{randal2}
L.~Randall and R.~Sundrum,
Phys.\ Rev.\ Lett.\  {\bf 83} (1999) 4690.


\bibitem{lukas}
A.~Lukas, B.~A.~Ovrut, K.~S.~Stelle and D.~Waldram,
Phys.\ Rev.\ D {\bf 59} (1999) 086001.


\bibitem{cohen}
A.~G.~Cohen and D.~B.~Kaplan,
Phys.\ Lett.\ B {\bf 470} (1999) 52.


\bibitem{Gherghetta:2000qi}
T.~Gherghetta and M.~Shaposhnikov,
Phys.\ Rev.\ Lett.\  {\bf 85} (2000) 240;

\cite{Gherghetta:2000jf}
T.~Gherghetta, E.~Roessl and M.~E.~Shaposhnikov,
Phys.\ Lett.\ B {\bf 491} (2000) 353.


\bibitem{Olasagasti:2000gx}
I.~Olasagasti and A.~Vilenkin,
Phys.\ Rev.\ D {\bf 62} (2000) 044014

\bibitem{Dvali:2000ty}
G.~R.~Dvali,
hep-th/0004057.

\bibitem{Gherghetta:2000jf}
T.~Gherghetta, E.~Roessl and M.~Shaposhnikov,
Phys.\ Lett.\ B {\bf 491} (2000) 353.

\bibitem{Arkani-Hamed:2000dc}
N.~Arkani-Hamed and M.~Schmaltz,
Phys.\ Rev.\ D {\bf 61} (2000) 033005.


\bibitem{Chen:2000at}
J.~Chen, M.~A.~Luty and E.~Ponton,
JHEP {\bf 0009} (2000) 012.

\bibitem{Randjbar-Daemi:2000ft}
S.~Randjbar-Daemi and M.~Shaposhnikov,
Phys.\ Lett.\ B {\bf 491} (2000) 329

\bibitem{Giovannini:2001hh}
M.~Giovannini, H.~Meyer and M.~E.~Shaposhnikov,
``Warped compactification on Abelian vortex in six dimensions,''
hep-th/0104118.

\bibitem{Ito:2001gk}
M.~Ito,
``Various types of five dimensional warp factor and effective Planck  scale,''
hep-th/0109140.

\bibitem{Nath:2001kr}
P.~Nath,
Part.\ Nucl.\ Lett.\  {\bf 104} (2001) 7.

\bibitem{rubn}
V.~A.~Rubakov,
``Large and infinite extra dimensions: An Introduction,''
hep-ph/0104152.

\bibitem{Dick:2001sc}
R.~Dick,
Class.\ Quant.\ Grav.\  {\bf 18} (2001) R1.

\bibitem{pol}
J.~Polchinski,
``TASI lectures on D-branes,''
hep-th/9611050.

\bibitem{stelle}
A.~Lukas, B.~A.~Ovrut, K.~S.~Stelle and D.~Waldram,
Nucl.\ Phys.\ B {\bf 552} (1999) 246

\bibitem{behr}
K.~Behrndt and M.~Cvetic,
Phys.\ Lett.\ B {\bf 475} (2000) 253


\bibitem{Chamblin:2000cj}
A.~Chamblin and G.~W.~Gibbons,
Phys.\ Rev.\ Lett.\  {\bf 84} (2000) 1090.

\bibitem{kallosh}
R.~Kallosh and A.~D.~Linde,
JHEP {\bf 0002} (2000) 005

\bibitem{guk}
K.~Behrndt and S.~Gukov,
Nucl.\ Phys.\ B {\bf 580} (2000) 225

\bibitem{cvetic}
K.~Behrndt and M.~Cvetic,
Phys.\ Rev.\ D {\bf 61} (2000) 101901

\bibitem{behr2}
K.~Behrndt,
Phys.\ Lett.\ B {\bf 487} (2000) 30

\bibitem{agata}
A.~Ceresole, G.~Dall'Agata, R.~Kallosh and A.~Van Proeyen,
``Hypermultiplets, domain walls and supersymmetric attractors,''
hep-th/0104056.

\bibitem{card}
K.~Behrndt, G.~Lopes Cardoso and D.~Lust,
Nucl.\ Phys.\ B {\bf 607} (2001) 391.

\bibitem{lop}
G.~Lopes Cardoso, G.~Dall'Agata and D.~Lust,
``de Sitter BPS domain wall solutions in four- and five-dimensional  gauged supergravity,''
hep-th/0104156.


\bibitem{DeAzcarraga:1989vh}
J.~A.~De Azcarraga and P.~K.~Townsend,
Phys.\ Rev.\ Lett.\  {\bf 62} (1989) 2579.


\bibitem{Achucarro:1987nc}
A.~Achucarro, J.~M.~Evans, P.~K.~Townsend and D.~L.~Wiltshire,
Phys.\ Lett.\ B {\bf 198} (1987) 441.

\bibitem{batt}
R.~A.~Battye and B.~Carter,
Phys.\ Lett.\ B {\bf 509} (2001) 331.


\bibitem{is}
W.~Israel, Nuovo Cimento {\bf B44} (1966) 1.


\bibitem{mis}
C.~W.~Misner, K.~S.~Throne, and J.~A.~Wheeler, 
``Gravitation'', Freeman and Company, San Francisco (1973).

\bibitem{tk}
V.~A.~Berezin, V.~A.~Kuzmin and I.~I.~Tkachev,
Phys.\ Rev.\ D {\bf 36} (1987) 2919.

\bibitem{Bonjour:1999kz}
F.~Bonjour, C.~Charmousis and R.~Gregory,
Class.\ Quant.\ Grav.\  {\bf 16} (1999) 2427;
see also 
D.~Garfinkle and R.~Gregory,
Phys.\ Rev.\ D {\bf 41} (1990) 1889.

\bibitem{Csaki:2000fc}
C.~Csaki, J.~Erlich, T.~J.~Hollowood and Y.~Shirman,
Nucl.\ Phys.\ B {\bf 581} (2000) 309

\bibitem{der}
A.~De Rujula, A.~Donini, M.~B.~Gavela and S.~Rigolin,
Phys.\ Lett.\ B {\bf 482} (2000) 195.


\bibitem{Binetruy:2000wn}
P.~Binetruy, J.~M.~Cline and C.~Grojean,
Phys.\ Lett.\ B {\bf 489} (2000) 403.

\bibitem{Alexander:2001ic}
S.~Alexander, Y.~Ling and L.~Smolin,
``A thermal instability for positive brane cosmological constant in the  Randall-Sundrum cosmologies,''
hep-th/0106097.


\bibitem{Kobayashi:2001jd}
S.~Kobayashi, K.~Koyama and J.~Soda,
``Thick brane worlds and their stability,''
hep-th/0107025.


\bibitem{Bouhmadi-Lopez:2001rf}
M.~Bouhmadi-Lopez and A.~Zhuk,
``Comments on conformal stability of brane-world models,''
hep-th/0107227.

\bibitem{Karch:2001ct}
A.~Karch and L.~J.~Randall,
JHEP {\bf 0105}, 008 (2001)

\bibitem{Ida:2001qw}
D.~Ida, T.~Shiromizu and H.~Ochiai,
``Semiclassical instability of the brane-world: Randall-Sundrum bubbles,''
hep-th/0108056.

\bibitem{ran3}
J.~Lykken and L.~J.~Randall,
JHEP {\bf 0006} (2000) 014.

\bibitem{witten55}
E.~Witten,
``The cosmological constant from the viewpoint of string theory,''
hep-ph/0002297.


\bibitem{Gregory:2000jc}
R.~Gregory, V.~A.~Rubakov and S.~M.~Sibiryakov,
Phys.\ Rev.\ Lett.\  {\bf 84} (2000) 5928.

\bibitem{mald}
J.~Maldacena,
Adv.\ Theor.\ Math.\ Phys.\  {\bf 2} (1998) 231
[Int.\ J.\ Theor.\ Phys.\  {\bf 38} (1998) 1113].



\bibitem{horava}
P.~Horava and E.~Witten,
Nucl.\ Phys.\ B {\bf 475} (1996) 94

\bibitem{Bergshoeff:2001ii}
E.~Bergshoeff, R.~Kallosh and A.~Van Proeyen,
Fortsch.\ Phys.\  {\bf 49} (2001) 625. 


\bibitem{rand5}
A.~Karch and L.~J.~Randall,
JHEP {\bf 0106} (2001) 063


\bibitem{Gherghetta:2001kr}
T.~Gherghetta and A.~Pomarol,
Nucl.\ Phys.\ B {\bf 586} (2000) 141;
Nucl.\ Phys.\ B {\bf 602} (2001) 3.

\bibitem{Goldberger:2000un}
W.~D.~Goldberger and M.~B.~Wise,
Phys.\ Lett.\ B {\bf 475} (2000) 275.

\bibitem{Sasaki:2000mi}
M.~Sasaki, T.~Shiromizu and K.~Maeda,
Phys.\ Rev.\ D {\bf 62} (2000) 024008

\bibitem{Kogan:2000wc}
I.~I.~Kogan, S.~Mouslopoulos, A.~Papazoglou, G.~G.~Ross and J.~Santiago,
Nucl.\ Phys.\ B {\bf 584} (2000) 313.

\bibitem{Lesgourgues:2000tj}
J.~Lesgourgues, S.~Pastor, M.~Peloso and L.~Sorbo,
Phys.\ Lett.\ B {\bf 489} (2000) 411.


\bibitem{Hawking:2000kj}
S.~W.~Hawking, T.~Hertog and H.~S.~Reall,
Phys.\ Rev.\ D {\bf 62} (2000) 043501.


\bibitem{Kim:2000dp}
J.~E.~Kim and B.~Kyae,
Phys.\ Lett.\ B {\bf 486} (2000) 165
[hep-th/0005139].


\bibitem{Csaki:2000mp}
C.~Csaki, M.~Graesser, L.~J.~Randall and J.~Terning,
Phys.\ Rev.\ D {\bf 62} (2000) 045015.


\bibitem{Santos:2001vr}
M.~G.~Santos, F.~Vernizzi and P.~G.~Ferreira,
Phys.\ Rev.\ D {\bf 64} (2001) 063506.


\bibitem{Brevik:2000vt}
I.~Brevik, K.~A.~Milton, S.~Nojiri and S.~D.~Odintsov,
``Quantum (in)stability 
of a brane-world AdS(5) universe at nonzero  temperature,''
hep-th/0010205.

\bibitem{Santos:2001nt}
M.~G.~Santos, F.~Vernizzi and P.~G.~Ferreira,
``Isotropization and instability of the brane,''
hep-ph/0103112.

\bibitem{Ghoroku:2001pi}
K.~Ghoroku and A.~Nakamura,
``Stability of Randall-Sundrum brane-world and tachyonic scalar,''
hep-th/0103071.

\bibitem{Mannheim:2001zy}
P.~D.~Mannheim,
Phys.\ Rev.\ D {\bf 63} (2001) 024018.

\bibitem{Arkani-Hamed:2001ds}
N.~Arkani-Hamed, M.~Porrati and L.~J.~Randall,
JHEP {\bf 0108} (2001) 017.

\bibitem{Gibbons:1987wg}
G.~W.~Gibbons and D.~L.~Wiltshire,
Nucl.\ Phys.\ B {\bf 287} (1987) 717.

\bibitem{Dvali:1997xe}
G.~R.~Dvali and M.~A.~Shifman,
Phys.\ Lett.\ B {\bf 396} (1997) 64.

\bibitem{rangrav}
S.~B.~Giddings, E.~Katz and L.~J.~Randall,
JHEP {\bf 0003} (2000) 023.


\bibitem{locgrav}
A.~Karch and L.~J.~Randall,
Phys.\ Rev.\ Lett.\  {\bf 87} (2001) 061601


\bibitem{Hebecker:2001nv}
A.~Hebecker and J.~March-Russell,
Nucl.\ Phys.\ B {\bf 608} (2001) 375.


\bibitem{Garriga:2000yh}
J.~Garriga and T.~Tanaka,
Phys.\ Rev.\ Lett.\  {\bf 84} (2000) 2778.


\bibitem{Binetruy:2000ut}
P.~Binetruy, C.~Deffayet and D.~Langlois,
Nucl.\ Phys.\ B {\bf 565} (2000) 269.


\bibitem{Binetruy:2000hy}
P.~Binetruy, C.~Deffayet, U.~Ellwanger and D.~Langlois,
Phys.\ Lett.\ B {\bf 477} (2000) 285.


\bibitem{Flanagan:2000cu}
E.~E.~Flanagan, S.~H.~Tye and I.~Wasserman,
Phys.\ Rev.\ D {\bf 62} (2000) 044039.


\bibitem{Csaki:1999jh}
C.~Csaki, M.~Graesser, C.~Kolda and J.~Terning,
Phys.\ Lett.\ B {\bf 462} (1999) 34.


\bibitem{Cline:1999wi}
J.~M.~Cline, G.~D.~Moore and G.~Servant,
Phys.\ Rev.\ D {\bf 60} (1999) 105035.


\bibitem{Randjbar-Daemi:2000cr}
S.~Randjbar-Daemi and M.~E.~Shaposhnikov,
Phys.\ Lett.\ B {\bf 492} (2000) 361.


\bibitem{kars}
L.~H.~Karsten,
Phys.\ Lett.\ B {\bf 104} (1981) 315.


\bibitem{kap}
D.~B.~Kaplan,
Phys.\ Lett.\ B {\bf 288} (1992) 342.

\bibitem{Narayanan:1994sk}
R.~Narayanan and H.~Neuberger,
Nucl.\ Phys.\ B {\bf 412} (1994) 574;
H.~Neuberger,
Phys.\ Rev.\ Lett.\  {\bf 81} (1998) 4060.

\bibitem{Randjbar-Daemi:1995sq}
S.~Randjbar-Daemi and J.~Strathdee,
Phys.\ Lett.\ B {\bf 348} (1995) 543;
Phys.\ Rev.\ D {\bf 51} (1995) 6617;
Nucl.\ Phys.\ B {\bf 443} (1995) 386;
Nucl.\ Phys.\ B {\bf 466} (1996) 335;
Nucl.\ Phys.\ B {\bf 461} (1996) 305;
Phys.\ Lett.\ B {\bf 402} (1997) 134;
C.~D.~Fosco and S.~Randjbar-Daemi,
Phys.\ Lett.\ B {\bf 354} (1995) 383.

\bibitem{Neuberger:2001nb}
For a recent review see: H.~Neuberger,
``Exact chiral symmetry on the lattice,''
hep-lat/0101006.


\bibitem{Kaplan:1996pe}
D.~B.~Kaplan and M.~Schmaltz,
Phys.\ Lett.\ B {\bf 368} (1996) 44.


\bibitem{Cullen:2000ef}
S.~Cullen, M.~Perelstein and M.~E.~Peskin,
Phys.\ Rev.\ D {\bf 62} (2000) 055012.


\bibitem{Hall:1999mk}
L.~J.~Hall and D.~R.~Smith,
Phys.\ Rev.\ D {\bf 60} (1999) 085008.

\bibitem{Perez-Lorenzana:2000hf}
A.~Perez-Lorenzana,
``Theories in more than four dimensions,''
hep-ph/0008333.


\bibitem{Hannestad:2001jv}
S.~Hannestad and G.~Raffelt,
Phys.\ Rev.\ Lett.\  {\bf 87} (2001) 051301.


\bibitem{Besancon:2001wp}
M.~Besancon,
hep-ph/0106165.


\bibitem{Han:1999sg}
T.~Han, J.~D.~Lykken and R.~Zhang,
Phys.\ Rev.\ D {\bf 59} (1999) 105006.

\bibitem{Antoniadis:2001ts}
I.~Antoniadis and K.~Benakli,
hep-ph/0108174.


\bibitem{Barger:1999jf}
V.~Barger, T.~Han, C.~Kao and R.~J.~Zhang,
Phys.\ Lett.\ B {\bf 461} (1999) 34.


\bibitem{Adloff:2000dp}
C.~Adloff {\it et al.}  [H1 Collaboration],
Phys.\ Lett.\ B {\bf 479} (2000) 358.

\bibitem{Sigurdsson:2001wz}
S.~Sigurdsson,
``Experimental hints of gravity in large extra dimensions?,''
astro-ph/0107169.


\bibitem{Giudice:1999ck}
G.~F.~Giudice, R.~Rattazzi and J.~D.~Wells,
Nucl.\ Phys.\ B {\bf 544} (1999) 3.


\bibitem{Benakli:1999ur}
K.~Benakli and S.~Davidson,
Phys.\ Rev.\ D {\bf 60} (1999) 025004.


\bibitem{Rizzo:2000cv}
T.~G.~Rizzo,
hep-ph/0011139.

\bibitem{Davoudiasl:2000jd}
H.~Davoudiasl, J.~L.~Hewett and T.~G.~Rizzo,
Phys.\ Rev.\ Lett.\  {\bf 84} (2000) 2080;
Phys.\ Lett.\ B {\bf 473} (2000) 43;
Phys.\ Rev.\ D {\bf 63} (2001) 075004;
Phys.\ Lett.\ B {\bf 493} (2000) 135;
``Gravi-burst: Super-GZK cosmic rays from localized gravity,''
hep-ph/0010066.

\bibitem{Murayama:1996ec}
H.~Murayama and M.~E.~Peskin,
Ann.\ Rev.\ Nucl.\ Part.\ Sci.\  {\bf 46} (1996) 533


\bibitem{Peskin:1999hj}
M.~E.~Peskin,
``Physics goals of the linear collider,''
hep-ph/9910521.


\bibitem{Kolb:2000se}
E.~W.~Kolb and M.~S.~Turner,
Eur.\ Phys.\ J.\ C {\bf 15} (2000) 125.


\bibitem{Riess:1998cb}
A.~G.~Riess {\it et al.}  [Supernova Search Team Collaboration],
Astron.\ J.\  {\bf 116} (1998) 1009.

\bibitem{Perlmutter:1999np}
S.~Perlmutter {\it et al.}  [Supernova Cosmology Project Collaboration],
Astrophys.\ J.\  {\bf 517} (1999) 565.


\bibitem{Mirabelli:1999rt}
E.~A.~Mirabelli, M.~Perelstein and M.~E.~Peskin,
Phys.\ Rev.\ Lett.\  {\bf 82} (1999) 2236.


\bibitem{Cheung:1999fj}
K.~Cheung,
``Mini-review on collider signatures for extra dimensions,''
hep-ph/0003306.


\bibitem{Rizzo:2001sd}
T.~G.~Rizzo,
``Probes of universal extra dimensions at colliders,''
hep-ph/0106336.


\bibitem{Fairbairn:2001ct}
M.~Fairbairn,
Phys.\ Lett.\ B {\bf 508} (2001) 335.

\bibitem{Cullen:1999hc}
S.~Cullen and M.~Perelstein,
Phys.\ Rev.\ Lett.\  {\bf 83} (1999) 268.

\bibitem{Chodos:1980vk}
A.~Chodos and S.~Detweiler,
Phys.\ Rev.\ D {\bf 21} (1980) 2167.


\bibitem{Freund:1982pg}
P.~G.~Freund,
Nucl.\ Phys.\ B {\bf 209} (1982) 146.

\bibitem{Shafi:1983hj}
Q.~Shafi and C.~Wetterich,
Phys.\ Lett.\ B {\bf 129} (1983) 387.


\bibitem{Rubin:1983ap}
M.~A.~Rubin and B.~D.~Roth,
Nucl.\ Phys.\ B {\bf 226} (1983) 444.

\bibitem{Sahdev:1984fp}
D.~Sahdev,
Phys.\ Lett.\ B {\bf 137} (1984) 155.



\bibitem{Randjbar-Daemi:1984jz}
S.~Randjbar-Daemi, A.~Salam and J.~Strathdee,
Phys.\ Lett.\ B {\bf 135} (1984) 388.


\bibitem{Kolb:1984fm}
E.~W.~Kolb and R.~Slansky,
Phys.\ Lett.\ B {\bf 135} (1984) 378.


\bibitem{Alvarez:1983kt}
E.~Alvarez and M.~Belen Gavela,
Phys.\ Rev.\ Lett.\  {\bf 51} (1983) 931.


\bibitem{Maeda:1984fq}
K.~Maeda,
Phys.\ Lett.\ B {\bf 138} (1984) 269.


\bibitem{Krori:1990gw}
K.~D.~Krori, P.~Borgohain and K.~Das,
Gen.\ Rel.\ Grav.\  {\bf 22} (1990) 791.

\bibitem{Bleyer:1995ic}
U.~Bleyer, M.~Mohazzab and M.~Rainer,
``Dynamics of dimensions in factor space cosmology,''
gr-qc/9508035.

\bibitem{Gunther:1998qr}
U.~Gunther and A.~Zhuk,
Class.\ Quant.\ Grav.\  {\bf 15} (1998) 2025.


\bibitem{Shiu:1998pa}
G.~Shiu and S.~H.~Tye,
Phys.\ Rev.\ D {\bf 58} (1998) 106007.


\bibitem{Dvali:1999cn}
G.~R.~Dvali and A.~Y.~Smirnov,
Nucl.\ Phys.\ B {\bf 563} (1999) 63.


\bibitem{Kaloper:1999sw}
N.~Kaloper and A.~D.~Linde,
Phys.\ Rev.\ D {\bf 59} (1999) 101303.


\bibitem{Cline:1999ts}
J.~M.~Cline, C.~Grojean and G.~Servant,
Phys.\ Rev.\ Lett.\  {\bf 83} (1999) 4245.


\bibitem{Chung:2000zs}
D.~J.~Chung and K.~Freese,
Phys.\ Rev.\ D {\bf 61} (2000) 023511.

\bibitem{Dvali:1999pa}
G.~R.~Dvali and S.~H.~Tye,
Phys.\ Lett.\ B {\bf 450} (1999) 72.

\bibitem{Arkani-Hamed:1999kq}
N.~Arkani-Hamed, S.~Dimopoulos, N.~Kaloper and J.~March-Russell,
``Early inflation and cosmology in theories with sub-millimeter  dimensions,''
hep-ph/9903239.


\bibitem{Melchiorri:1999km}
A.~Melchiorri, F.~Vernizzi, R.~Durrer and G.~Veneziano,
Phys.\ Rev.\ Lett.\  {\bf 83} (1999) 4464.




\bibitem{Arkani-Hamed:2000gq}
N.~Arkani-Hamed, S.~Dimopoulos, N.~Kaloper and J.~March-Russell,
Nucl.\ Phys.\ B {\bf 567} (2000) 189.

\bibitem{Riotto:2000kn}
A.~Riotto,
Phys.\ Rev.\ D {\bf 61} (2000) 123506.


\bibitem{Chung:2000xg}
D.~J.~Chung and K.~Freese,
Phys.\ Rev.\ D {\bf 62} (2000) 063513.

\bibitem{Delgado:2000sv}
A.~Delgado, A.~Pomarol and M.~Quiros,
JHEP {\bf 0001} (2000) 030.

\bibitem{Akama:2000vz}
K.~Akama and T.~Hattori,
Mod.\ Phys.\ Lett.\ A {\bf 15} (2000) 2017.

\bibitem{Hannestad:2001nq}
S.~Hannestad,
Phys.\ Rev.\ D {\bf 64} (2001) 023515.

\bibitem{Starkman:2001xu}
G.~D.~Starkman, D.~Stojkovic and M.~Trodden,
Phys.\ Rev.\ D {\bf 63} (2001) 103511;
``Homogeneity, flatness and 'large' extra dimensions,''
hep-th/0106143.

\bibitem{Steigman:1979kw}
G.~Steigman,
Ann.\ Rev.\ Nucl.\ Part.\ Sci.\  {\bf 29} (1979) 313.


\bibitem{Kolb:1986sj}
E.~W.~Kolb, M.~J.~Perry and T.~P.~Walker,
Phys.\ Rev.\ D {\bf 33} (1986) 869.

\bibitem{weinbc}
S.~Weinberg, ``Gravitation and Cosmology'', 
John Wiley \& Sons Inc. (1972).


\bibitem{Kaloper:1999sm}
N.~Kaloper,
Phys.\ Rev.\ D {\bf 60} (1999) 123506.

\bibitem{Candelas:1984ae}
P.~Candelas and S.~Weinberg,
Nucl.\ Phys.\ B {\bf 237} (1984) 397.


\bibitem{Krasnikov:1987jj}
N.~V.~Krasnikov,
Phys.\ Lett.\ B {\bf 193} (1987) 37.


\bibitem{Taylor:1990wr}
T.~R.~Taylor,
Phys.\ Lett.\ B {\bf 252} (1990) 59.

\bibitem{Kikkawa:1985qc}
K.~Kikkawa, T.~Kubota, S.~Sawada and M.~Yamasaki,
Nucl.\ Phys.\ B {\bf 260} (1985) 429;
Phys.\ Lett.\ B {\bf 144} (1984) 365.

\bibitem{Okada:1985cv}
Y.~Okada,
Phys.\ Lett.\ B {\bf 150} (1985) 103.


\bibitem{Koikawa:1985et}
T.~Koikawa and M.~Yoshimura,
Phys.\ Lett.\ B {\bf 150} (1985) 107.

\bibitem{Gleiser:1985bs}
M.~Gleiser, S.~Rajpoot and J.~G.~Taylor,
Annals Phys.\  {\bf 160} (1985) 299.

\bibitem{Okada:1986sf}
Y.~Okada,
Nucl.\ Phys.\ B {\bf 264} (1986) 197.

\bibitem{Accetta:1986vq}
F.~S.~Accetta, M.~Gleiser, R.~Holman and E.~W.~Kolb,
Nucl.\ Phys.\ B {\bf 276} (1986) 501.

\bibitem{Maeda:1986bq}
K.~Maeda,
Class.\ Quant.\ Grav.\  {\bf 3} (1986) 233;
Class.\ Quant.\ Grav.\  {\bf 3} (1986) 651;
Phys.\ Lett.\ B {\bf 186} (1987) 33.

\bibitem{Sokolowski:1987dh}
L.~M.~Sokolowski and Z.~A.~Golda,
Phys.\ Lett.\ B {\bf 195} (1987) 349.


\bibitem{Szydlowski:1988ct}
M.~Szydlowski,
Phys.\ Lett.\ B {\bf 215} (1988) 711.

\bibitem{Berezin:1989zz}
V.~A.~Berezin, G.~Domenech, M.~L.~Levinas, C.~O.~Lousto and N.~D.~Umerez,
Gen.\ Rel.\ Grav.\  {\bf 21} (1989) 1177.


\bibitem{Kubyshin:1989mj}
Y.~A.~Kubyshin, V.~A.~Rubakov and I.~Tkachev,
Int.\ J.\ Mod.\ Phys.\ A {\bf 4} (1989) 1409.

\bibitem{Demianski:1990tb}
M.~Demianski, M.~Szydlowski and J.~Szczesny,
Gen.\ Rel.\ Grav.\  {\bf 22} (1990) 1217.

\bibitem{Amendola:1990iw}
L.~Amendola, E.~W.~Kolb, M.~Litterio and F.~Occhionero,
Phys.\ Rev.\ D {\bf 42} (1990) 1944.

\bibitem{Holman:1991hg}
R.~Holman, E.~W.~Kolb, S.~L.~Vadas and Y.~Wang,
Phys.\ Rev.\ D {\bf 43} (1991) 995.

\bibitem{Bleyer:1995uv}
U.~Bleyer and A.~Zhuk,
Class.\ Quant.\ Grav.\  {\bf 12} (1995) 89.

\bibitem{Carugno:1996wn}
E.~Carugno, M.~Litterio, F.~Occhionero and G.~Pollifrone,
Phys.\ Rev.\ D {\bf 53} (1996) 6863.

\bibitem{Banks:1994sg}
T.~Banks and M.~Dine,
Phys.\ Rev.\ D {\bf 50} (1994) 7454.


\bibitem{Arkani-Hamed:2001kx}
N.~Arkani-Hamed, S.~Dimopoulos and J.~March-Russell,
Phys.\ Rev.\ D {\bf 63} (2001) 064020.

\bibitem{Dienes:1999hx}
K.~R.~Dienes, E.~Dudas, T.~Gherghetta and A.~Riotto,
Nucl.\ Phys.\ B {\bf 543} (1999) 387.


\bibitem{Huey:2000jx}
G.~Huey, P.~J.~Steinhardt, B.~A.~Ovrut and D.~Waldram,
Phys.\ Lett.\ B {\bf 476} (2000) 379.


\bibitem{Goldberger:1999uk}
W.~D.~Goldberger and M.~B.~Wise,
Phys.\ Rev.\ Lett.\  {\bf 83} (1999) 4922.

\bibitem{Mersini:2001ee}
L.~Mersini,
Mod.\ Phys.\ Lett.\ A {\bf 16} (2001) 1583.

\bibitem{Gunther:2000jj}
U.~Gunther and A.~Zhuk,
Phys.\ Rev.\ D {\bf 61} (2000) 124001;
Class.\ Quant.\ Grav.\  {\bf 18} (2001) 1441.

\bibitem{Garriga:2001jb}
J.~Garriga, O.~Pujolas and T.~Tanaka,
Nucl.\ Phys.\ B {\bf 605} (2001) 192.

\bibitem{Mahanta:2000ig}
U.~Mahanta,
``Radion stabilization in the Randall-Sundrum model with quadratic and  quartic potentials,''
hep-ph/0006350.

\bibitem{KalyanaRama:2000dz}
S.~Kalyana Rama,
Phys.\ Lett.\ B {\bf 495} (2000) 176.

\bibitem{Gen:2000nu}
U.~Gen and M.~Sasaki,
``Radion on the de Sitter brane,''
gr-qc/0011078.


\bibitem{Gibbons:2001tf}
G.~W.~Gibbons, R.~Kallosh and A.~D.~Linde,
JHEP {\bf 0101} (2001) 022.


\bibitem{Bae:2000pd}
S.~Bae and H.~S.~Lee,
``Constraints on the radion from vacuum stability in the Randall-Sundrum  theory,''
hep-ph/0011275.

\bibitem{Chacko:2001em}
Z.~Chacko and P.~J.~Fox,
Phys.\ Rev.\ D {\bf 64} (2001) 024015.

\bibitem{Randjbar-Daemi:1985fs}
S.~Randjbar-Daemi and M.~H.~Sarmadi,
Phys.\ Lett.\ B {\bf 151} (1985) 343.

\bibitem{Feruglio:2001nf}
F.~Feruglio,
``Grand unification in extra dimensions and proton decay,''
hep-ph/0105321.

\bibitem{Altarelli:2001qj}
G.~Altarelli and F.~Feruglio,
Phys.\ Lett.\ B {\bf 511} (2001) 257.


\bibitem{Aranda:2001ma}
A.~Aranda and C.~D.~Carone,
Phys.\ Rev.\ D {\bf 63} (2001) 075012.

\bibitem{Adams:2001za}
F.~C.~Adams, G.~L.~Kane, M.~Mbonye and M.~J.~Perry,
Int.\ J.\ Mod.\ Phys.\ A {\bf 16} (2001) 2399.


\bibitem{Huber:2001ie}
S.~J.~Huber and Q.~Shafi,
Phys.\ Lett.\ B {\bf 498} (2001) 256.


\bibitem{Shiozawa:1998si}
M.~Shiozawa {\it et al.}  [Super-Kamiokande Collaboration],
Phys.\ Rev.\ Lett.\  {\bf 81} (1998) 3319.


\bibitem{Masiero:2000fr}
A.~Masiero, M.~Peloso, L.~Sorbo and R.~Tabbash,
Phys.\ Rev.\ D {\bf 62} (2000) 063515.

\bibitem{Mirabelli:2000ks}
E.~A.~Mirabelli and M.~Schmaltz,
Phys.\ Rev.\ D {\bf 61} (2000) 113011.

\bibitem{Arkani-Hamed:2000za}
N.~Arkani-Hamed, Y.~Grossman and M.~Schmaltz,
Phys.\ Rev.\ D {\bf 61} (2000) 115004.


\bibitem{Jackiw:1976fn}
R.~Jackiw and C.~Rebbi,
Phys.\ Rev.\ D {\bf 13} (1976) 3398.


\bibitem{Weinberg:1981eu}
E.~J.~Weinberg,
Phys.\ Rev.\ D {\bf 24} (1981) 2669.


\bibitem{Nielsen:1973cs}
H.~B.~Nielsen and P.~Olesen,
Nucl.\ Phys.\ B {\bf 61} (1973) 45.



\bibitem{Kolb:1996jt}
E.~W.~Kolb, A.~D.~Linde and A.~Riotto,
Phys.\ Rev.\ Lett.\  {\bf 77} (1996) 4290.



\bibitem{Chung:1999rq}
D.~J.~Chung, E.~W.~Kolb and A.~Riotto,
Phys.\ Rev.\ D {\bf 60} (1999) 063504.


\bibitem{Giudice:2001ex}
G.~F.~Giudice, E.~W.~Kolb and A.~Riotto,
Phys.\ Rev.\ D {\bf 64} (2001) 023508.

\bibitem{Nanopoulos:1979gx}
D.~V.~Nanopoulos and S.~Weinberg,
Phys.\ Rev.\ D {\bf 20} (1979) 2484.

\bibitem{Mazumdar:2001nw}
A.~Mazumdar and A.~Perez-Lorenzana,
``Affleck-Dine baryogensis in large extra dimensions,''
hep-ph/0103215;
R.~Allahverdi, K.~Enqvist, A.~Mazumdar and A.~Perez-Lorenzana,
``Baryogenesis in theories with large extra spatial dimensions,''
hep-ph/0108225.


\bibitem{Ibanez:1999it}
L.~E.~Ibanez and F.~Quevedo,
JHEP {\bf 9910} (1999) 001.


\bibitem{Dvali:2001qr}
G.~Dvali, S.~Randjbar-Daemi and R.~Tabbash,
%
``The origin of spontaneous 
symmetry breaking in theories with large  extra dimensions,''
hep-ph/0102307.

\bibitem{Coleman:1973jx}
S.~R.~Coleman and E.~Weinberg,
Phys.\ Rev.\ D {\bf 7} (1973) 1888.

\bibitem{barbieri}
 R.\ Barbieri,
L.\ J.\ Hall and Y.\ Nomura,
Phys.\ Rev.\ D {\bf 63} (2001) 105007.
For further explanation see 
A.~Masiero, C.~A.~Scrucca, M.~Serone and L.~Silvestrini,
``Non-local symmetry breaking in Kaluza-Klein theories,''
hep-ph/0107201.


\bibitem{st}
M.~Shaposhnikov and P.~Tinyakov,
Phys.\ Lett.\ B {\bf 515} (2001) 442.


\bibitem{arkani}
N.~Arkani-Hamed, A.~G.~Cohen and H.~Georgi,
Phys.\ Lett.\ B {\bf 513} (2001) 232.

\bibitem{Hall:2001zb}
L.~J.~Hall, Y.~Nomura and D.~R.~Smith,
``Gauge-Higgs unification in higher dimensions,''
hep-ph/0107331;

\bibitem{Antoniadis:2001cv}
I.~Antoniadis, K.~Benakli and M.~Quiros,
``Finite Higgs mass without supersymmetry,''
hep-th/0108005.


\bibitem{Hatanaka:1998yp}
N.~V.~Krasnikov,
Phys.\ Lett.\ B {\bf 273} (1991) 246.

\bibitem{hata2}
H.~Hatanaka, T.~Inami and C.~S.~Lim,
Mod.\ Phys.\ Lett.\ A {\bf 13} (1998) 2601.

\bibitem{manton}
N.~S.~Manton,
Nucl.\ Phys.\ B {\bf 158} (1979) 141.

\bibitem{bachas}
C.~Bachas,
``A way to break supersymmetry,''
hep-th/9503030;
``Magnetic supersymmetry breaking,''
hep-th/9509067.


\bibitem{b2}
R.~Blumenhagen, L.~G\"{o}rlich, B.~K\"{o}rs and D.~L\"{u}st,
JHEP {\bf 0010}, (2000) 006;
Fortsch.\ Phys.\ {\bf 49}, (2001) 591.



\bibitem{Sakamoto:2001gn}
M.~Sakamoto and S.~Tanimura,
``Spontaneous breaking of the 
C, P, and rotational symmetries 
by topological defects in extra two dimensions,''
hep-th/0108208.

\bibitem{hata3}
H.~Hatanaka,
Prog.\ Theor.\ Phys.\  {\bf 102} (1999) 407.


\bibitem{host}Y.~Hosotani,
Annals Phys.\  {\bf 190} (1989) 233.


\bibitem{gs}
M.\ Green and J.\ Schwarz,
Phys. Lett {\bf 149B}\ (1984) 117.


\bibitem{sss3}
S.\ Randjbar-Daemi, Abdus Salam, E.\ Sezgin, and
J.\ A.\ Strathdee, Phys. Lett {\bf 151B}\ (1985) 351.


\bibitem{witten2}
E.\ Witten, Phys. Lett {\bf 117B}\ (1982) 324.



\bibitem{witten5}
E.~Witten,
Nucl.\ Phys.\ B {\bf 195} (1982) 481.


\bibitem{Schellekens:1985iu}
A.~N.~Schellekens,
Nucl.\ Phys.\ B {\bf 250} (1985) 252.

\bibitem{Frieman:1985xs}
J.~A.~Frieman and E.~W.~Kolb,
Phys.\ Rev.\ Lett.\  {\bf 55} (1985) 1435.

\bibitem{ssez}
``Supergravities in Diverse Dimensions'', v.2 ,
(Edited by 
M.~Abdus Salam and E.~Sezgin), 
Elsevier Science Publishers B.V., and 
World Sientific Publishing CO. PTE. LTD. (1989), 
and references therein.


\bibitem{Dienes:1998vh}
K.~R.~Dienes, E.~Dudas and T.~Gherghetta,
Phys.\ Lett.\ B {\bf 436} (1998) 55;
Nucl.\ Phys.\ B {\bf 537} (1999) 47.

















\end{thebibliography}
\end{document}